\documentclass[%
nofootinbib,
 amsmath,amssymb,
 aps,
pre,
twocolumn,
]{revtex4-2}

\usepackage{newtxtext,newtxmath}  
\usepackage{graphicx}
\usepackage{booktabs}
\usepackage{dcolumn} 
\usepackage{bm}
\usepackage{microtype}
\usepackage{xcolor}
\usepackage{float}
\usepackage{physics} 
\definecolor{darkblue}{rgb}{0,0,0.6}
\definecolor{apsblue}{rgb}{0.18,0.19,0.57}
\usepackage[colorlinks,linkcolor=apsblue,citecolor=apsblue,urlcolor=apsblue]{hyperref}

\begin{document}

\title{
  Response of a Model Glass to Athermal Quasistatic Pinching
}

\author{Takumi Nagasawa}
\author{Kirsten Martens}
\author{Jean-Louis Barrat}
\author{Misaki Ozawa}

\affiliation{LIPhy, Université Grenoble Alpes, CNRS, 38000 Grenoble, France}

\date{\today}

\begin{abstract}
We numerically investigate the mechanical response of amorphous solids to localized force dipoles, referred to as pinching, using athermal quasistatic simulations of model glasses with varying degrees of stability. We employ a control parameter, the imposed extension, corresponding to the increase in rest length between the two pinched particles relative to their initial separation. Increasing this extension allows us to continuously tune the system from the elastic to the plastic regime. For small extensions, the response remains elastic. In this regime, the displacement field induced by pinching is well described by linear elasticity and exhibits a long-range power-law decay consistent with dipolar forcing. Averaged responses display anisotropic, quadrupolar-like patterns, with quantitative agreement between simulations and analytical predictions. This continuum description remains valid down to particle-scale distances. As the imposed extension increases, the response becomes plastic. Pinching can then trigger either localized or system-spanning rearrangements, depending on glass stability. Well-annealed glasses exhibit localized plastic events, whereas poorly annealed systems display delocalized cascades. We introduce a method to extract the principal axis of plastic deformation and analyze the associated displacement fields and plastic activity. Overall, our results demonstrate that pinching provides a minimal local probe of amorphous solids. The resulting response, governed by both glass stability and imposed extension, offers insight into the interplay between elasticity, elementary rearrangements, and the emergence of collective plasticity.
\end{abstract}

\maketitle

\section{Introduction}

Amorphous solids such as metallic glasses, colloids, granular materials, and biological tissues are characterized by a disordered packing of their constituent particles. Despite the huge variation in particle size, from nanometers to centimeters, and the fact that their microscopic interactions are completely different, their mechanical behavior under external loading exhibits remarkably similar features~\cite{argon1979plastic, rodney2011modeling, bonn2017yield}. These universal aspects motivate the search for theoretical descriptions of the mechanical response of amorphous solids ~\cite{baret2002extremal} and yield stress fluids~\cite{picard2005slow}, for a detailed review see Ref.~\cite{RevModPhys.90.045006}.

When an amorphous solid is deformed, plastic events occur as localized, irreversible particle rearrangements~\cite{argon1979plastic}. These events are accompanied by a redistribution of stress in the form of long-range elastic fields, often described by Eshelby’s solution of linear elasticity~\cite{eshelby1957determination, picard2004elastic}. Because these elastic interactions extend over large distances, one plastic event can trigger others, which in turn may induce further events. In this way, a cascade or avalanche of plasticity can develop~\cite{talamali2011avalanches}. Under sufficiently large deformations, such avalanches may span the entire system, leading to the yielding of the bulk material~\cite{berthier2025yielding}. Thus, localized rearrangements together with their associated Eshelby fields constitute the elementary building blocks of the mechanical response of amorphous solids.
Most previous studies have focused on global deformation modes such as shear~\cite{maloney2006amorphous,lemaitre2007plastic,tsamados2009local, puosi2016plastic, PhysRevE.89.042302, albaret2016mapping}, tensile loading~\cite{richard2021brittle}, or compression~\cite{ding2024anomalous}. In bulk samples, these loading modes generally produce shear deformation~\cite{argon1979plastic}, which is one of the most fundamental modes of deformation and is relevant to many practical applications.

\begin{figure}[t]
    \centering
    \includegraphics[width=1.0\linewidth]{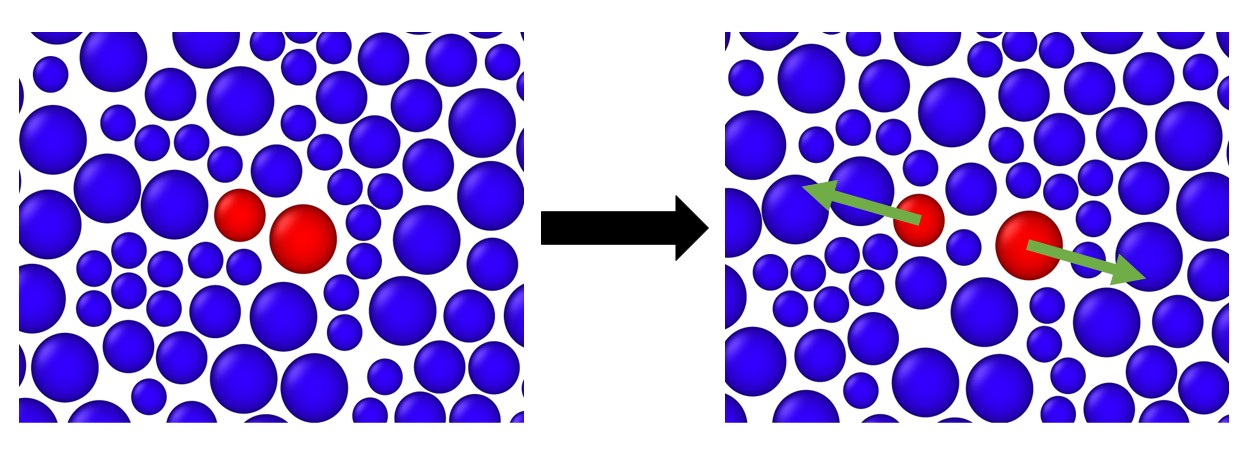}
    \caption{Schematic illustration of the pinching simulation. A randomly selected pair of particles (shown in red) is subjected to a force dipole. The two particles, initially in contact and forming a local cage (left), are separated by pinching, leading to cage breaking in the form of a localized plastic event (right).
}
    \label{fig:placeholder}
\end{figure}

More recently, another type of deformation process has attracted attention: the application of a force dipole to two contacting particles~\cite{lerner2018characteristic,lerner2018characteristic, rainone2020pinching, rainone2020statistical, ji2025role,richard2025rigidity}, as illustrated in Fig.~\ref{fig:placeholder}. For weak dipole forces, the system responds elastically, producing only a small relative displacement between the two particles. By contrast, sufficiently strong dipole forces can break the local cage structure of the glass, triggering a localized plastic rearrangement. This "pinching" process generates displacement fields similar to those predicted by linear elasticity.
Previous studies have mainly focused either on the purely elastic response~\cite{lerner2018characteristic,rainone2020pinching,rainone2020statistical,kapteijns2021does} or on the single plastic event directly induced by the applied dipole~\cite{ji2025role}. For instance, pinching has been used to probe local softness through its elastic response~\cite{rainone2020pinching}, and it has also been proposed as a way to identify localized excitations in supercooled liquids at low temperatures~\cite{ji2025role}.
Beyond amorphous solids, a related force-dipole protocol has recently been applied to supercooled liquids to probe the breakdown of the Navier--Stokes description and its connection to growing dynamical heterogeneity~\cite{maeda2025flow}, further illustrating the broad utility of force dipoles as local probes of glassy systems.

Another important aspect of amorphous solids is the dependence of their stability on the preparation protocol~\cite{rodney2011modeling,patinet2016connecting,ozawa2018random}. Since amorphous solids are nonequilibrium materials, their properties are strongly influenced by how they are prepared. Well-annealed systems, such as slowly cooled metallic glasses, typically settle into deeper energy minima and are more resistant to shear deformation~\cite{fielding2000aging, berthier2025yielding}. In these materials, plastic events organize into large avalanches only under substantial external loading~\cite{ singh2020brittle, berthier2025yielding}. In contrast, poorly annealed systems, such as rapidly quenched glasses or colloidal pastes, are less stable: their plastic events are less correlated in the transient dynamics towards failure or flow and lead to more ductile macroscopic stress-strain relations~\cite{divoux2024ductile}. It is therefore natural to investigate how pinching-induced rearrangements depend on the stability of the material~\cite{rainone2020pinching,kapteijns2021does,ji2025role}.
However, how the response to pinching evolves from the elastic to the plastic regime remains largely unexplored. In particular, it remains unclear whether a local pinching event can trigger additional plastic rearrangements away from the pinching region, and how this process depends on glass stability. 

In this paper, we investigate the elementary pinching process in detail using linear elasticity theory and numerical simulations. Within the framework of linear elasticity, we theoretically establish the connection between pinching, modeled as a single force dipole, and a shear transformation, modeled as a pair of force dipoles~\cite{pica2024local}. We then numerically analyze the spatial structure of the displacement fields induced by pinching and examine their dependence on glass stability and system size, with particular attention to finite-size effects arising from the long-range nature of elasticity. The numerical results are compared quantitatively with the predictions of linear elasticity. In addition, we assess whether the initial pinching event can induce further plastic events.
Our results provide insight into the fundamental role of pinching as a local probe of amorphous solids. We also discuss possible connections to elementary processes in glassy dynamics, such as localized excitations and structural changes induced by X-ray irradiation~\cite{ruta2017hard,dallari2023stochastic,PhysRevX.13.041031}.

The paper is organized as follows. Predictions from linear elasticity theory are presented in Sec.~\ref{sec:theory}. The simulation model and numerical methods are described in Sec.~\ref{sec:methods}. We then present the simulation results for the elastic and plastic responses to pinching in Sec.~\ref{sec:simulation}. Finally, we summarize our findings and discuss their implications in Sec.~\ref{sec:conclusion}.

\section{Linear elasticity theory} 
\label{sec:theory}

In the general elasto-plastic description of deformed glasses, the external driving is assumed to induce local irreversible deformations, to which the surrounding medium responds by an elastic deformation that can be described within the framework of linear elastic theory. In this section, we recall briefly this general framework, apply it  to the particular case of a local "pinching" event, and compare this situation to the more frequently studied case of local shear transformations~\cite{picard2004elastic}. The predictions from linear elasticity for a pinching event will serve as a reference for direct 
comparison with molecular dynamics simulations, as presented in the 
following sections.

\subsection{Navier-Lamé equations}

In isotropic elastic materials, the elastic displacement $\mathbf{u}(\mathbf{x})$  in response to an applied force density $\mathbf{f}(\mathbf{x})$ follows the Navier Lam\'e equations~\cite{landau2012theory},
\begin{equation}
    (\lambda + \mu)\,
    \frac{\partial^2 u_j}{\partial x_i \partial x_j}
    + \mu \,
    \frac{\partial^2 u_i}{\partial x_j \partial x_j}
    + f_i = 0 .
    \label{eq:NL_equation}
\end{equation}
where $\lambda$ and $\mu$ are the usual Lam\'e coefficients, and indexes $i$,$j$ refer to Cartesian coordinates. Summation over repeated indexes is used throughout. 

The corresponding strain and stress tensors are expressed as 
\begin{equation}
    \epsilon_{ij} = \frac{1}{2}
    \left(
    \frac{\partial u_i}{\partial x_j}
    +
    \frac{\partial u_j}{\partial x_i}
    \right).
    \label{eq:def_strain}
\end{equation}
and
\begin{equation}
    \sigma_{ij}
    = \lambda \, \epsilon_{kk} \, \delta_{ij}
    + 2 \mu \, \epsilon_{ij},
    \label{eq:Hooke}
\end{equation}
respectively.

The general solution of Eq.~(\ref{eq:NL_equation}) involves introducing the corresponding Green's function
$G_{ij}(\mathbf{x})$ defined  through
\begin{equation}
    u_i(\mathbf{x})
    = \int d\mathbf{x}' \,
    G_{ij}(\mathbf{x}-\mathbf{x}')
    f_j(\mathbf{x}').
    \label{eq:Green_def}
\end{equation}
This Green's function is most conveniently expressed in Fourier space and reads
\begin{equation}
    \tilde G_{ij}(\mathbf q)
    =
    \frac{1}{\mu q^2}
    \left(
        \delta_{ij}
        -
        \frac{\lambda+\mu}{\lambda+2\mu}
        \frac{q_i q_j}{q^2}
    \right) 
    \label{eq:G_q_d}
\end{equation}
for wavevectors ${\bf q} \neq 0$.

While this solution is valid for any spatial dimension, the corresponding real space expression depends on the dimension. We now specialize to the two dimensional case for which the inversion to real space yields: 
\begin{equation}
    G_{ij}(\mathbf x)
    =
    \frac{1}{8\pi\mu}
    \left[
        -(3-\nu)\delta_{ij}
        \ln \left( \frac{r}{r_0} \right)
        + (1+\nu)
        \frac{x_i x_j}{r^2}
    \right].
\end{equation}
In this expression, $r_0$ is a short distance cutoff that does not influence the final results, and we have introduced the Poisson ratio $\nu= \lambda/(\lambda+2\mu)$. 
In what follows, we will frequently use the spatial derivative of
$G_{ij}(\mathbf x)$,
\begin{eqnarray}
\frac{\partial G_{ij}(\mathbf x)}{\partial x_k}
&=&
\frac{1}{8\pi \mu}
\left[
-(3-\nu)
\frac{x_k \delta_{ij}}{r^2}
\right. \nonumber \\
&&
\left.
+
(1+\nu)
\frac{(x_i \delta_{jk} + x_j \delta_{ik}) r^2
- 2 x_i x_j x_k}{r^4}
\right].
\label{eq:derivative_G}
\end{eqnarray}

\subsection{Pinching and shear transformation events}

In the following, we will be interested in the far field response of the elastic medium to  force distributions that are strongly localized in space, with a zero total force. Under these conditions, a general approach consists in introducing a multipolar expansion of the response
\begin{eqnarray}
u_i(\mathbf x)
&=&
\sum_{n=0}^\infty
\frac{(-1)^n}{n!}
\frac{\partial^n G_{ij}(\mathbf x)}
{\partial x_{i_1}\partial x_{i_2}\cdots \partial x_{i_n}}
\int d\mathbf x' 
f_j(\mathbf x') \, x_{i_1}' x_{i_2}' \cdots x_{i_n}'
\nonumber \\
&=&
G_{ij}(\mathbf x)
\int d\mathbf x' f_j(\mathbf x')
-
\frac{\partial G_{ij}(\mathbf x)}{\partial x_k}
\int d\mathbf x'  f_j(\mathbf x') \, x_k'
+ \cdots . \nonumber \\
\label{eq:u_expansion}
\end{eqnarray}
\label{sec:pinching_standard}

Limiting the expansion to the second term, we introduce the dipole moment tensor 
$P_{ij}$,
\begin{equation}
    P_{ij}
    =
    \int d\mathbf x' \,
    f_i(\mathbf x') \, x_j',
    \label{eq:def_dipole_moment}
\end{equation}
which, if nonzero, will dominate the far field response, noting that the first term vanishes because the total force is zero.

\begin{figure}[htbp]
    \centering
    \includegraphics[width=0.6\linewidth]{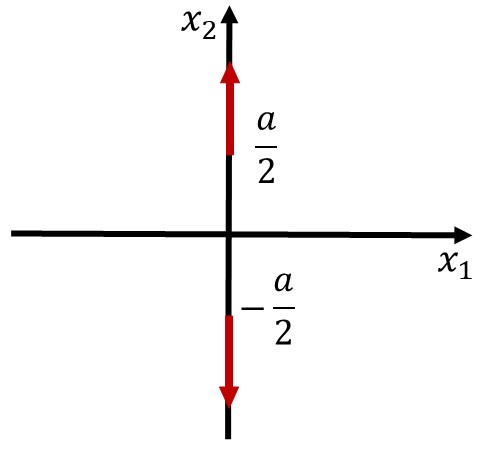}
\caption{Schematic representation of pinching. Two equal and opposite forces, shown by red arrows, are separated by a distance~$a$.}    \label{fig:pinching}
\end{figure}

We now consider the specific cases of pinching and of shear transformations. Specifically, we define pinching as consisting of two point forces of opposite sign,
separated by a distance $a$ along the $x_2$ direction and of
magnitude $F$.
The corresponding body-force density can be written as
\begin{equation}
    f_i(\mathbf x)
    =
    F \delta_{i2} \left[
        \delta(x_1)\delta\!\left(x_2-\frac{a}{2}\right)
        -
        \delta(x_1)\delta\!\left(x_2+\frac{a}{2}\right)
    \right] .
\end{equation}
and is illustrated schematically 
in Fig.~\ref{fig:pinching}.

The corresponding dipole force tensor is 
\begin{equation}
    \mathrm P =
    \begin{bmatrix}
        P_{11} & P_{12} \\
        P_{21} & P_{22}
    \end{bmatrix}
    =
    \begin{bmatrix}
        0 & 0 \\
        0 & aF
    \end{bmatrix},
    \label{eq:P_pinching}
\end{equation}
and the corresponding far field solution for the displacement field reads:
\begin{equation}
u_i(\mathbf x) = 
-
\frac{\partial G_{i2}(\mathbf x)}{\partial x_2} a F
\label{eq:u_far_field}
\end{equation}
with the explicit expressions
for  $u_1$ and $u_2$ given by
\begin{eqnarray}
u_1(x_1,x_2)
&=&
\frac{(1+\nu)}{8\pi\mu}
\frac{x_1(x_2^2-x_1^2)}{r^4}\, aF,
\nonumber \\
u_2(x_1,x_2)
&=&
\frac{x_2}{8\pi\mu}
\frac{(3-\nu)x_2^2-(3\nu-1)x_1^2}{r^4}\, aF,
\label{eq:u_pin_theory_cartesian}
\end{eqnarray}
where $r=\sqrt{x_1^2+x_2^2}$.

It is also convenient to express the result using polar coordinates,
$x_1=r\cos\theta$ and $x_2=r\sin\theta$, yielding:
\begin{eqnarray}
u_1(r,\theta)
&=&
\frac{(1+\nu)}{8\pi\mu}
\frac{\cos\theta}{r}
(2\sin^2\theta-1)\, aF,
\nonumber \\
u_2(r,\theta)
&=&
\frac{\sin\theta}{8\pi\mu}
\frac{2(1+\nu)\sin^2\theta-(3\nu-1)}{r}\, aF .
\label{eq:u_pin_theory_polar}
\end{eqnarray}
These  theoretical predictions will be compared with
molecular simulation results in Sec.~\ref{sec:simulation}.

\begin{figure}[htbp]
    \centering
    \includegraphics[width=0.6\linewidth]{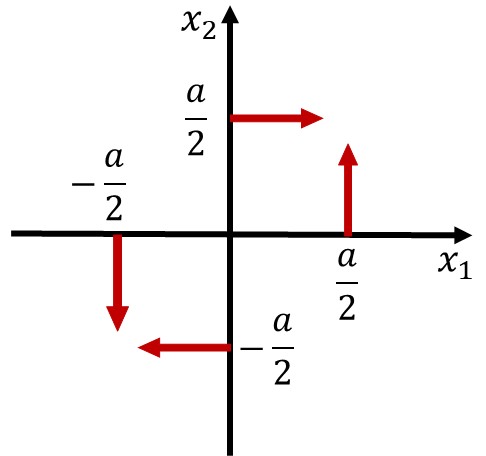}
    \caption{Schematic representation of a shear transformation. 
}
    \label{fig:picard}
\end{figure}

It is interesting to compare the pinching event to the shear transformations that were introduced in the work of Picard \textit{et al.}~\cite{picard2004elastic}, in the context of a driving induced by an external shear deformation. These authors proposed that the elementary flow event could be modeled as a 
"shear transformation" described  by a pair of perpendicular force dipoles, schematically illustrated in  Fig.~2 of
Ref.~\cite{picard2004elastic} or Fig.~\ref{fig:picard} of this paper.

This configuration corresponds to the body-force distribution
\begin{equation}
\begin{aligned}
f_i(\mathbf x)
&=
F \delta_{i1}
\left[
\delta(x_1)\delta\!\left(x_2-\frac{a}{2}\right)
-
\delta(x_1)\delta\!\left(x_2+\frac{a}{2}\right)
\right]
\\
&\quad
+
F \delta_{i2}
\left[
\delta\!\left(x_1-\frac{a}{2}\right)\delta(x_2)
-
\delta\!\left(x_1+\frac{a}{2}\right)\delta(x_2)
\right].
\end{aligned}
\end{equation}

In this case,  the dipole moment $P_{ij}$ is given by the symmetric, traceless matrix 
\begin{equation}
    \mathrm P =
    \begin{bmatrix}
        0 & aF \\
        aF & 0
    \end{bmatrix}.
\end{equation}

We then express the displacement
field components as
\begin{eqnarray}
u_1({\bf x})
&=&
\frac{x_2}{4\pi \mu}
\frac{(\nu+3)x_1^2+(1-\nu)x_2^2}{r^4}
\, aF
\nonumber \\
&=&
\frac{aF}{4 \pi \mu r}
\sin \theta
\left[
(\nu+3)\cos^2 \theta
+
(1-\nu)\sin^2 \theta
\right],
\nonumber \\
u_2({\bf x})
&=&
\frac{x_1}{4\pi \mu}
\frac{(\nu+3)x_2^2+(1-\nu)x_1^2}{r^4}
\, aF
\nonumber \\
&=&
\frac{aF}{4\pi \mu r}
\cos \theta
\left[
(\nu+3)\sin^2 \theta
+
(1-\nu)\cos^2 \theta
\right],
\label{eq:u_pinch_cont}
\end{eqnarray}
where the first expressions correspond to the Cartesian
coordinates, while the second expressions are written in
polar coordinates.

We note that, due to the symmetry of the dipole tensor, this solution exhibits a symmetry between the two main diagonals, i.e., 
$u_2(r, \theta+\pi/2)=-u_1(r,\theta)$, irrespective of the Poisson ratio $\nu$. As a result, the solution will exhibit 4  "lobes" of identical intensity and classically described as  "qadrupolar symmetry", a somewhat misleading vocabulary in view of its dipolar nature. This feature is most often used to describe the shear stress response, 
$\sigma_{12}(\mathbf x)=2\mu \epsilon_{12}(\mathbf x)$, which is  expressed (under   the incompressible condition $\nu \to 1$) as 
\begin{equation}
\sigma_{12}({\bf x})
=
\frac{\cos(4\theta)}{\pi r^2}
\, aF .
\label{eq:famous_Picard}
\end{equation}

\begin{figure}[htbp]
    \centering
    \includegraphics[width=0.6\linewidth]{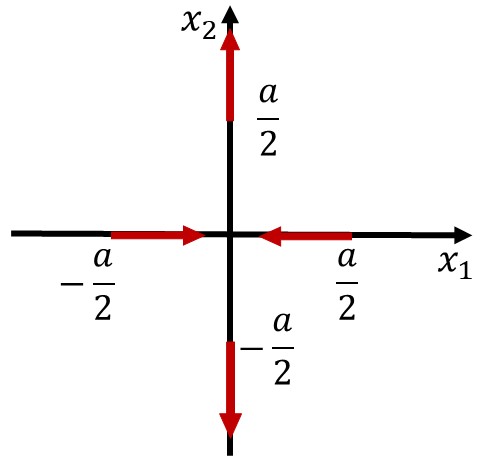}
    \caption{Schematic representation of pinching-contraction
}
    \label{fig:two_dipoles}
\end{figure}

In order to contrast pinching and shear transformation, it is interesting to consider the shear transformation in a frame rotated by $\pi/4$, a configuration that we will describe as "pinching-contraction"  schematically shown
in Fig.~\ref{fig:two_dipoles}.

In this frame, the dipole moment is given by
\begin{equation}
    \mathrm P =
    \begin{bmatrix}
        -aF & 0 \\
        0 & aF
    \end{bmatrix},
    \label{eq:P_pin_con}
\end{equation}
and the  displacement field components are
\begin{eqnarray}
u_1({\bf x})
&=&
\frac{x_1}{2\pi \mu}
\frac{(\nu x_2^2 - x_1^2)}{r^4}
\, aF
\nonumber \\
&=&
\frac{aF}{2 \pi \mu r}
\cos \theta
\left(
\nu \sin^2 \theta - \cos^2 \theta
\right),
\nonumber \\
u_2({\bf x})
&=&
\frac{x_2}{2\pi \mu}
\frac{(x_2^2 - \nu x_1^2)}{r^4}
\, aF
\nonumber \\
&=&
\frac{aF}{2 \pi \mu r}
\sin \theta
\left(
\sin^2 \theta - \nu \cos^2 \theta
\right) .
\label{eq:u_pinch_cont}
\end{eqnarray}
with the lobes now aligned with the coordinate axis, and the shear stress in an incompressible medium  now given by
\begin{equation}
\sigma_{12}({\bf x})
=
\frac{\sin(4\theta)}{\pi r^2}
\, aF .
\label{eq:stres_pinch_cont}
\end{equation}

We can now easily compare the simple pinching with the pinching-contraction case, equivalent to a shear transformation. In particular, in the incompressible limit, $\nu \to 1$, using the pinching displacement field $u_i^{\rm pin}$ given by Eq.~(\ref{eq:u_pin_theory_polar}) and the pinching-contraction displacement field $u_i^{\rm pin-con}$ given by Eq.~(\ref{eq:u_pinch_cont}), we immediately obtain
\begin{equation}
    u_i^{\rm pin}(\mathbf x)
    =
    \frac{1}{2}\,
    u_i^{\rm pin\mbox{-}con}(\mathbf x)
    \qquad (\textrm{when} \quad \nu \to 1).
    \label{eq:u_pin_pin_con_observation}
\end{equation}

This can be interpreted as follows. In the incompressible limit,
where the volume is preserved, a pinching-only forcing already
induces an effective contraction (to satisfy incompressibility).
However, its overall magnitude is exactly one half of the case
where pinching and contraction are explicitly imposed by external
forces. This can also be seen by decomposing the dipole moment matrix for pinching in Eq.~(\ref{eq:P_pinching}) into its isotropic and traceless parts,
\begin{equation}
    \mathrm P =
    \begin{bmatrix}
        0 & 0 \\
        0 & aF
    \end{bmatrix}
    =
    \begin{bmatrix}
        aF/2 & 0 \\
        0 & aF/2
    \end{bmatrix}
    +
    \begin{bmatrix}
        -aF/2 & 0 \\
        0 & aF/2
    \end{bmatrix}.
\end{equation}
Here the first term corresponds to a local isotropic dilation, and the second one to a "pinching-contraction" event. In an incompressible medium the response to a local isotropic dilation  vanishes, and only the second term contributes.

As a result, we can expect that pinching events will be very similar to shear transformations in nearly incompressible materials. However in the 
general compressible case, we expect that the response will loose its "quadrupolar" symmetry, and the relative intensity of the lobes will become sensitive to the Poisson ratio.

\section{Simulation methods} 
\label{sec:methods}

\subsection{Model}
We study a two-dimensional model glass consisting of a ternary mixture of particles of three different sizes: small (S), medium (M), and large (L)~\cite{sharma2024selecting,sharma2026interpretability}. Following the doping strategy introduced in Ref.~\cite{parmar2020ultrastable}, we add the additional species (M) to a widely studied two-dimensional binary glass former composed of S and L particles~\cite{falk1998dynamics,PhysRevE.97.033001}, in order to enhance the efficiency of swap Monte Carlo simulations~\cite{PhysRevX.7.021039}.
The composition ratio of small (S), medium (M), and large (L) particles is $11\!:\!5\!:\!9$.
The number density is $\rho = N/L^2 = 1.024$, where $N$ is the total number of particles and $L$ is the linear size of the simulation box.
We denote the set of particle positions by ${\bf r}^N = ({\bf r}_1, {\bf r}_2, \ldots, {\bf r}_N)$.

Particles interact via the Lennard--Jones potential
\[
v_{\alpha\beta}(r)=4\varepsilon_{\alpha\beta}\left[\left(\frac{\sigma_{\alpha\beta}}{r}\right)^{12}-\left(\frac{\sigma_{\alpha\beta}}{r}\right)^6\right],
\]
where $\alpha,\beta=\mathrm{S},\mathrm{M},\mathrm{L}$. The potential is modified to be twice continuously differentiable at the cutoff~\cite{PhysRevE.97.033001}.
The parameters $\sigma_{\alpha\beta}$ and $\varepsilon_{\alpha\beta}$ are given as follows: $\sigma_{\rm LL}=2 \sin(\pi/5)\simeq 1.18$, $\sigma_{\rm SS}=2 \sin(\pi/10) \simeq 0.62$, $\sigma_{\rm LS}=1$,
$\sigma_{\rm LM}=(\sigma_{\rm LL}+\sigma_{\rm LS})/2$, $\sigma_{\rm MS}=(\sigma_{\rm LS}+\sigma_{\rm SS})/2$, and $\sigma_{\rm MM}=(\sigma_{\rm LL}+\sigma_{\rm SS})/2$.
Similarly, $\varepsilon_{\rm LL}=1/2$, $\varepsilon_{\rm SS}=1/2$,  $\varepsilon_{\rm LS}=1$, $\varepsilon_{\rm LM}=(\varepsilon_{\rm LL}+\varepsilon_{\rm LS})/2$, $\varepsilon_{\rm MS}=(\varepsilon_{\rm LS}+\varepsilon_{\rm SS})/2$, and $\varepsilon_{\rm MM}=(\varepsilon_{\rm LL}+\varepsilon_{\rm SS})/2$. All particles have the same mass, which is set to unity. 
We use $\sigma_{\rm LS}$ and $\varepsilon_{\rm LS}$ as the units of length and energy, respectively.
The total potential energy is denoted by $U({\bf r}^N)$.

\subsection{Glass preparation}

We prepare glass samples by rapidly quenching equilibrated supercooled liquids~\cite{sastry1998signatures}. To investigate the dependence of the mechanical response to pinching on glass stability, we generate glasses with a wide range of stabilities, characterized by the parent thermal equilibrium temperature $T_{\rm ini}$ prior to quenching.
Thermally equilibrated configurations are prepared at $T_{\rm ini}=0.25$, $0.35$, and $1.0$ using the swap Monte Carlo (SMC) algorithm and subsequently quenched to their inherent structures (IS) using the FIRE 2.0 algorithm~\cite{PhysRevLett.97.170201,GUENOLE2020109584}. The resulting samples correspond to stable glasses ($T_{\rm ini}=0.25$), moderately annealed glasses ($T_{\rm ini}=0.35$), and poorly annealed glasses ($T_{\rm ini}=1.0$).

We consider system sizes $N=1000$, $4000$, $16000$, and $64000$, corresponding to box lengths $L \simeq 31.2$, $62.5$, $125.0$, and $250.0$, respectively. For each of the system sizes $N=1000$, $4000$, and $16000$, we prepare 20 independent glass configurations, and for $N=64000$ we prepare 10 independent configurations.

\subsection{Pinching protocols}

Starting from glass samples at zero temperature, we select randomly two neighboring particles and apply pinching very slowly, in an athermal quasistatic manner, which we refer to as athermal quasistatic pinching (AQP). Below we describe this protocol in detail.

For an initial configuration (before pinching) denoted by ${\bf r}_0^N$, we randomly select a particle (denoted $i$) and identify one of its neighbors (denoted $j$) within a cutoff distance of $1.4$. We then compute the initial distance between the two particles in the initial configuration,
\[
\ell_0 = |{\bf r}_{0i} - {\bf r}_{0j}|.
\]

We next add an external pinching potential to the total potential energy $U({\bf r}^N)$, such that the effective energy $E({\bf r}^N)$ becomes
\begin{equation}
E({\bf r}^N) = U({\bf r}^N) + K\bigl(|{\bf r}_i - {\bf r}_j| - \ell\bigr)^2,
\label{eq:potential_length}
\end{equation}
where $\ell$ is the externally controlled separation~\cite{ji2025role}.
Starting from $\ell=\ell_0$, we increase $\ell$ in steps of $d \ell = 0.01$. After each increment, we minimize the energy given by Eq.~(\ref{eq:potential_length}) using the FIRE algorithm. The spring constant is fixed to $K=10$. This procedure is repeated until the total extension $\Delta \ell=\ell - \ell_0$ reaches $2.0$. The protocol described above is analogous to athermal quasistatic shear simulations performed under strain-controlled conditions. We therefore refer to it as the {\it length-controlled protocol}. We also study a {\it force-controlled protocol}, which is the analog of stress-controlled loading in uniform shear simulations. The results for this case are presented in Appendix~\ref{sec:force_control}.

For each initial configuration, AQP simulations were performed 50 times (with independent random selections of particle pairs and pinching) for $N=1000$, $4000$, and $16000$, and 500 times for $N=64000$, resulting in a total of $5000$ pinching simulations for each system size. We denote by $\langle \cdots \rangle$ averages over these independent pinching realizations.

\section{Simulation results}
\label{sec:simulation}

\begin{figure*}
\includegraphics[width=0.66\columnwidth]{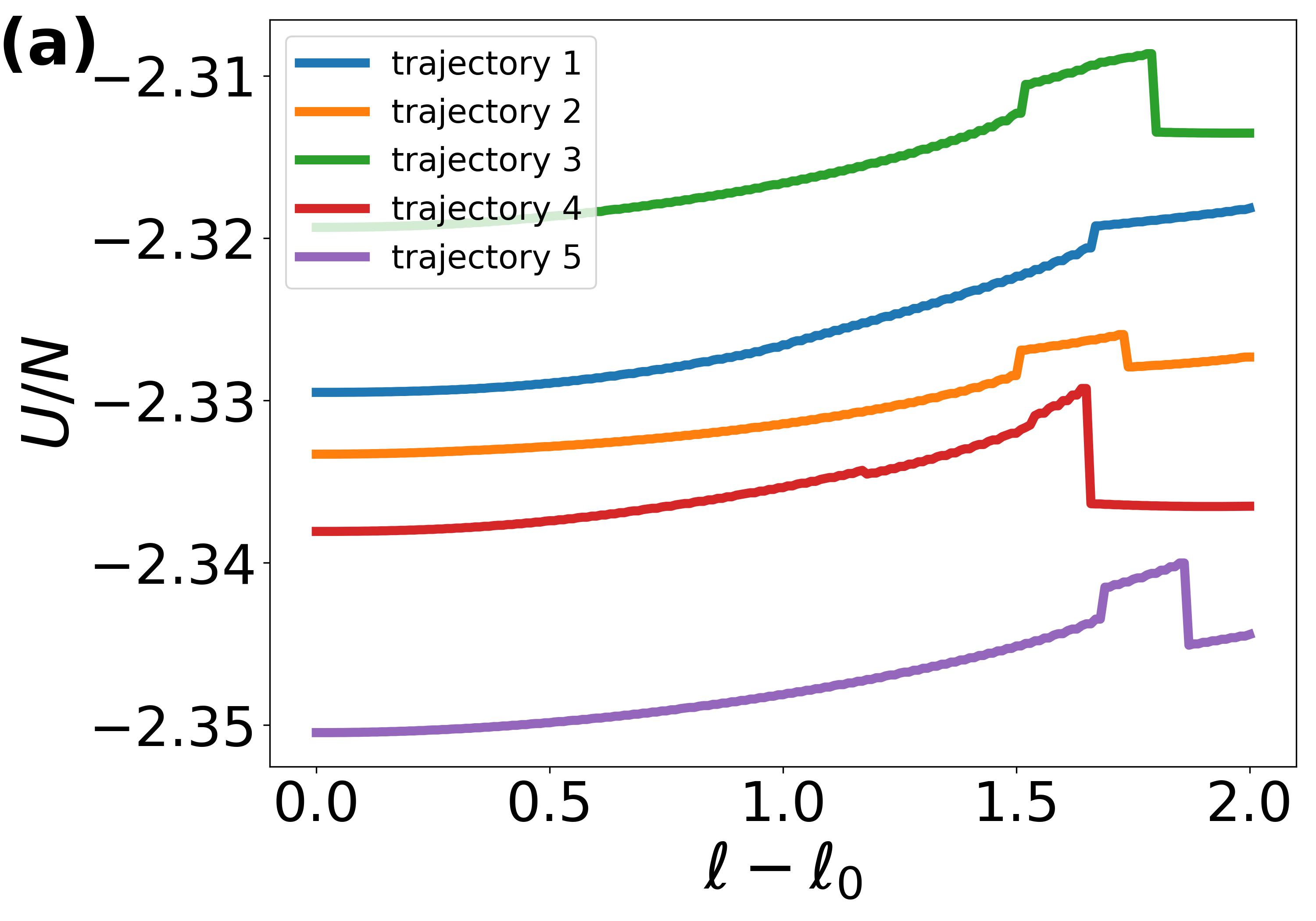}
\includegraphics[width=0.66\columnwidth]{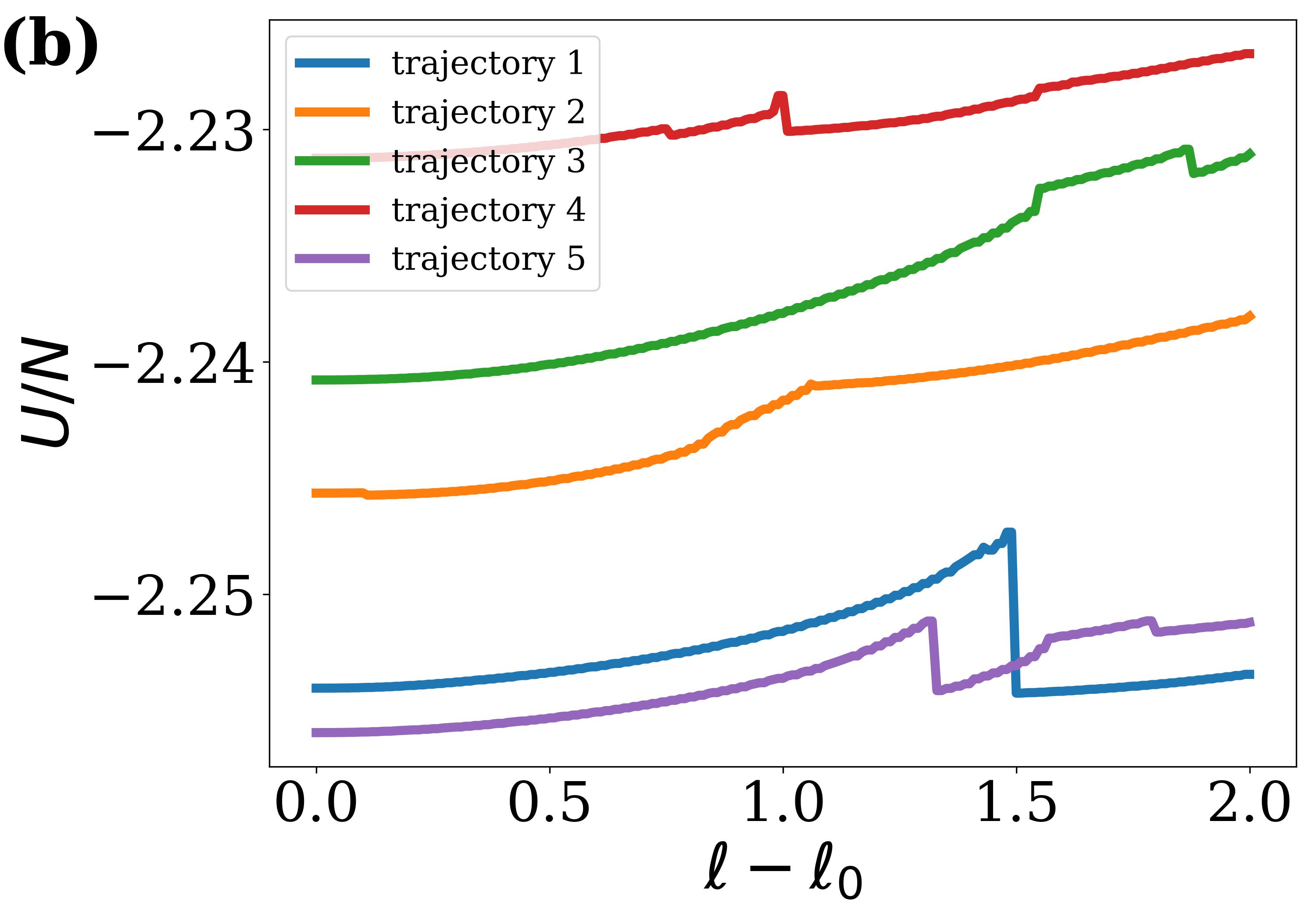}
\includegraphics[width=0.66\columnwidth]{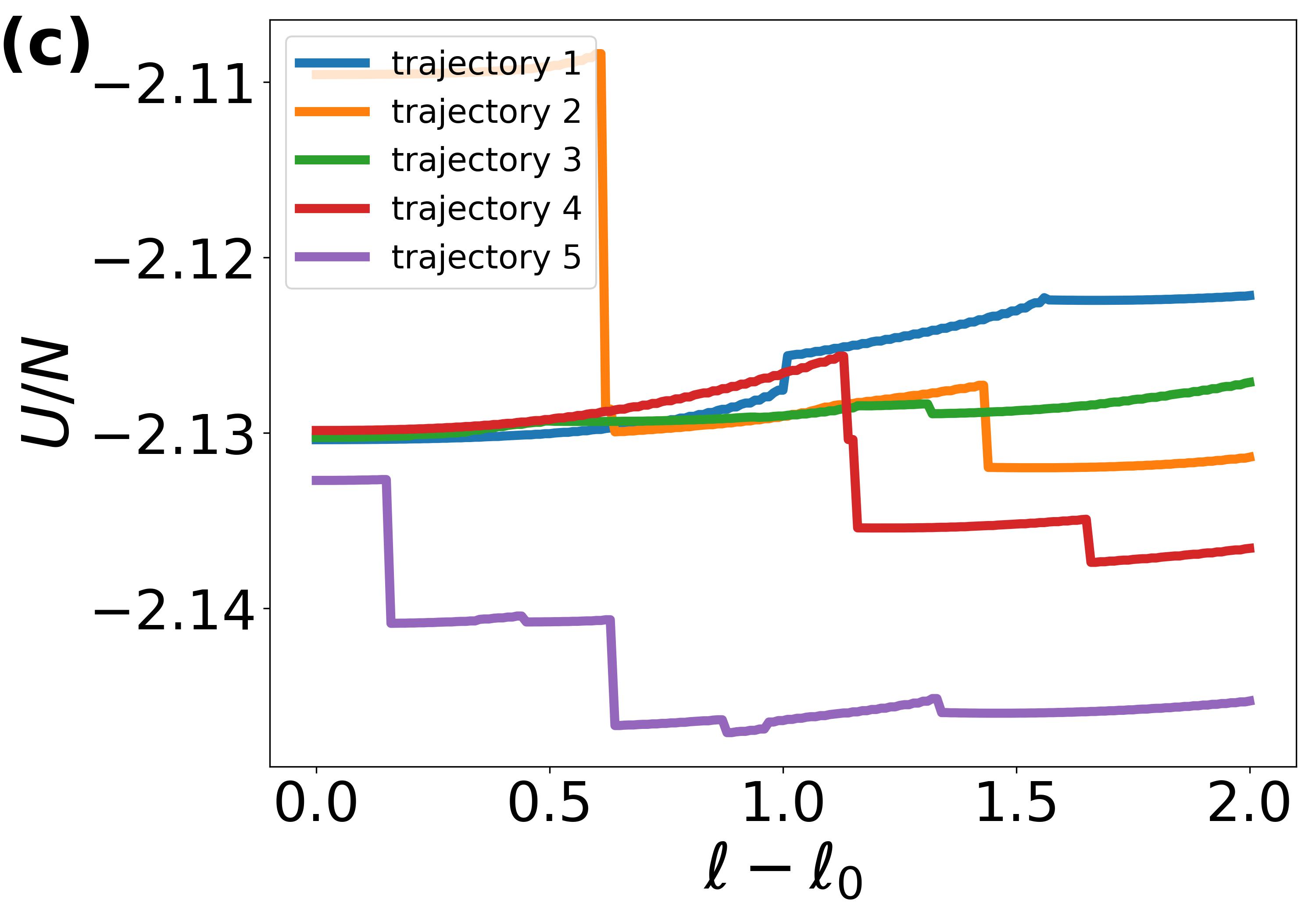}
\includegraphics[width=0.66\columnwidth]{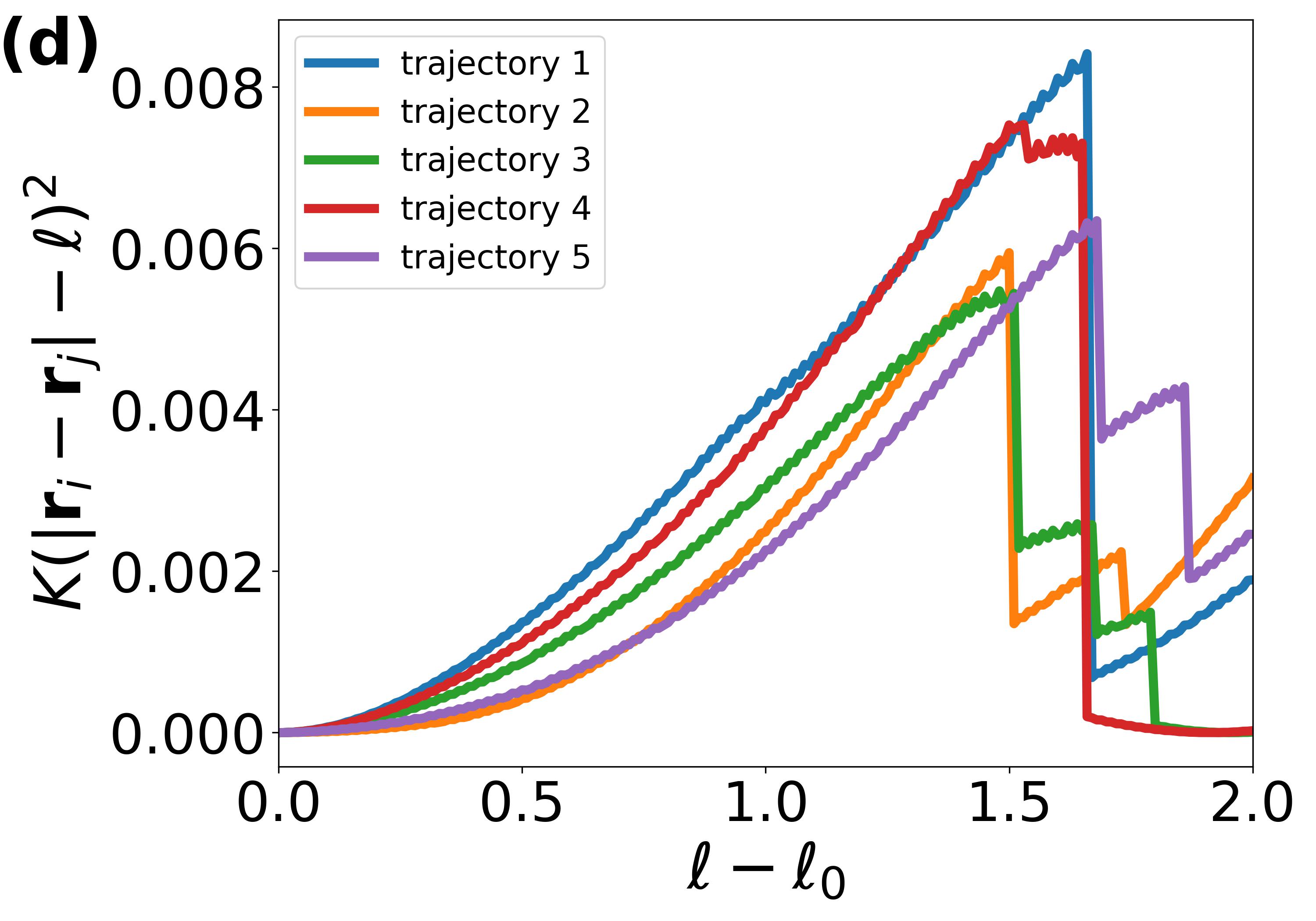}
\includegraphics[width=0.66\columnwidth]{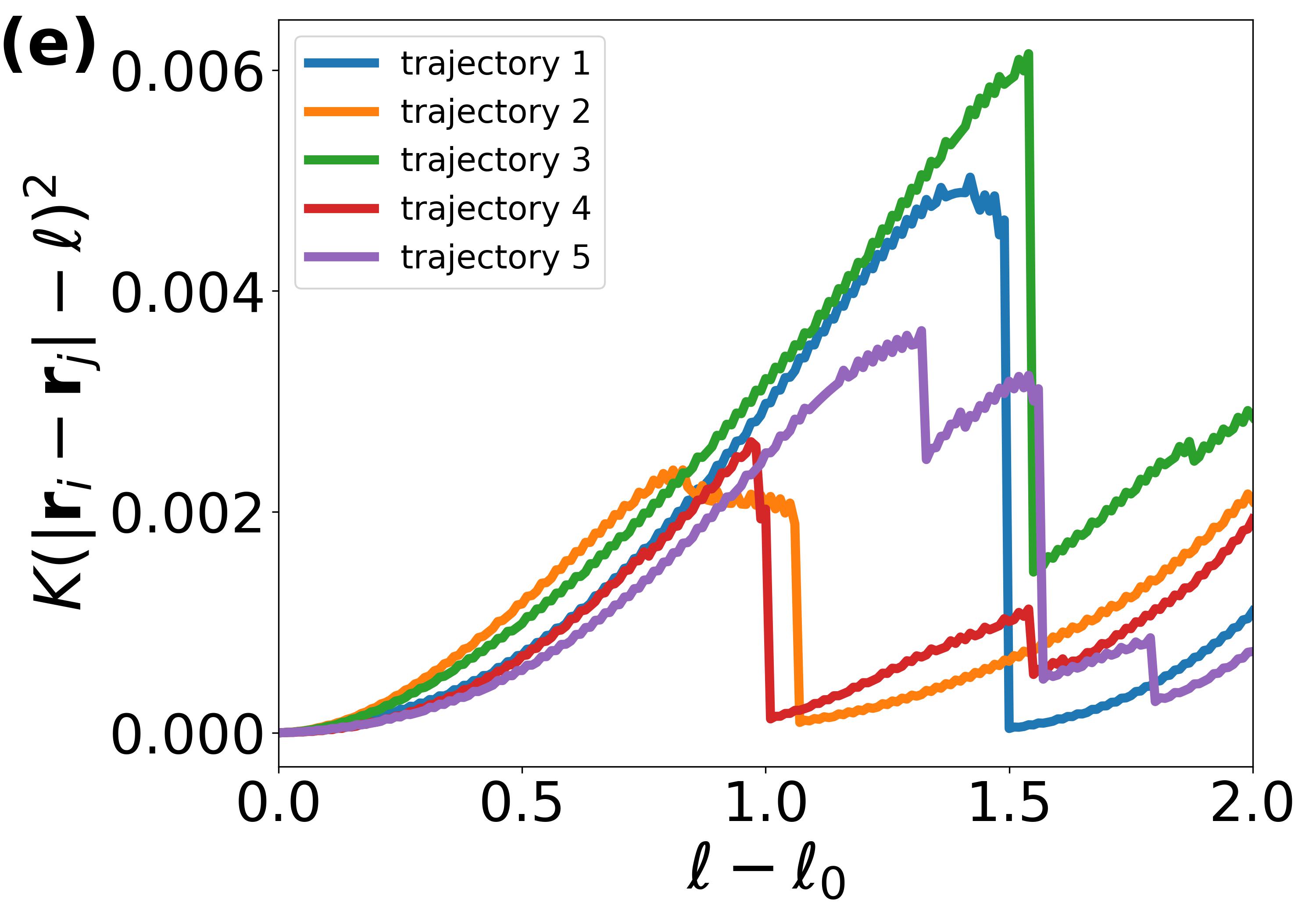}
\includegraphics[width=0.66\columnwidth]{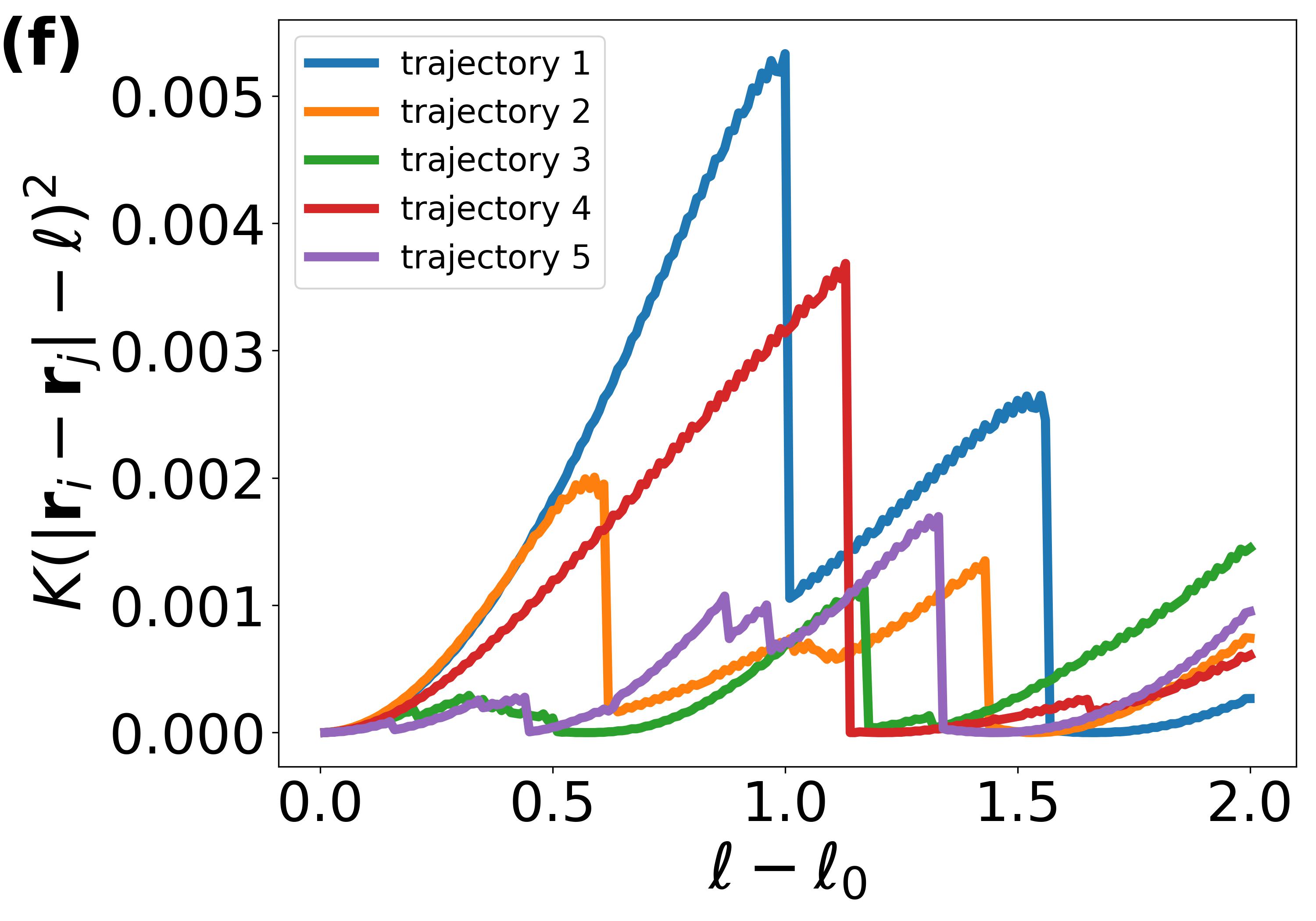}
\includegraphics[width=0.66\columnwidth]{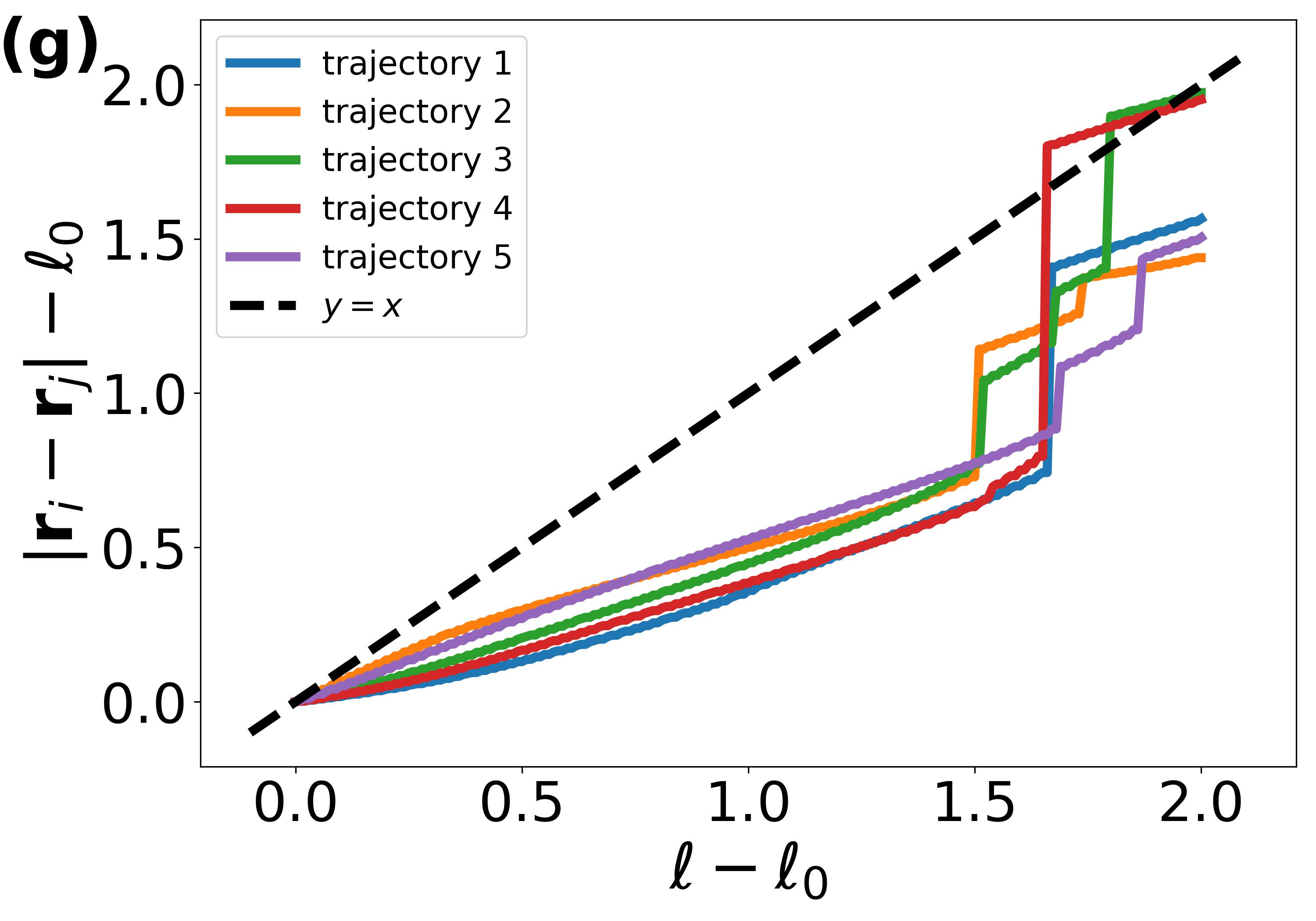}
\includegraphics[width=0.66\columnwidth]{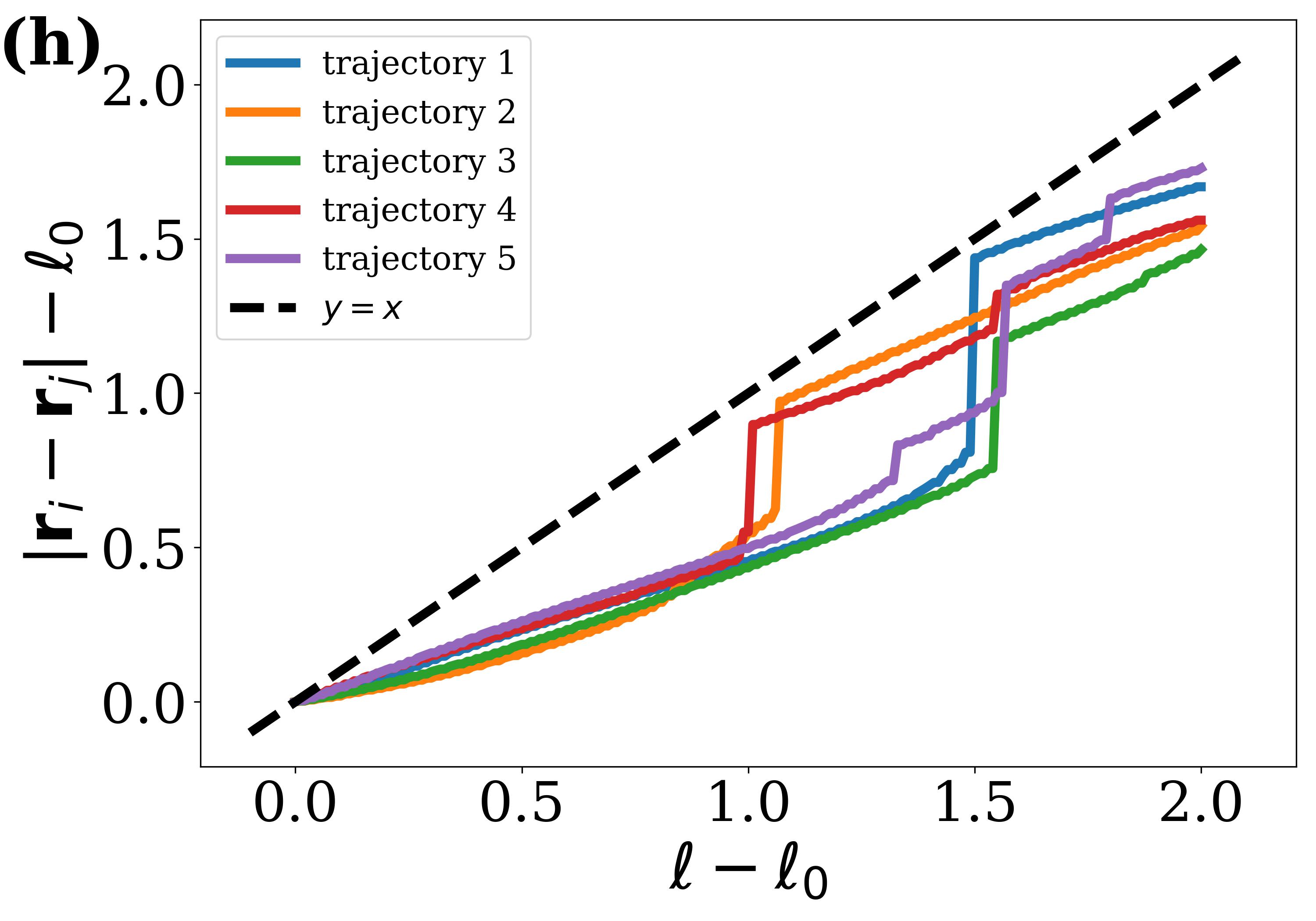}
\includegraphics[width=0.66\columnwidth]{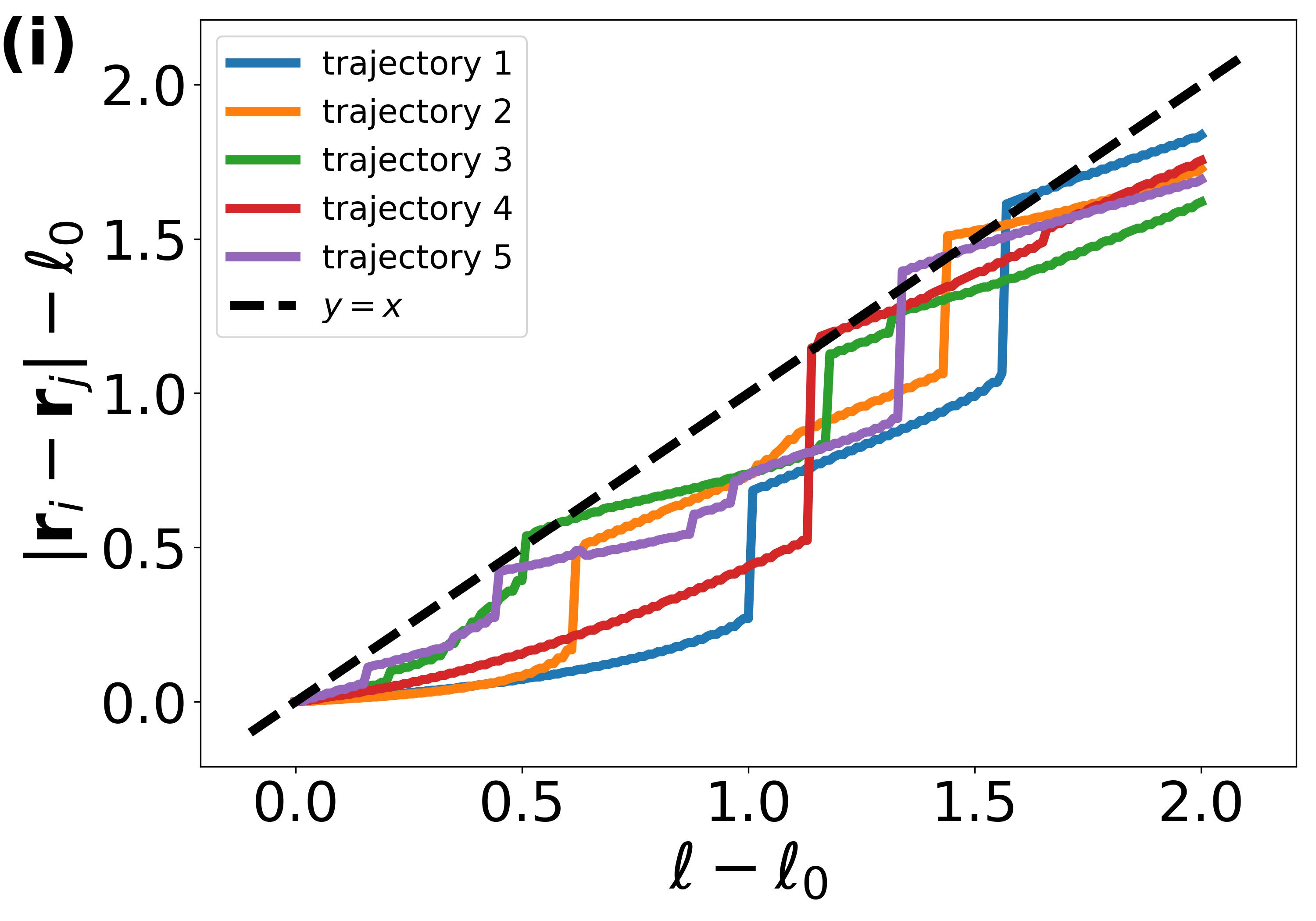}
\caption{(a–c): Potential energy $U/N$ as a function of the control parameter increment $\Delta\ell=\ell-\ell_0$ for five pinching trajectories obtained with the length-controlled protocol at $T_{\rm ini}=0.25$ (a), $0.35$ (b), and $1.0$ (c). (d-f): Spring energy $K\bigl(|{\bf r}_i - {\bf r}_j| - \ell\bigr)^2$ as a function of the control parameter increment $\Delta\ell$ for five pinching trajectories obtained with the length-controlled protocol at $T_{\rm ini}=0.25$ (d), $0.35$ (e), and $1.0$ (f).
(g–i): The corresponding actual separation increment $|{\bf r}_i-{\bf r}_j|-\ell_0$ as a function of $\Delta\ell=\ell-\ell_0$ for $T_{\rm ini}=0.25$ (g), $0.35$ (h), and $1.0$ (i). The dashed straight lines correspond to $y=x$.
}
\label{fig:individual_length_control}
\end{figure*}

\subsection{Individual pinching trajectories}

We present simulation results for individual pinching trajectories.

Figures~\ref{fig:individual_length_control}(a--c) show the evolution of the potential energy
$U(\mathbf{r}^N)/N$ as a function of the increment in the controlled length,
$\Delta \ell=\ell-\ell_0$, for stable (a), moderately annealed (b), and poorly annealed (c) glasses,
respectively.

For the stable glasses shown in Fig.~\ref{fig:individual_length_control}(a),
$U(\mathbf{r}^N)/N$ initially increases smoothly and approximately quadratically up to
$\Delta \ell \approx 1$, corresponding to an elastic response. Discontinuous changes
signaling plastic events appear only around $\Delta \ell \approx 1.5$.
These discontinuities are more clearly visible in the spring-energy term shown in
Fig.~\ref{fig:individual_length_control}(d), where the elastic energy accumulated as
$\Delta \ell$ increases is released through noticeable drops.

We emphasize that increasing the controlled length $\Delta \ell=\ell-\ell_0$ (a parameter inside the effective energy $E({\bf r}^N)$) does not imply that
the actual distance between the two pinched particles follows the imposed value exactly.
Figure~\ref{fig:individual_length_control}(g) shows the evolution of the actual extension,
$|\mathbf{r}_i-\mathbf{r}_j|-\ell_0$, as a function of $\Delta \ell$.
In the elastic regime, $|\mathbf{r}_i-\mathbf{r}_j|-\ell_0$ remains considerably smaller
than $\Delta \ell$. At a plastic event, however, the actual extension increases abruptly
and approaches the imposed extension more closely.

\begin{figure}[H]
\centering
\begin{minipage}[b]{0.32\columnwidth}
    \centering
    \includegraphics[width=\columnwidth]{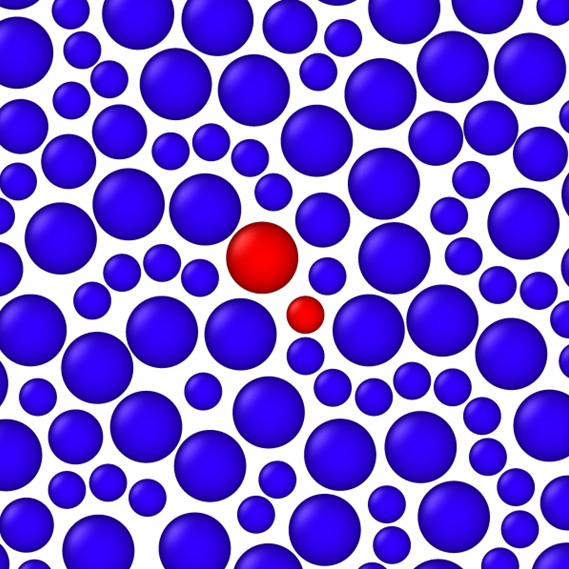}
\end{minipage}
\begin{minipage}[b]{0.32\columnwidth}
    \centering
    \includegraphics[width=\columnwidth]{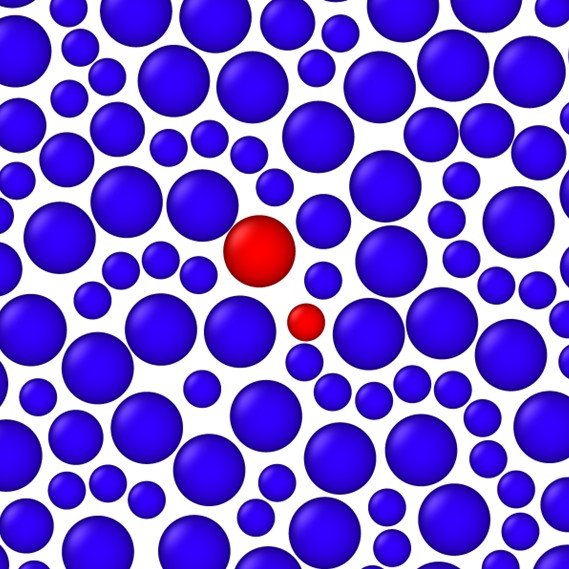}
\end{minipage}
\begin{minipage}[b]{0.32\columnwidth}
    \centering
    \includegraphics[width=\columnwidth]{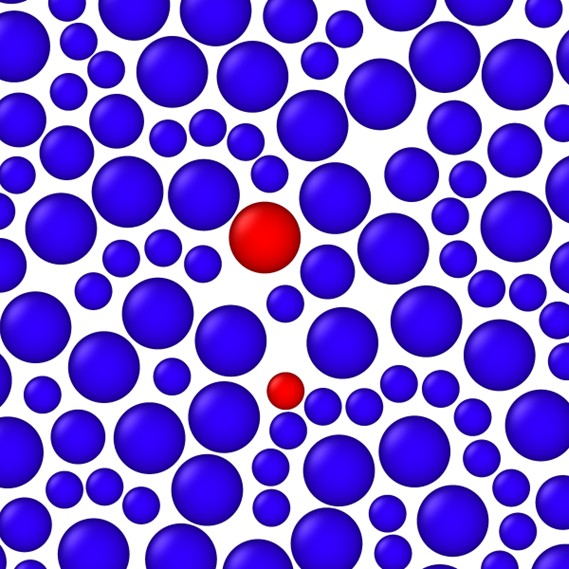}
\end{minipage}

\vspace{0.2cm} 

\begin{minipage}[b]{0.32\columnwidth}
    \centering
    \includegraphics[width=\columnwidth]{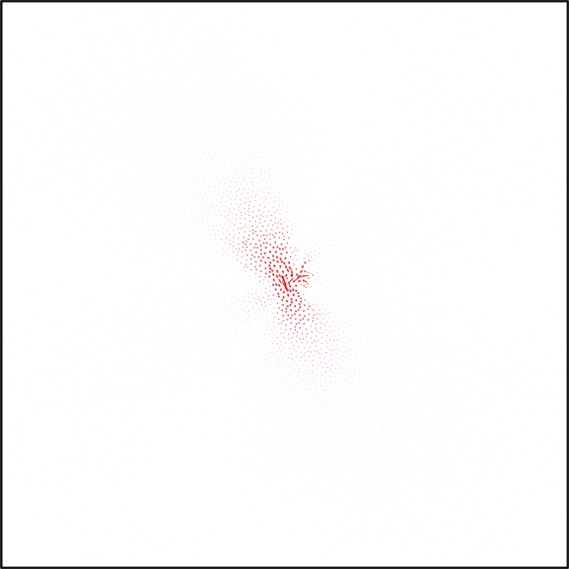}
\end{minipage}
\begin{minipage}[b]{0.32\columnwidth}
    \centering
    \includegraphics[width=\columnwidth]{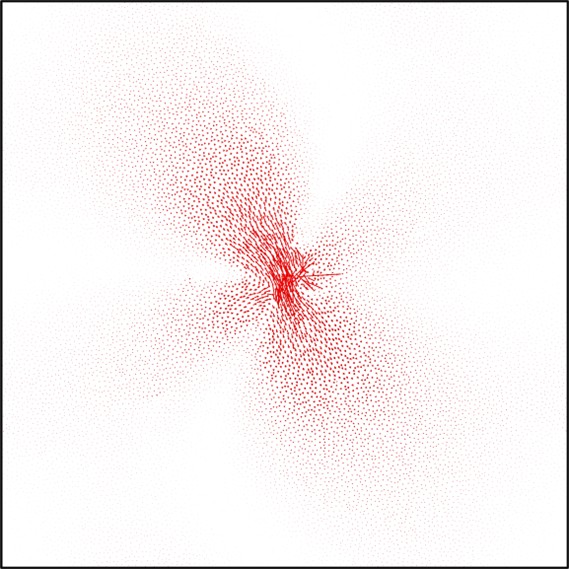}
\end{minipage}
\begin{minipage}[b]{0.32\columnwidth}
    \centering
    \includegraphics[width=\columnwidth]{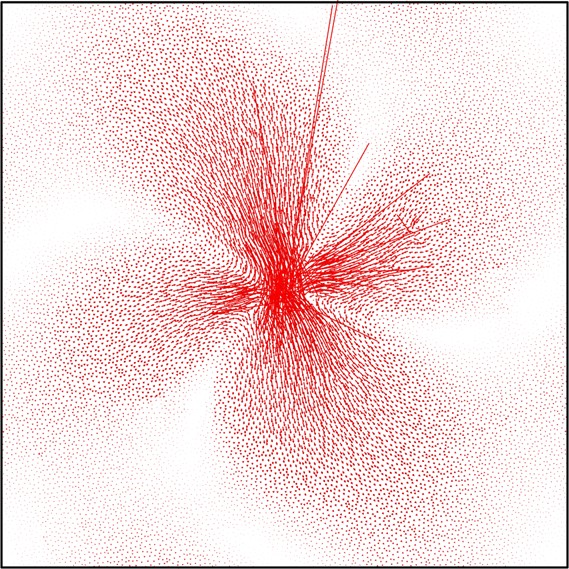}
\end{minipage}
\caption{Particle visualizations (top row) and corresponding displacement fields (bottom row) at $T_{\rm ini} = 0.25$ for $\Delta\ell = 0.5,1.0,2.0$ (from left to right). particle-visualization is only focused local environment. Displacement field is scaled $\times50$ for clarity.
}
\label{fig:snapshot_Tini0.25}
\end{figure}

Representative real-space snapshots, shown using
both particle-based and displacement-vector visualizations, are presented in
Fig.~\ref{fig:snapshot_Tini0.25}. At the small extension
$\Delta \ell=0.5$, the two selected particles remain in contact and form a local
cage, while the surrounding particles exhibit an elastic displacement field.
At $\Delta \ell=1.0$, the two particles are still barely in contact, and the
elastic displacements are amplified. At $\Delta \ell=2.0$, the two particles
are sufficiently separated for other particles to enter the space between them,
indicating a cage-breaking plastic event. This plastic event induces a
quadrupolar, Eshelby-like displacement field~\cite{lerner2018characteristic,rainone2020statistical}, which is analyzed further in the
next section.

For the moderately annealed glasses, plastic events occur at smaller values of
$\Delta \ell$ along some trajectories, as evidenced by the discontinuities in
$U(\mathbf{r}^N)/N$ and in the spring energy shown in
Figs.~\ref{fig:individual_length_control}(b) and
\ref{fig:individual_length_control}(e), respectively. Concomitantly, the actual
extension $|\mathbf{r}_i-\mathbf{r}_j|-\ell_0$ tends to follow the imposed
extension $\Delta \ell$ more closely, as shown in
Fig.~\ref{fig:individual_length_control}(h).

\begin{figure}[H]
\centering
\begin{minipage}[b]{0.32\columnwidth}
    \centering
    \includegraphics[width=\columnwidth]{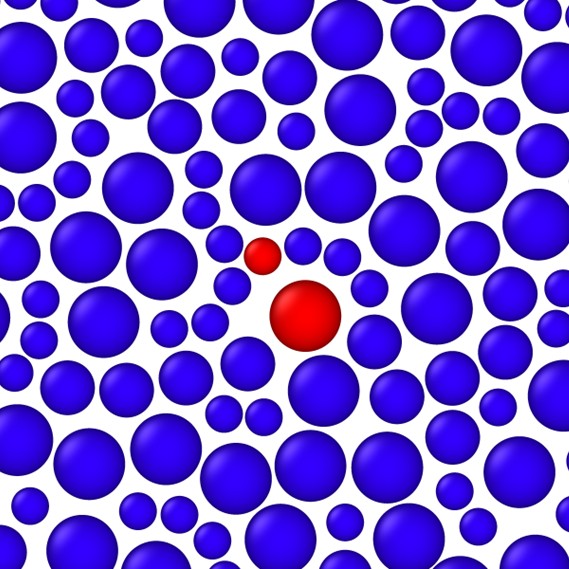}
\end{minipage}
\begin{minipage}[b]{0.32\columnwidth}
    \centering
    \includegraphics[width=\columnwidth]{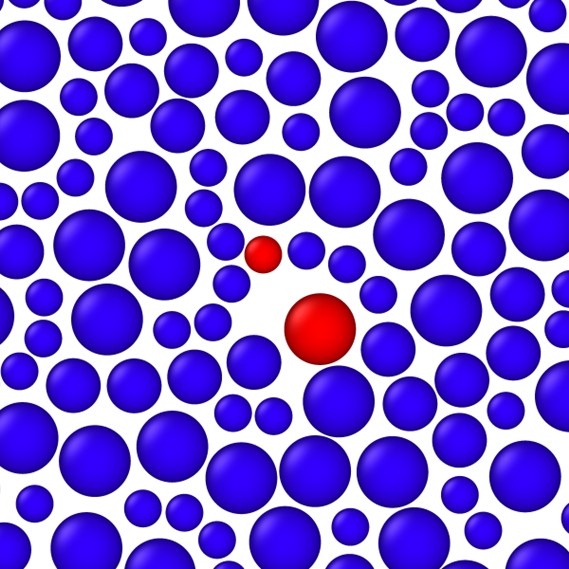}
\end{minipage}
\begin{minipage}[b]{0.32\columnwidth}
    \centering
    \includegraphics[width=\columnwidth]{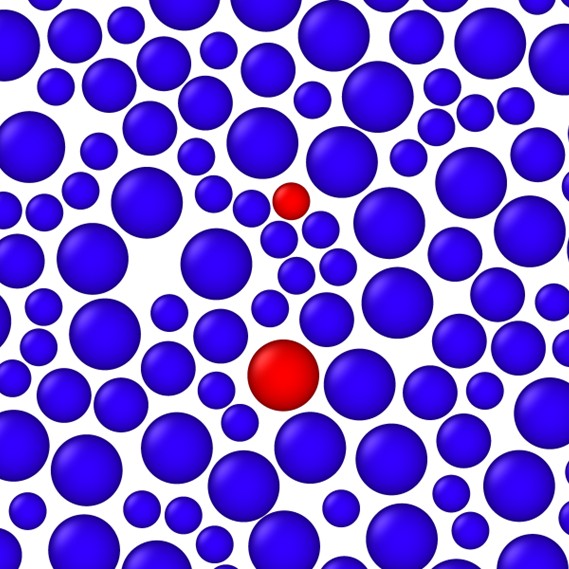}
\end{minipage}

\vspace{0.2cm} 

\begin{minipage}[b]{0.32\columnwidth}
    \centering
    \includegraphics[width=\columnwidth]{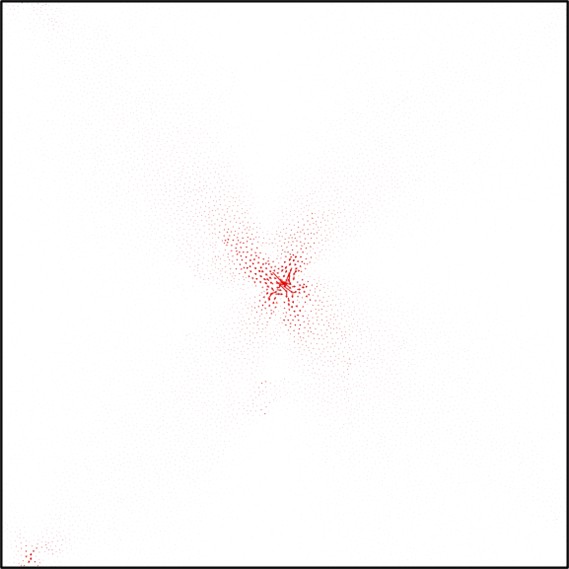}
\end{minipage}
\begin{minipage}[b]{0.32\columnwidth}
    \centering
    \includegraphics[width=\columnwidth]{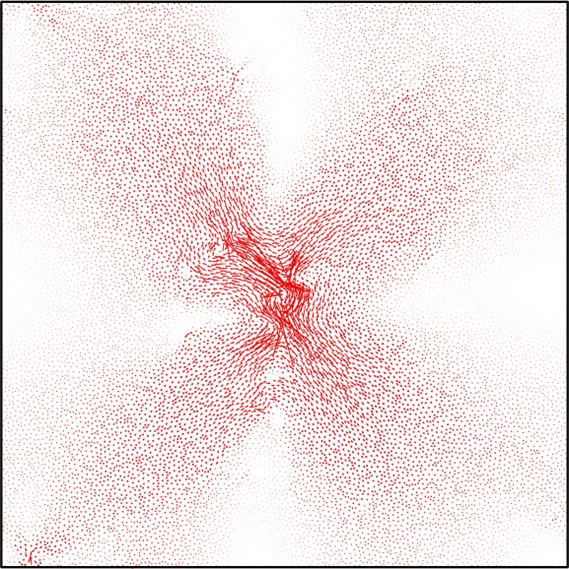}
\end{minipage}
\begin{minipage}[b]{0.32\columnwidth}
    \centering
    \includegraphics[width=\columnwidth]{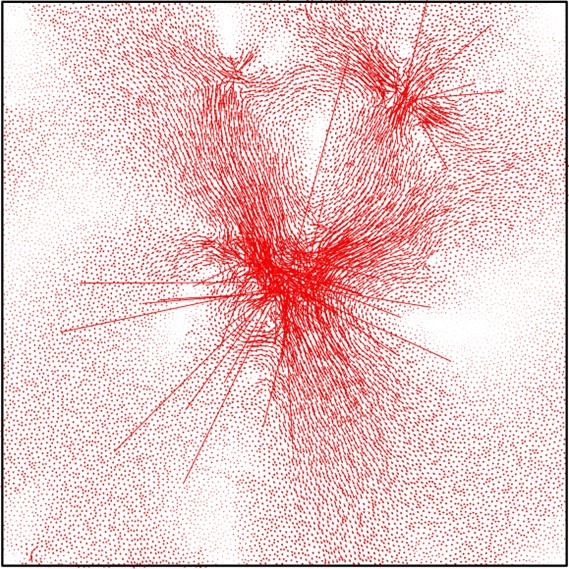}
\end{minipage}
\caption{Particle visualizations (top row) and corresponding displacement fields (bottom row) at $T_{\rm ini} = 0.35$ for $\Delta\ell = 0.5,1.0,2.0$ (from left to right). particle-visualization is only focused local environment. Displacement field is scaled $\times50$ for clarity.
}
\label{fig:snapshot_Tini0.35}
\end{figure}

The corresponding real-space snapshots are shown in
Fig.~\ref{fig:snapshot_Tini0.35}, where a plastic event is already visible at
$\Delta \ell=1.0$. Interestingly, at the larger extension
$\Delta \ell=2.0$, additional plastic events occur outside the local pinching
region. These events are triggered in distant soft regions~\cite{manning2011vibrational,ding2014soft} by the long-range
elastic displacement field generated by the plastic rearrangement near the
pinched particles. Such secondary events are more readily activated because the
glass is only moderately annealed.

Finally, we turn our attention to the poorly annealed glasses. As shown by
$U(\mathbf{r}^N)/N$ and the spring energy in
Figs.~\ref{fig:individual_length_control}(c) and
\ref{fig:individual_length_control}(f), respectively, plastic events, signaled
by discontinuities, occur at relatively small values of $\Delta \ell$.
Consequently, the actual extension follows the imposed extension more closely, as shown in \ref{fig:individual_length_control}(i).

\begin{figure}[H]
\centering
\begin{minipage}[b]{0.32\columnwidth}
    \centering
    \includegraphics[width=\columnwidth]{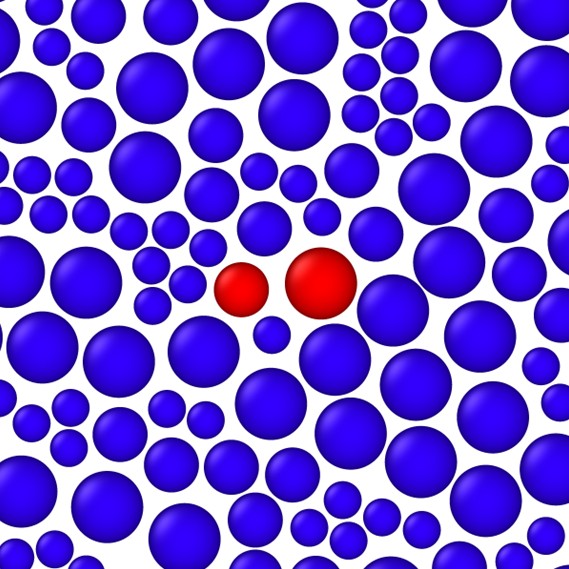}
\end{minipage}
\begin{minipage}[b]{0.32\columnwidth}
    \centering
    \includegraphics[width=\columnwidth]{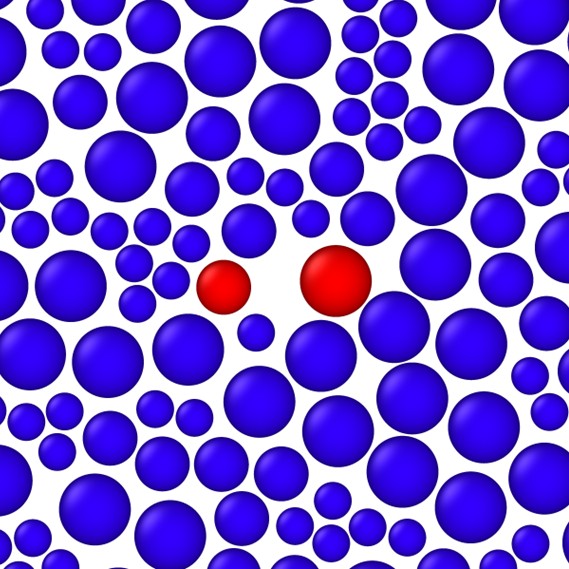}
\end{minipage}
\begin{minipage}[b]{0.32\columnwidth}
    \centering
    \includegraphics[width=\columnwidth]{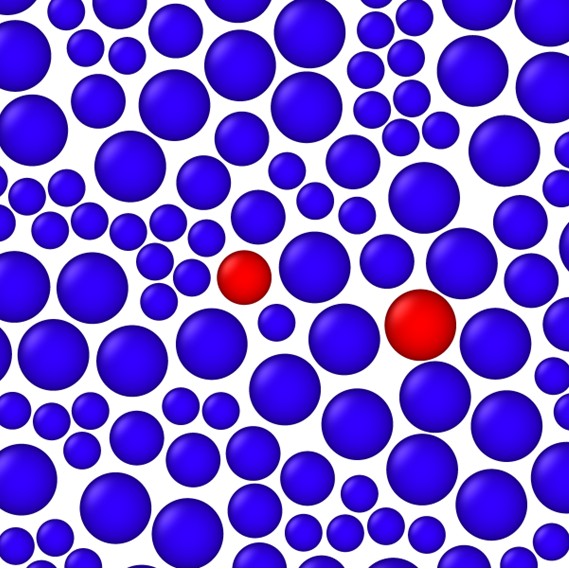}
\end{minipage}

\vspace{0.2cm} 

\begin{minipage}[b]{0.32\columnwidth}
    \centering
    \includegraphics[width=\columnwidth]{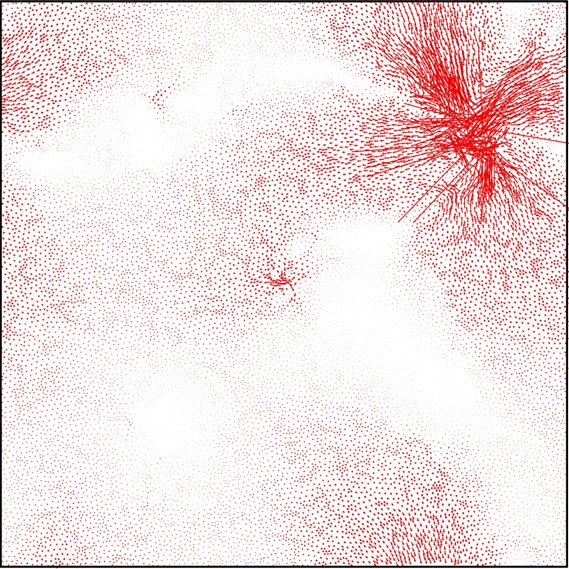}
\end{minipage}
\begin{minipage}[b]{0.32\columnwidth}
    \centering
    \includegraphics[width=\columnwidth]{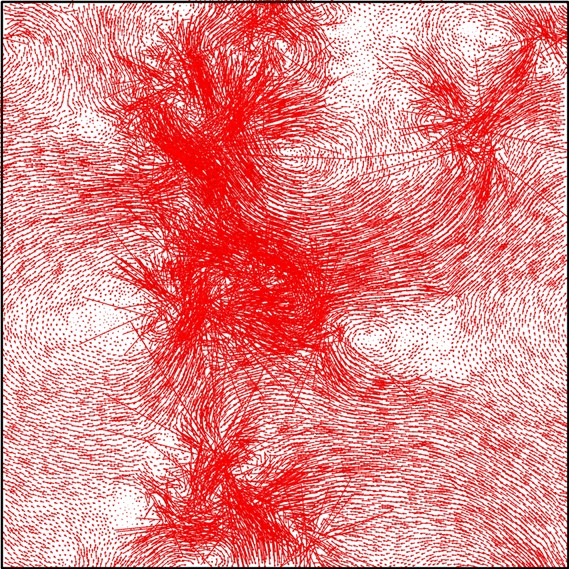}
\end{minipage}
\begin{minipage}[b]{0.32\columnwidth}
    \centering
    \includegraphics[width=\columnwidth]{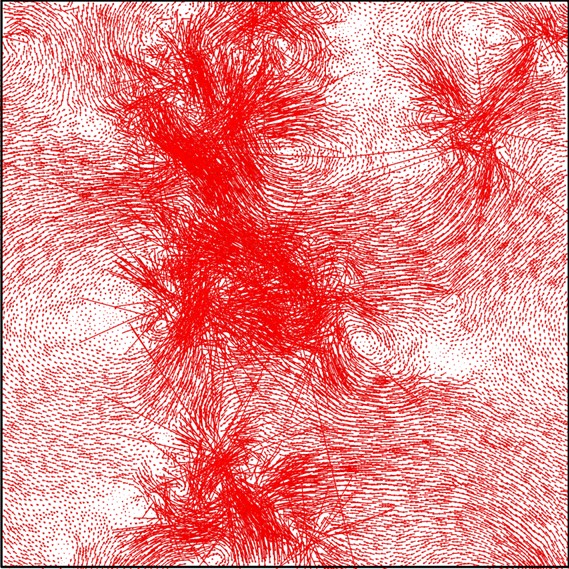}
\end{minipage}
\caption{Particle visualizations (top row) and corresponding displacement fields (bottom row) at $T_{\rm ini} = 1.0$ for $\Delta\ell = 0.5,1.0,2.0$ (from left to right). particle-visualization is only focused local environment. Displacement field is scaled $\times 50$ for clarity.
}
\label{fig:snapshot_Tini1.0}
\end{figure}

Representative real-space snapshots are shown in
Fig.~\ref{fig:snapshot_Tini1.0}. Even at small $\Delta \ell$, a plastic event is
immediately triggered in the far field by the elastic response generated near
the pinched particles. Further increases in $\Delta \ell$ activate additional
plastic events, producing an avalanche-like response. This behavior is
consistent with the presence of many soft regions in poorly annealed glasses.

\subsection{Averaged responses: Elastic regime}

\begin{figure*}
\includegraphics[width=0.6\columnwidth]{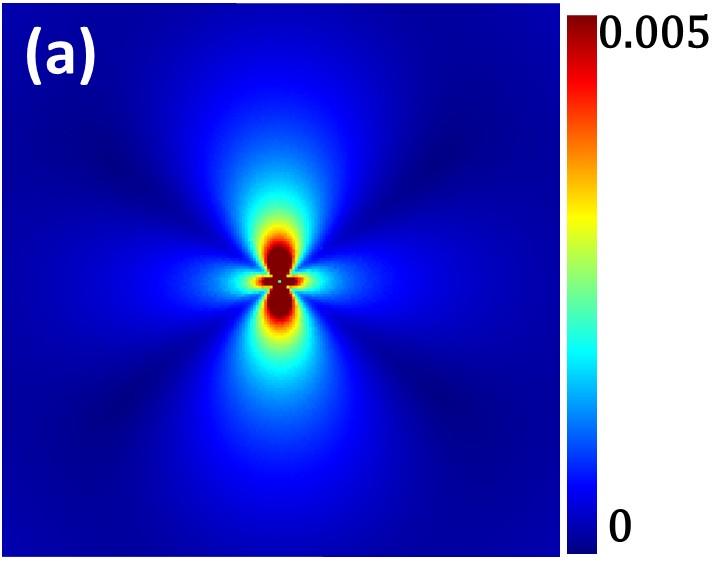}
\includegraphics[width=0.6\columnwidth]{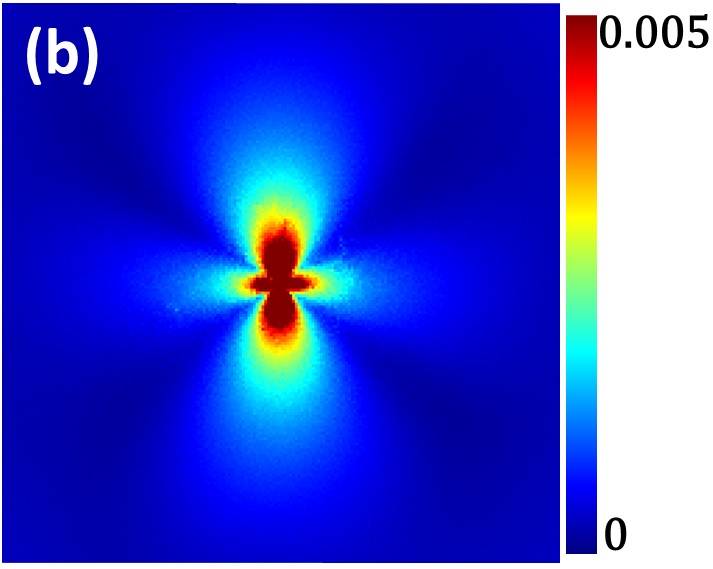}
\includegraphics[width=0.6\columnwidth]{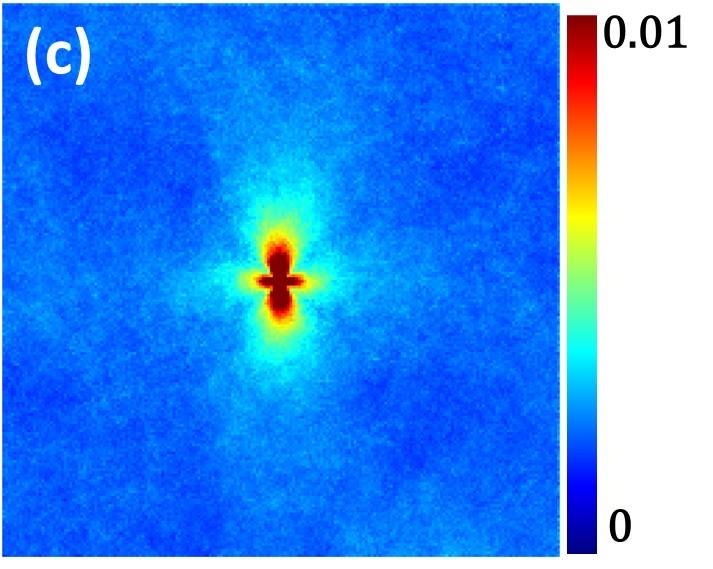}
\includegraphics[width=0.6\columnwidth]{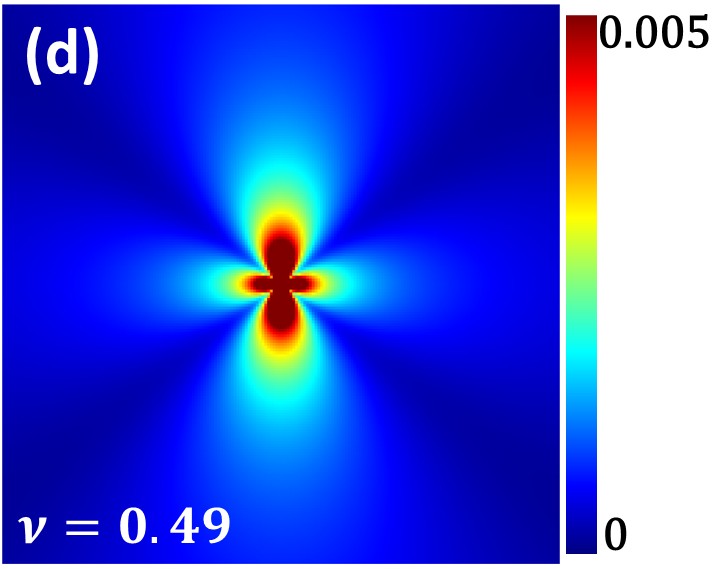}
\includegraphics[width=0.6\columnwidth]{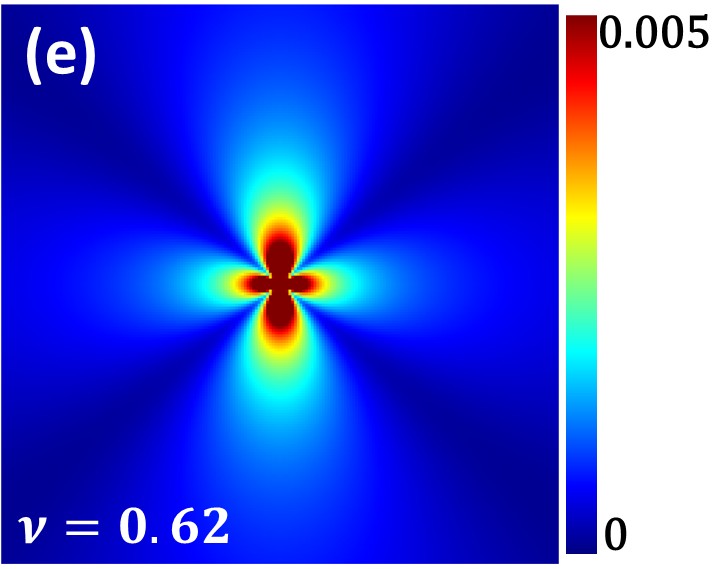}
\includegraphics[width=0.6\columnwidth]{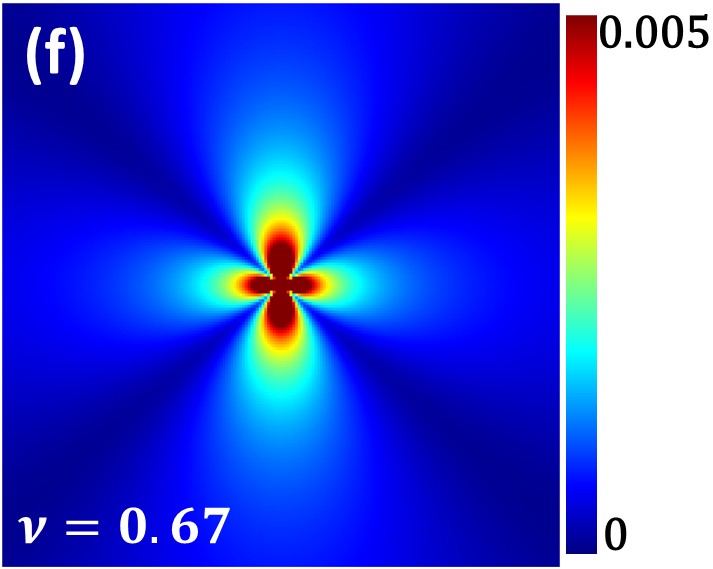} 
\caption{
(a--c) Magnitude of the displacement field induced by athermal quasistatic pinching at $\Delta \ell = 0.5$ for (a) stable glasses, $T_{\rm ini}=0.25$, (b) moderately annealed glasses, $T_{\rm ini}=0.35$, and (c) poorly annealed glasses, $T_{\rm ini}=1.0$. The results were obtained for $N=64000$ and averaged over $1000$ independent pinching realizations for (a), (b), and $5000$ independent pinching realizations for (c). 
(d--f) Corresponding predictions from linear elasticity theory for (d) stable, (e) moderately annealed, and (f) poorly annealed glasses. The Poisson ratio used in each theoretical calculation was measured independently from athermal quasistatic compression and shear simulations, and its value is indicated in white.}
\label{fig:mean_response_length_control}
\end{figure*}

In individual pinching simulations, we observed displacement fields with a quadrupole-like structure, as expected from elasticity. We now examine the mean response obtained by averaging over many independent pinching simulations and compare it with the prediction of linear elasticity.

We first focus on $\Delta \ell=0.5$, for which plastic rearrangements and cage-breaking events are rare, except in poorly annealed glasses. We therefore refer to this regime operationally as the ``elastic regime.''

Figure~\ref{fig:mean_response_length_control}(a--c) shows the magnitude of the displacement field induced by pinching, averaged over $1000$ independent pinching realizations, for stable glasses (a), moderately annealed glasses (b), and $1000$ independent pinching realizations for poorly annealed glasses (c). To perform the averaging, we rotate each displacement field such that the vector connecting the two pinched particles in the initial configuration is aligned along a common direction.

The averaged response exhibits a four-lobed structure, with two stronger lobes oriented along the pinching direction and two weaker lobes oriented perpendicular to it. This structure is clearly visible in stable glasses and remains well defined in moderately annealed glasses. In poorly annealed glasses, however, the signal is considerably more diffuse because of inelastic effects and/or additional plastic rearrangements triggered during the pinching process.

We next compare the simulation results with the predictions of linear elasticity in Fig.~\ref{fig:mean_response_length_control}(d--f), for stable, moderately annealed, and poorly annealed glasses, respectively. The theoretical displacement fields are calculated using $u_1$ and $u_2$ from Eqs.~\eqref{eq:u_pin_theory_cartesian}, or equivalently Eqs.~\eqref{eq:u_pin_theory_polar}. The Poisson ratio $\nu$, which controls the relative amplitudes of the lobes parallel and perpendicular to the pinching direction, is measured independently from athermal quasistatic shear and compression simulations. 
In particular, increasing $\nu$ enhances the lobes perpendicular to the pinching direction. In the incompressible limit, $\nu \to 1$, the parallel and perpendicular lobes have equal magnitudes.
Since the effective dipole strength $aF$ is not independently known, the elastic prediction is determined only up to an overall multiplicative factor. We therefore set the dimensionless prefactor $aF/(\mu\sigma_{\rm LS}^{2})=1$, where $\sigma_{\rm LS}$ is our
unit of length.

We find remarkably good agreement between the simulations and linear-elasticity predictions, particularly for stable and moderately annealed glasses. This agreement indicates that linear elasticity accurately describes the response even at distances comparable to the particle scale and for a highly localized perturbation. The poorer agreement observed for poorly annealed glasses is expected, because multiple plastic or inelastic rearrangements occur in this case and are not captured by standard linear elasticity.

Having examined the mean response in terms of its angular dependence and compared it with the theoretical prediction, we now turn to its radial dependence. A hallmark of linear elasticity is the power-law decay of the displacement field with distance. To characterize this decay, we consider the mean displacement magnitude at distance $r$ from the pinched particles, $u(r)$, the probability distribution of the displacement magnitude, $P(u)$, defined as
\begin{align}
    u(r)
    =
    \frac{1}{N_{\Omega(r)}}
    \left\langle
    \sum_{i\in\Omega(r)}
    \left|
    {\bf r}_{i}(\Delta\ell)-{\bf r}_{i}(0)
    \right|
    \right\rangle,
\end{align}
and
\begin{align}
    P(u)
    =
    \frac{1}{N}
    \left\langle
    \sum_{i=1}^{N}
    \delta\left(
    u-
    \left|
    {\bf r}_{i}(\Delta\ell)-{\bf r}_{i}(0)
    \right|
    \right)
    \right\rangle ,
\end{align}
where $\delta(\cdot)$ is the delta function.
Here, $\Omega(r)$ denotes the annular region spanning the interval
$\left[r,r+d r\right)$, where $r$ is measured from the center of the
two pinched particles, and $N_{\Omega(r)}$ is the number of particles
contained in this annulus.
We set $d r=0.488$.

In a general spatial dimension $d$, linear elasticity predicts
\begin{align}
    u(r) \sim r^{-(d-1)},
\end{align}
which implies
\begin{align}
    P(u) \sim u^{-\frac{2d-1}{d-1}}.
\end{align}
In the present two-dimensional case, these predictions reduce to
\begin{align}
    u(r) \sim r^{-1},
    \qquad
    P(u) \sim u^{-3},
\end{align}
consistent with Eqs.~\eqref{eq:u_pin_theory_polar}.

\begin{figure*}
\includegraphics[width=0.68\columnwidth]{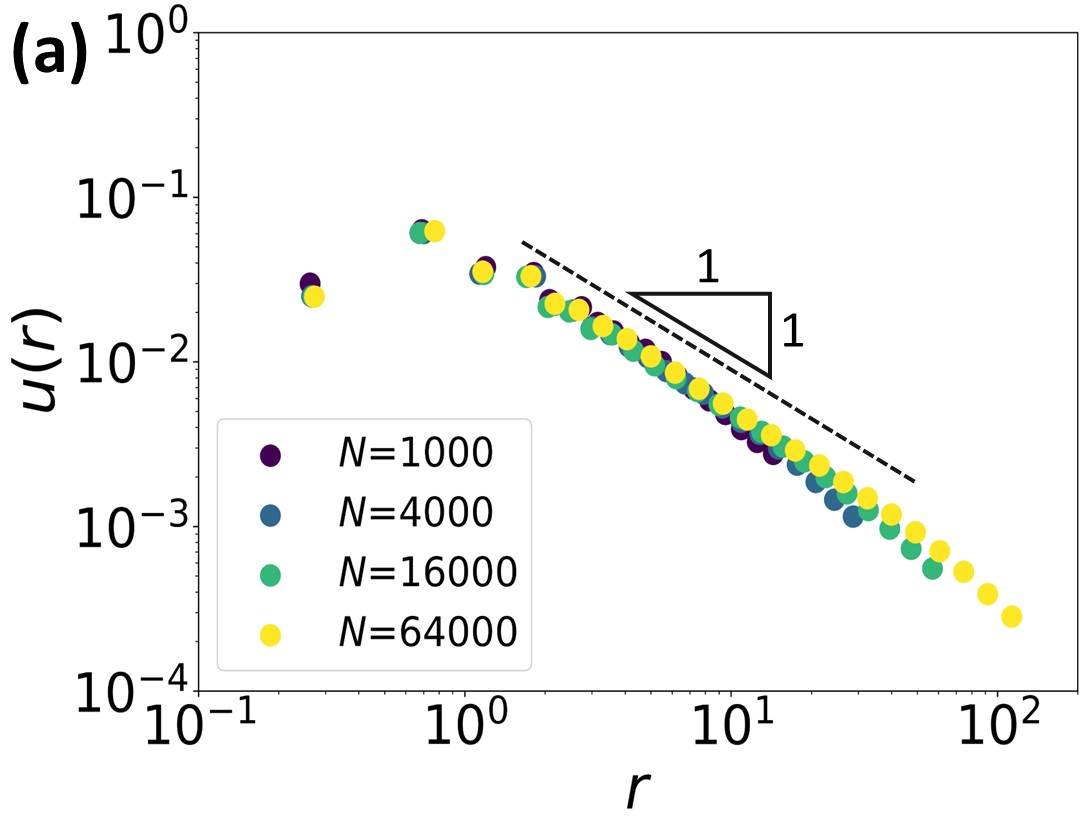}
\includegraphics[width=0.68\columnwidth]{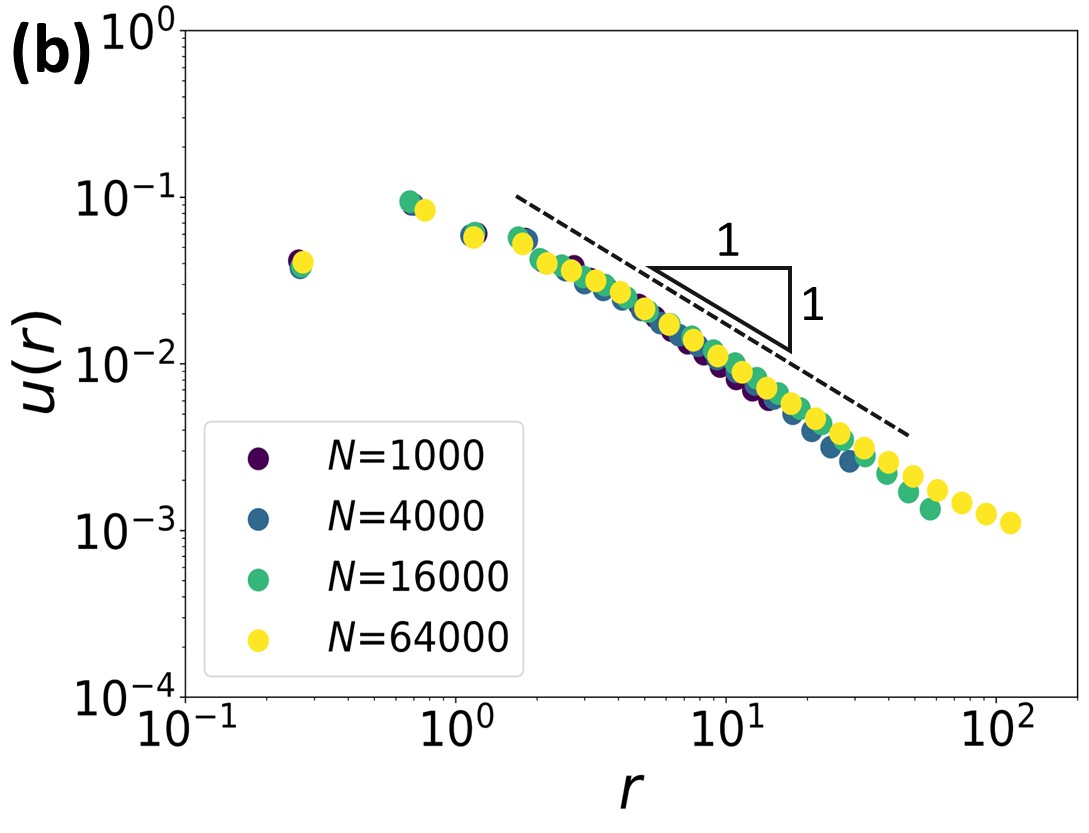}
\includegraphics[width=0.68\columnwidth]{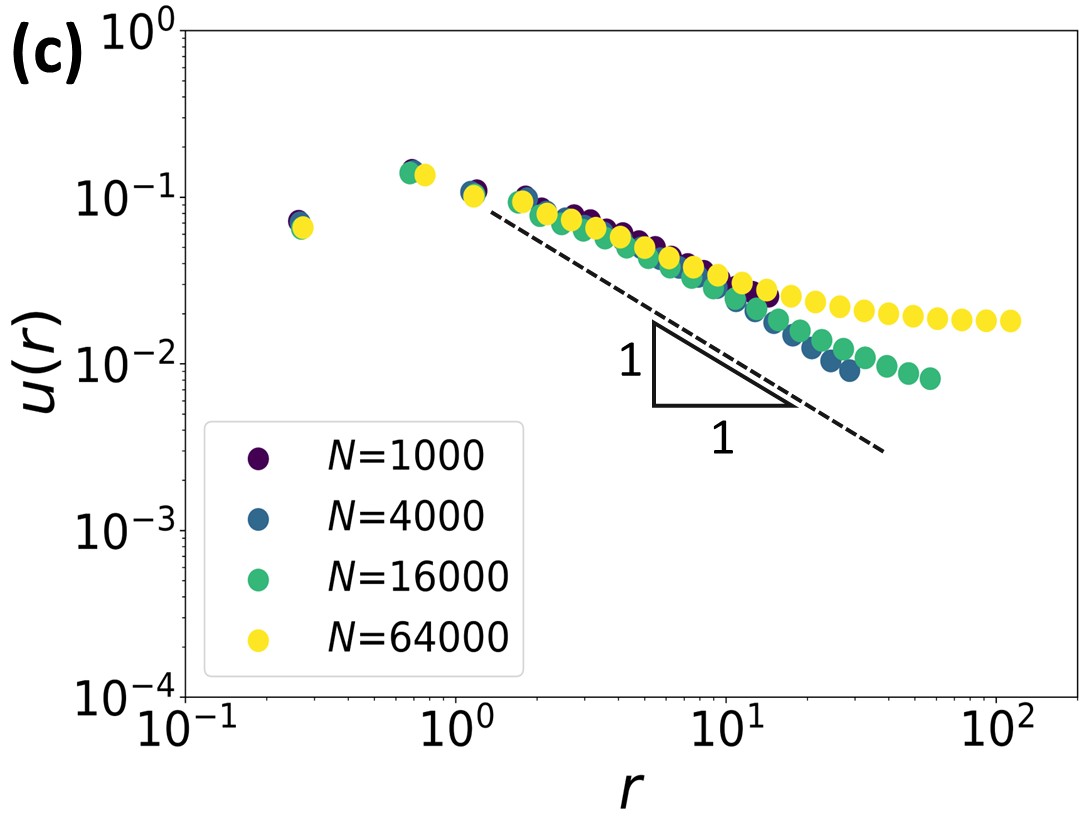}
\includegraphics[width=0.68\columnwidth]{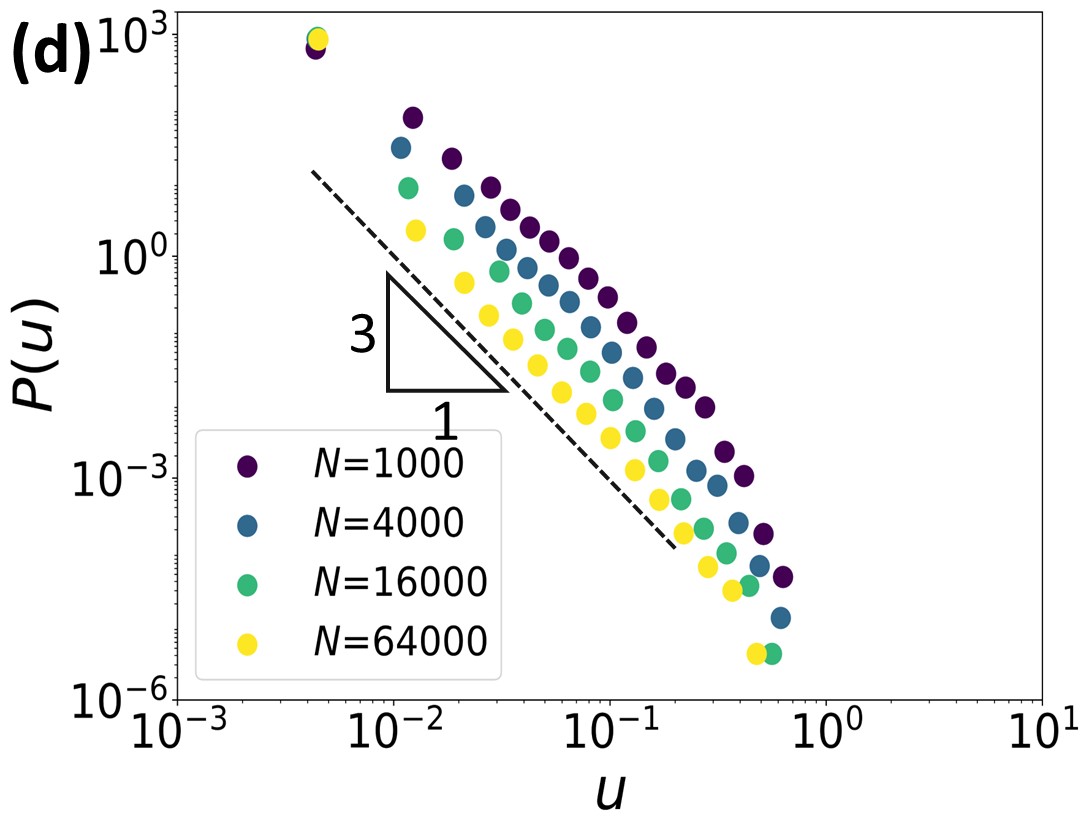}
\includegraphics[width=0.68\columnwidth]{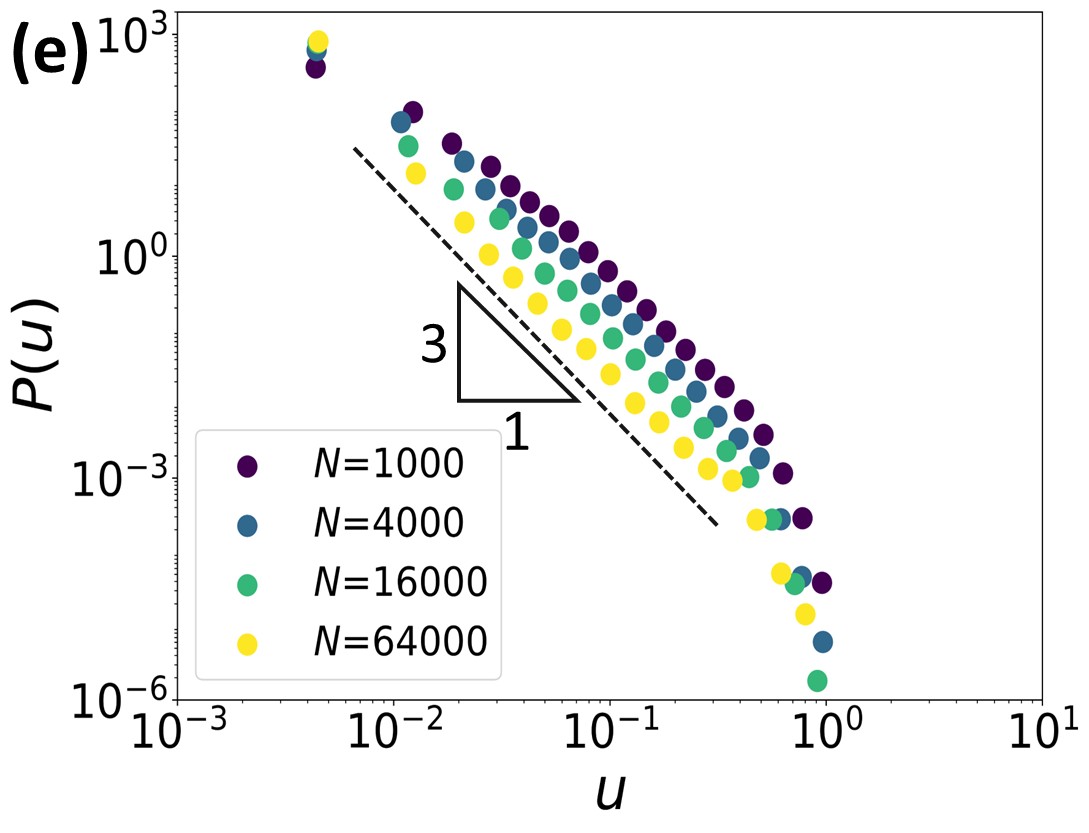}
\includegraphics[width=0.68\columnwidth]{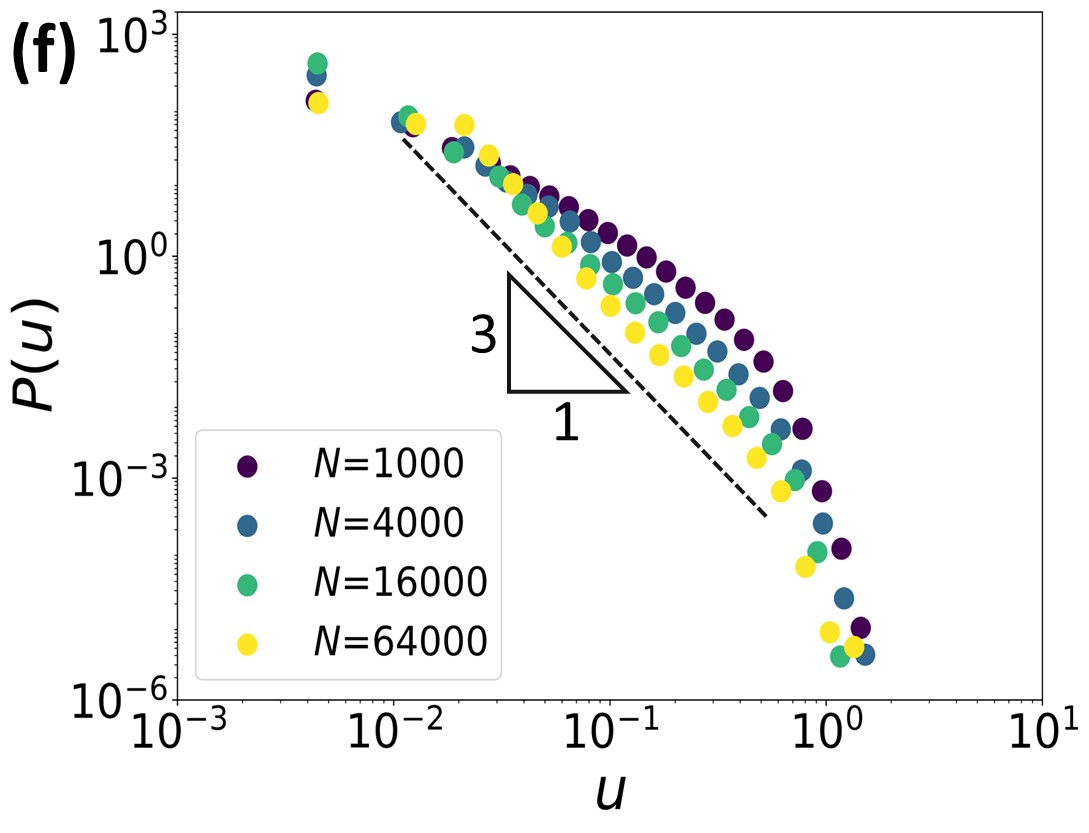}
\caption{
(a--c) Mean radial decay of the displacement magnitude, $u(r)$, induced by pinching at $\Delta\ell=0.5$ for (a) stable glasses, (b) moderately annealed glasses, and (c) poorly annealed glasses, for system sizes $N=1000$, $4000$, $16000$, and $64000$. The dashed line indicates the linear-elasticity prediction, $u(r)\sim r^{-1}$. 
(d--f) Probability distribution of the displacement magnitude, $P(u)$, induced by pinching at $\Delta\ell=0.5$ for (d) stable glasses, (e) moderately annealed glasses, and (f) poorly annealed glasses, for the same system sizes. The dashed line indicates the predicted scaling, $P(u)\sim u^{-3}$.}
\label{fig:decays_length_control}
\end{figure*}

In Figs.~\ref{fig:decays_length_control}(a--c), we show $u(r)$ for stable (a), moderately annealed (b), and poorly annealed (c) glasses for several system sizes. For stable glasses, we observe a clear power-law decay whose spatial range increases with system size, consistently with the linear-elasticity prediction, $u(r)\sim r^{-1}$. Remarkably, this scaling regime sets in already at distances of only a few particle diameters.
Moderately annealed glasses exhibit a similar behavior. However, for the largest system size, a slight upturn of $u(r)$ is observed at large $r$, possibly due to weak inelastic effects or rare plastic events that perturb the purely elastic displacement field~\cite{lemaitre2021anomalous}. This effect is considerably stronger in poorly annealed glasses: as the system size increases, the upturn at large $r$ becomes more pronounced, presumably because of the increasing occurrence of plastic events and/or inelastic responses, leading to clear deviations from the prediction of linear elasticity.

We note that previous work on marginal networks and jammed
particle packings identified a characteristic length beyond
which the response to a local force dipole recovers the
continuum-elastic scaling, with this length growing upon
approaching unjamming or an elastic
instability~\cite{lerner2014breakdown}. Although an analogous
crossover length may be defined operationally in our LJ-type
glasses, its relation to the length scale identified in marginal
networks is not straightforward.

We next turn our attention to the probability distribution of the displacement magnitude, $P(u)$. In Figs.~\ref{fig:decays_length_control}(d--f), we show $P(u)$ for stable (d), moderately annealed (e), and poorly annealed (f) glasses for several system sizes.

For stable glasses, the data exhibit a clear scaling regime, $P(u)\sim u^{-3}$, whose range extends as the system size increases, consistently with the prediction of linear elasticity. A similar trend is observed for moderately annealed glasses. Remarkably, the $u^{-3}$ scaling remains clearly visible even for poorly annealed glasses for large system sizes. Unlike $u(r)$, which is measured relative to the pinching center and is therefore sensitive to additional plastic events and inelastic effects occurring elsewhere in the system, $P(u)$ does not retain information about the spatial location of each displacement. It therefore captures the distribution of displacement magnitudes generated by both the original pinching event and additional plastic rearrangements throughout the system. Therefore, the observation of the $P(u)\sim u^{-3}$ scaling alone does not necessarily indicate a purely elastic response.

The results presented above are based on the length-controlled protocol.
In Appendix~\ref{sec:force_control}, we additionally investigate the
force-controlled protocol using relatively small imposed forces,
corresponding approximately to the elastic regime of the
length-controlled protocol. We find qualitatively the same elastic
response in terms of the individual pinching trajectories, the angular
dependence and radial decay of the displacement field, and the
probability distribution of the displacement magnitude.
These results indicate that the elastic response to pinching is robust
with respect to the choice of the control protocol.

\subsection{Averaged responses: Plastic regime}

We next study the averaged response at a sufficiently large pinching
extension, $\Delta \ell = 2.0$. At this extension, cage breaking occurs
at the pinched particles even in stable glass samples, indicating that
the response has entered the plastic regime. We therefore refer to this
case as the ``plastic regime.''

Special care is required when choosing the principal axis used to
average the displacement fields over different pinching realizations.
In the plastic regime, we still observe quadrupolar, Eshelby-like
displacement fields, as illustrated in
Figs.~\ref{fig:snapshot_Tini0.25} and
\ref{fig:snapshot_Tini0.35}. However, the orientation of the
quadrupolar pattern fluctuates from one realization to another and can
deviate appreciably from the initial pinching direction~\cite{xu2021atomic}. In particular,
the direction of the two lobes with the largest displacement magnitude
is not necessarily aligned with the vector connecting the two pinched
particles in the initial configuration.

We therefore introduce a method to determine the principal axis
directly from the displacement field, as detailed in
Appendix~\ref{sec:principal_axis}. For each pinching realization, we
then rotate the coordinate system such that the extracted principal
axis $\phi_*$ is aligned along a common direction, and recompute the
averaged displacement field.

\begin{figure*}
\includegraphics[width=0.6\columnwidth]{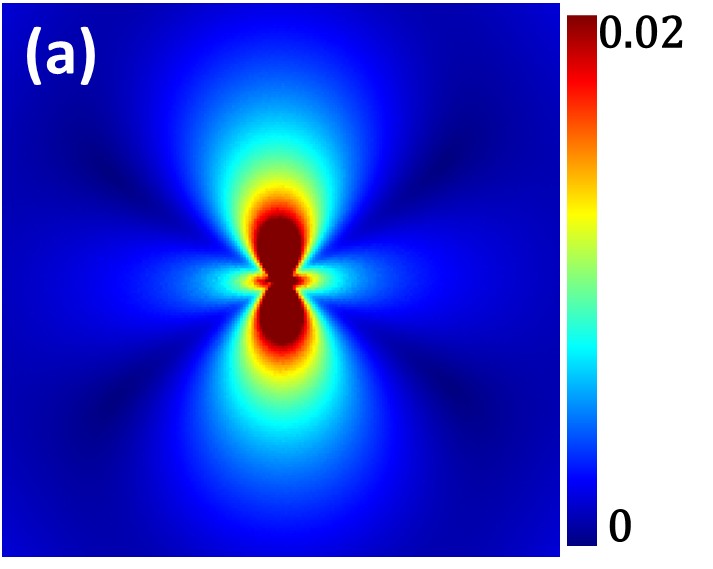}
\includegraphics[width=0.6\columnwidth]{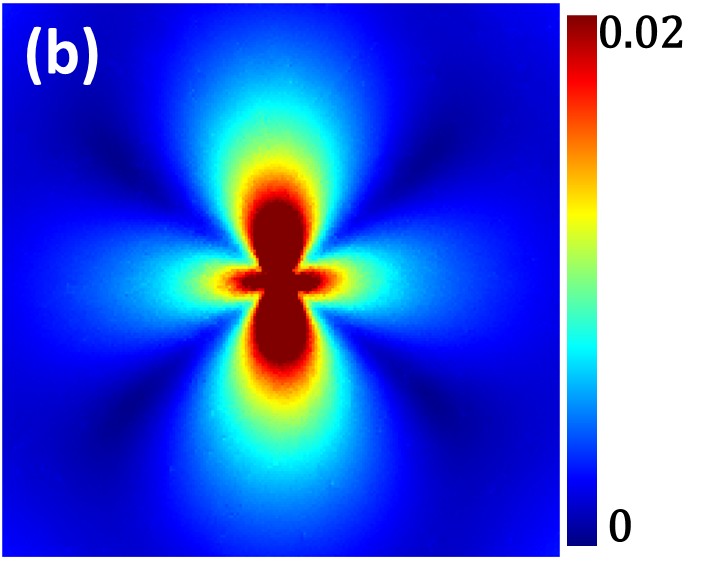}
\includegraphics[width=0.6\columnwidth]{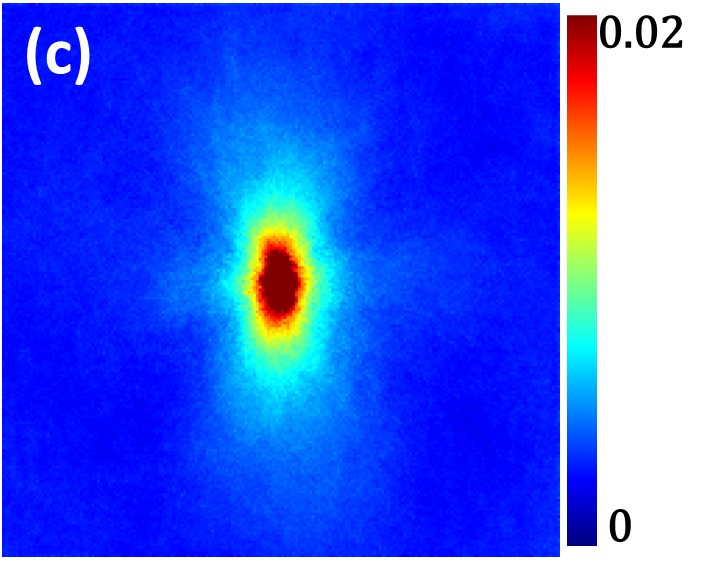}
\caption{(a--c) Magnitude of the displacement field induced by athermal quasistatic pinching at $\Delta \ell = 2.0$ for (a) stable glasses, $T_{\rm ini}=0.25$, (b) moderately annealed glasses, $T_{\rm ini}=0.35$, and (c) poorly annealed glasses, $T_{\rm ini}=1.0$. The results were obtained for $N=64000$ and averaged over $1000$ independent pinching realizations for (a), (b) and $5000$ for (c).}
\label{fig:average_anglar_plastic}
\end{figure*}

Figure~\ref{fig:average_anglar_plastic} shows the magnitude of the
averaged displacement field in the plastic regime for (a) stable,
(b) moderately annealed, and (c) poorly annealed glasses, respectively. 
From these data, we see that the response remains well described by linear elasticity theory, with a larger magnitude than in the case of $\Delta \ell = 0.5$ shown in Fig.~\ref{fig:mean_response_length_control}, except for the poorly annealed glass, for which significant deviations are observed. Therefore, except in the poorly annealed case, the induced plastic events, either at the origin (i.e., near the pinched particles) or additionally triggered in the far field, do not significantly modify the averaged response.
These results can be understood as follows. The elastic energy imposed by the pinching is released through a plastic event and becomes essentially zero around $\Delta \ell \approx 1.5$--$2.0$, as shown in Figs.~\ref{fig:individual_length_control}(d--f). The resulting plastic rearrangement can be described in terms of a local eigenstrain (or plastic strain), which generates a long-range stress redistribution and the associated Eshelby-like displacement field, reflecting the modification of the local equilibrium configuration. This situation is analogous to that in mechanically sheared systems, where external macroscopic loading induces local plastic events, commonly described as shear transformations.

\begin{figure*}
\includegraphics[width=0.68\columnwidth]{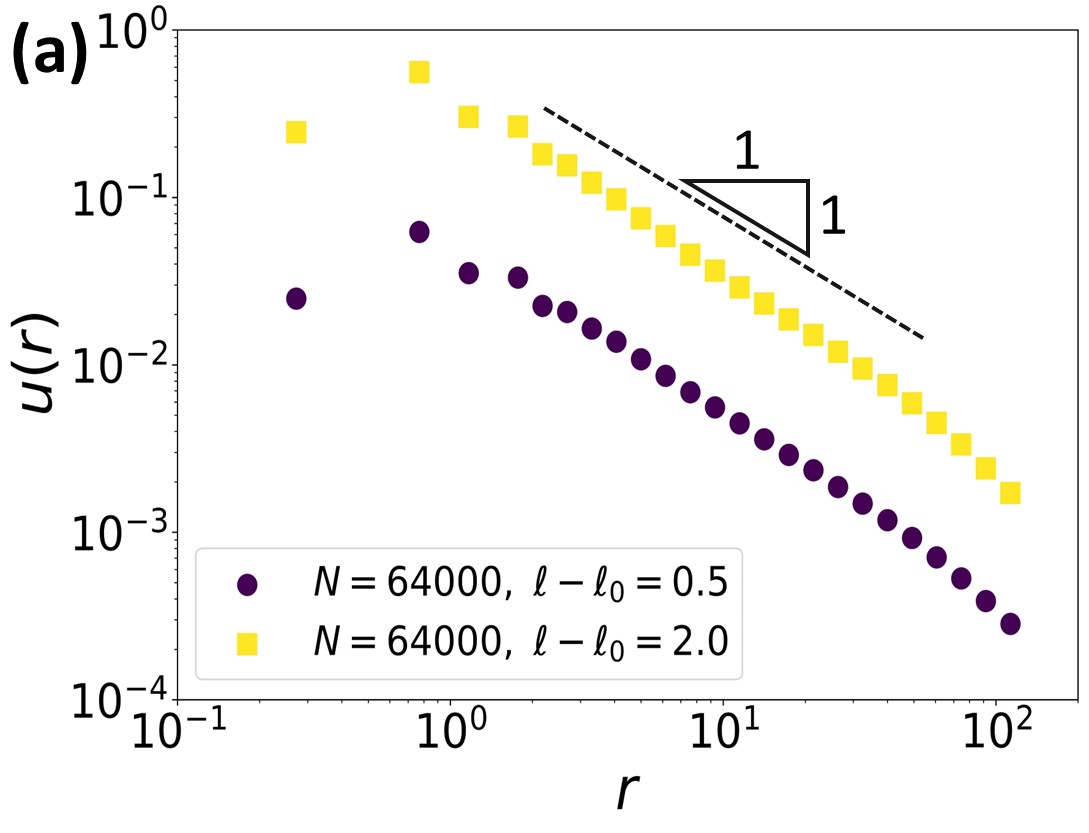}
\includegraphics[width=0.68\columnwidth]{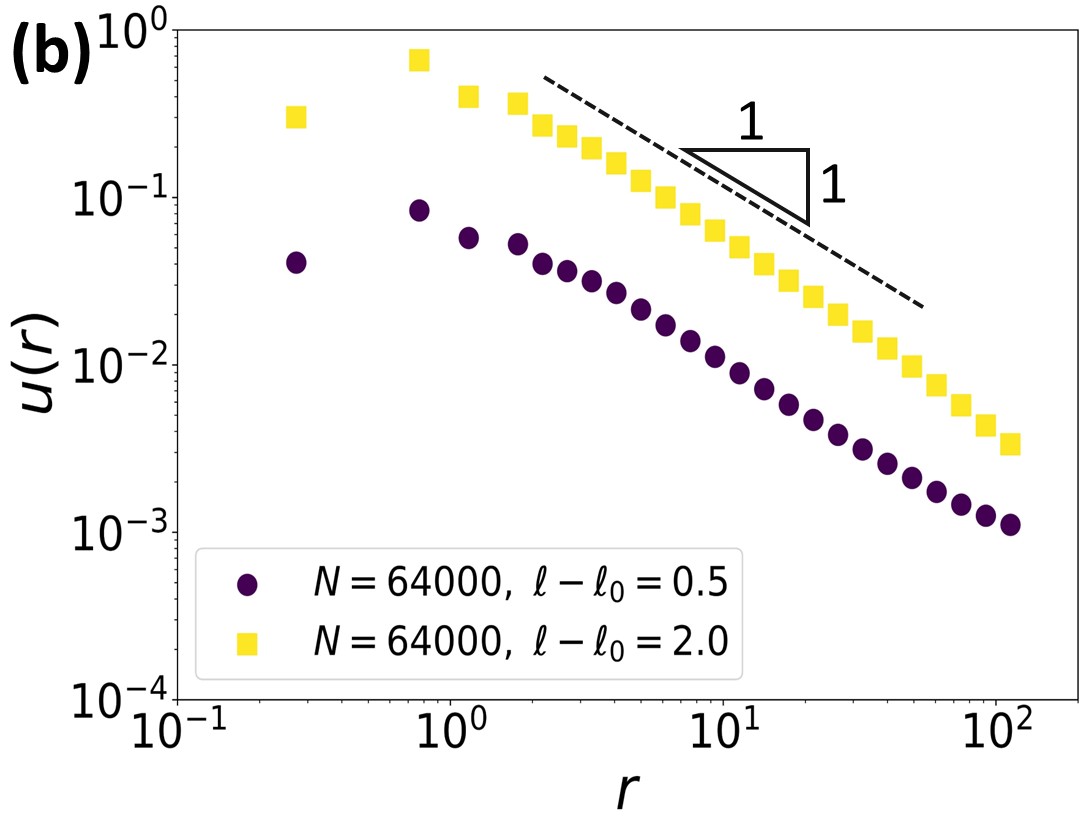}
\includegraphics[width=0.68\columnwidth]{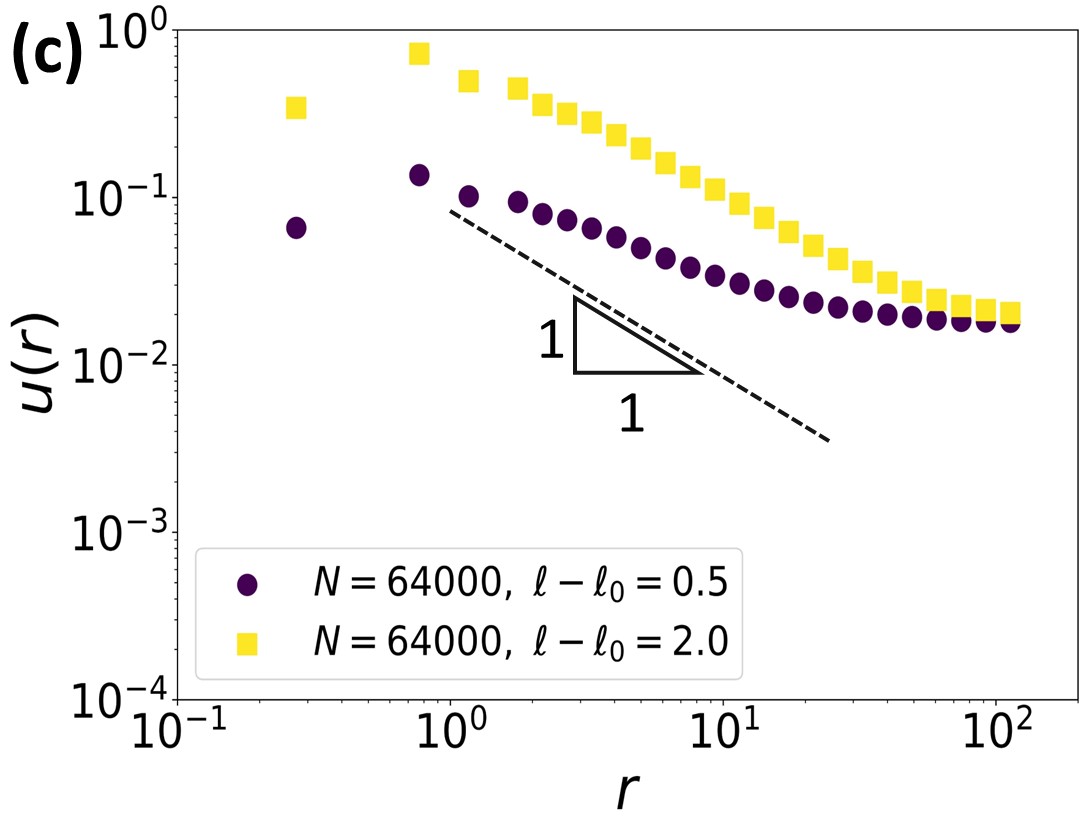}
\includegraphics[width=0.68\columnwidth]{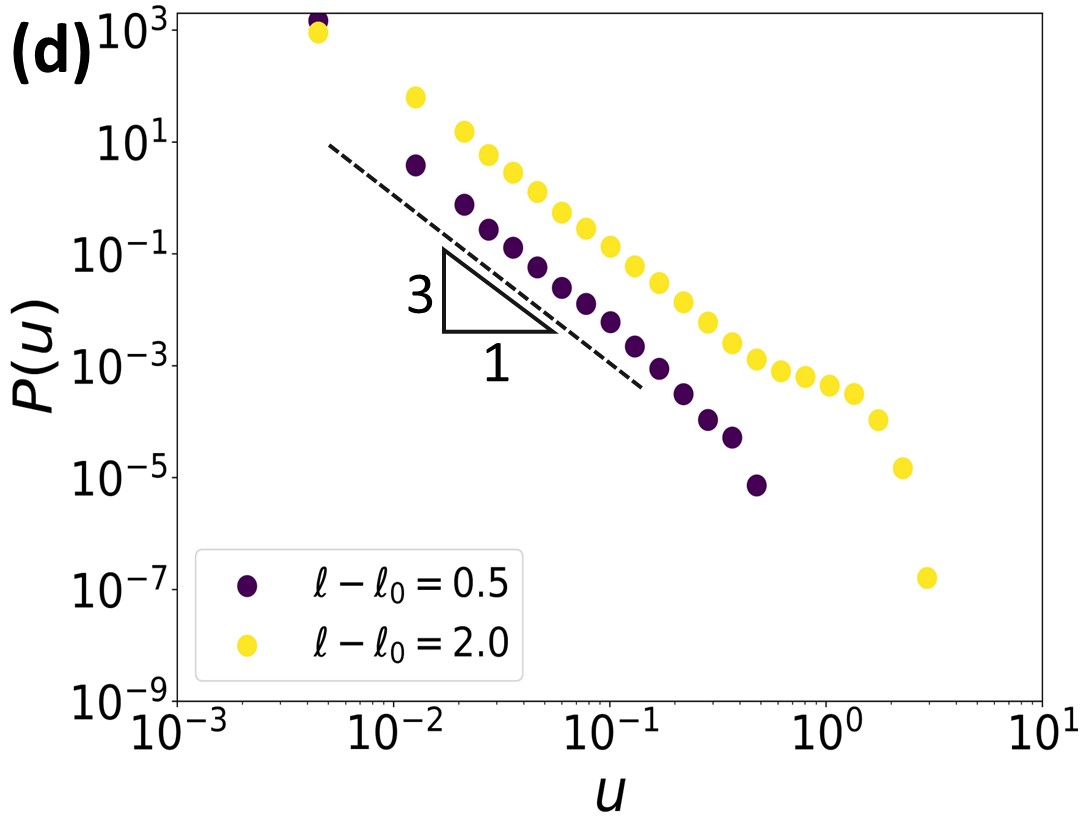}
\includegraphics[width=0.68\columnwidth]{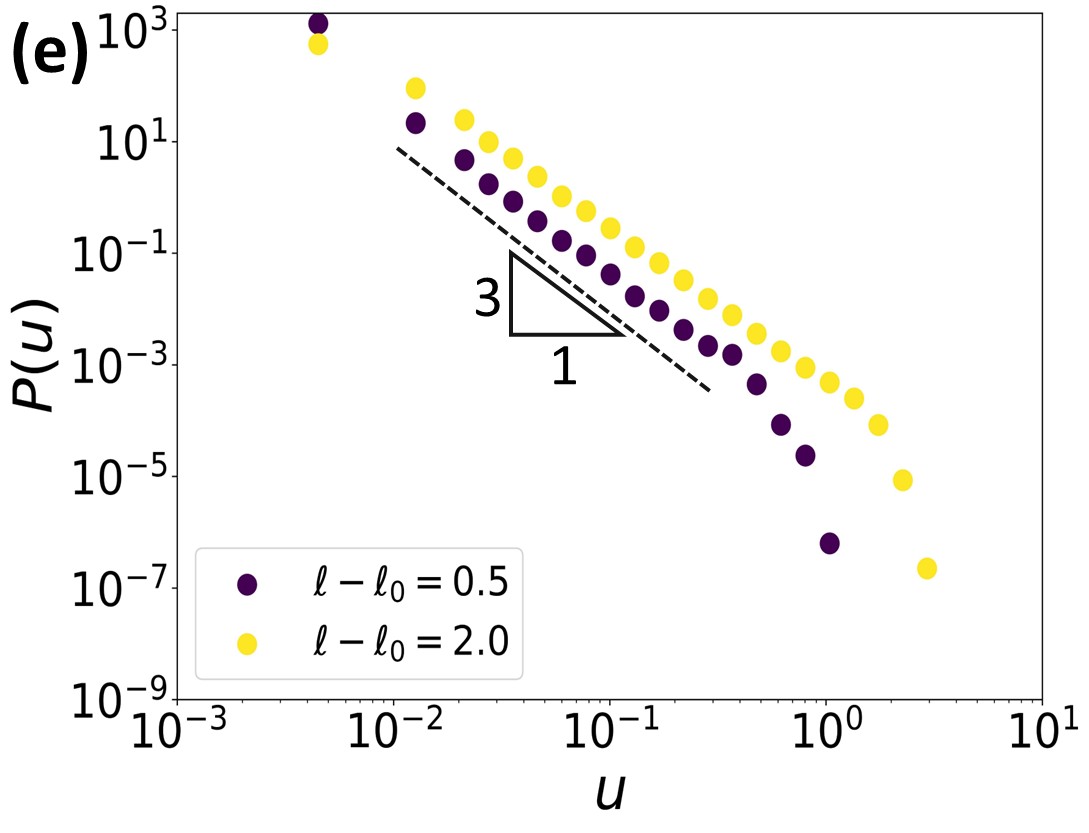}
\includegraphics[width=0.68\columnwidth]{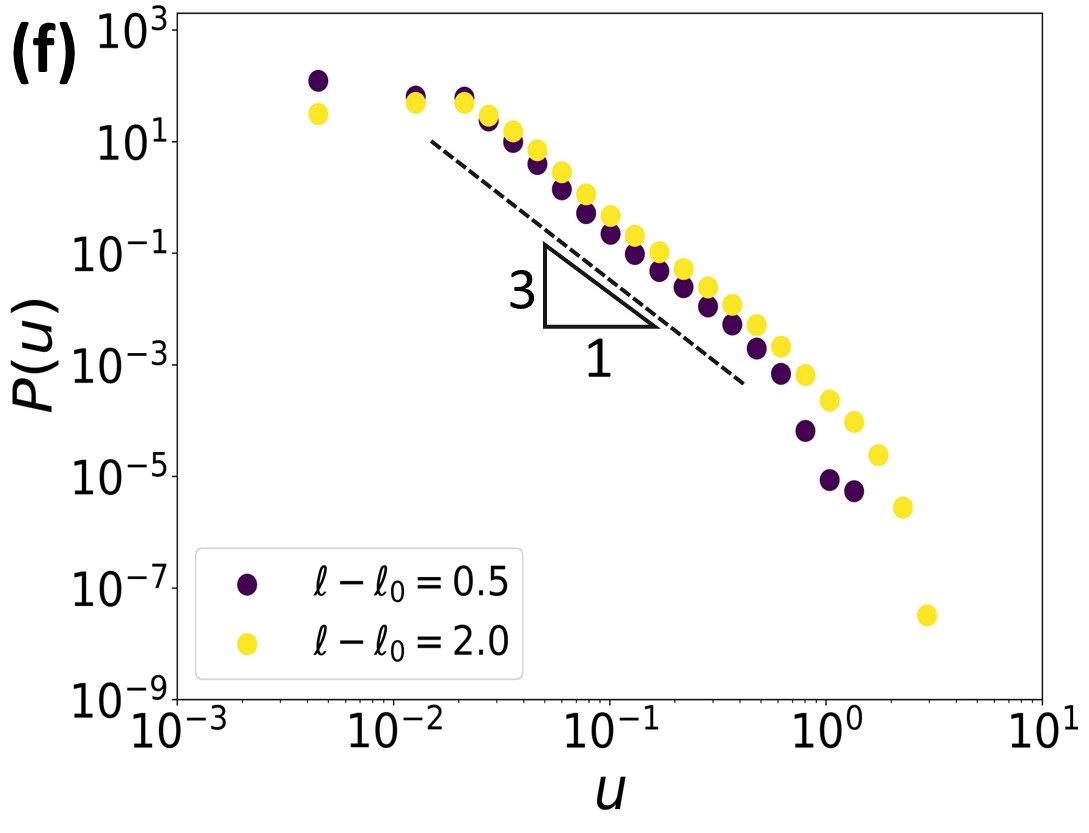}
\caption{(a--c) Mean radial decay of the displacement magnitude, $u(r)$,
comparing the elastic regime, $\Delta\ell=0.5$, and the plastic regime,
$\Delta\ell=2.0$, for (a) stable glasses, (b) moderately annealed
glasses, and (c) poorly annealed glasses, with system size $N=64000$.
The dashed line indicates the linear-elasticity prediction,
$u(r)\sim r^{-1}$.
(d--f) Probability distribution of the displacement magnitude, $P(u)$,
comparing $\Delta\ell=0.5$ and $\Delta\ell=2.0$ for (d) stable glasses,
(e) moderately annealed glasses, and (f) poorly annealed glasses, with
$N=64000$. The dashed line indicates the predicted scaling,
$P(u)\sim u^{-3}$.}
\label{fig:decays_length_control_2}
\end{figure*}

We then examine the radial decay of the mean displacement magnitude, $u(r)$, and the probability distribution, $P(u)$, in the plastic regime ($\Delta \ell = 2.0$) in Fig.~\ref{fig:decays_length_control_2}, for the largest system size, $N=64000$, and compare them with the corresponding data in the elastic regime ($\Delta \ell = 0.5$). Except for $u(r)$ in poorly annealed glasses, all data for $\Delta \ell = 2.0$ are consistent with the scaling predicted by linear elasticity, while exhibiting larger amplitudes than those for $\Delta \ell = 0.5$. These results further support the conclusion that pinching-induced plastic events generate Eshelby-like displacement fields.

Additionally induced plastic events, which are particularly prominent in poorly annealed glasses, will be discussed in more detail in the next subsection. From the behavior of $u(r)$ shown in Figs.~\ref{fig:decays_length_control} and \ref{fig:decays_length_control_2}, we see that these events strongly affect the response of poorly annealed glasses and can dominate over the elastic signal at large distances. Interestingly, this effect is even more pronounced for $\Delta \ell = 0.5$, where fewer plastic events occur, but the elastic contribution to the displacement field is also smaller.

\subsection{Pinching-induced plastic events}

We have observed that additional plastic events can be induced
outside the central pinching region by the long-range elastic
displacement field generated by the primary plastic rearrangement.
This effect is particularly pronounced in poorly annealed glasses.
We now characterize these pinching-induced plastic events more
quantitatively.

To quantify local plastic activity, we use the non-affine
displacement measure $D^2_{\rm min}$~\cite{falk1998dynamics}. For particle $i$, it is defined as
\begin{eqnarray}
    D^2_{{\rm min},i}(\Delta \ell)
    &=&
    \frac{1}{n_i}
    \sum_{j\in\mathcal{N}_i}
    \left|
    \left[
    {\bf r}_j(\Delta \ell)-{\bf r}_i(\Delta \ell)
    \right]
    \right.
    \nonumber \\
    &&
    \left.
    -
    (I+E_*)
    \left[
    {\bf r}_j(0)-{\bf r}_i(0)
    \right]
    \right|^2 .
    \label{eq:D2min}
\end{eqnarray}
Here, $\mathcal{N}_i$ denotes the set of neighbors of particle $i$
in the initial configuration, $n_i$ is the number of such neighbors,
$I$ is the identity tensor, and $E_*$ is the best-fit local affine
deformation tensor. Thus, $D^2_{\rm min}$ measures the component
of the local particle motion that cannot be described by an affine
deformation. We define the neighbors of particle $i$ as particles
located within a cutoff distance $r_{\rm min}$ in the initial
configuration, where $r_{\rm min}=1.4$ is chosen near the first
minimum of the radial distribution function.

We next construct a coarse-grained $D^2_{\rm min}$ field,
$\overline{D^2_{\rm min}}({\bf r},\Delta\ell)$, by assigning the
particle-level $D^2_{\rm min}$ values to a regular spatial grid with
spacing $\xi=2$. Maps of
$\overline{D^2_{\rm min}}({\bf r},\Delta\ell)$ reveal a strong
dependence of the spatial organization of plastic activity on glass
stability. In stable glasses, the plastic response is concentrated
around the pinched particles and their immediate neighbors, and
remains strongly localized even as $\Delta\ell$ increases. In
contrast, poorly annealed glasses exhibit additional plastic events
away from the pinching center. As $\Delta\ell$ increases, these
secondary events become stronger and spread over increasingly large
regions of the system. This behavior indicates that the plastic
response becomes progressively more delocalized with decreasing
glass stability, consistently with the individual displacement fields
discussed above.

\begin{figure*}
\includegraphics[width=0.6\columnwidth]{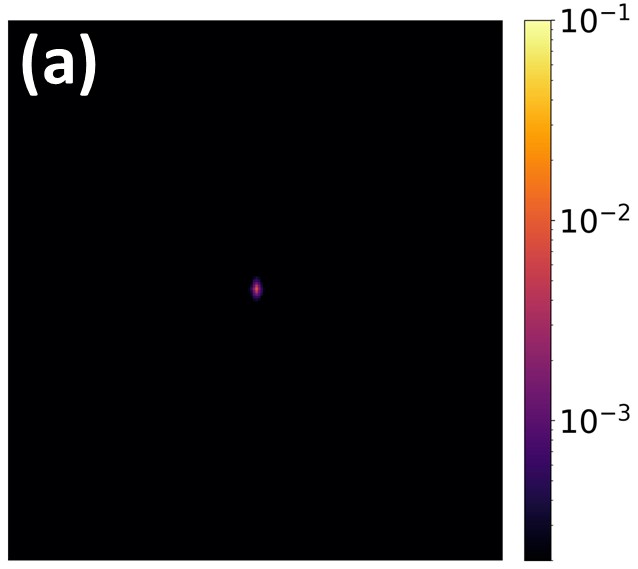}
\includegraphics[width=0.6\columnwidth]{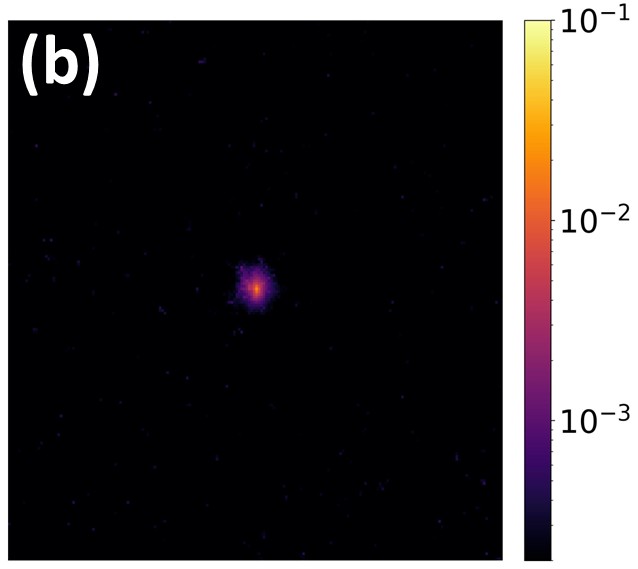}
\includegraphics[width=0.6\columnwidth]{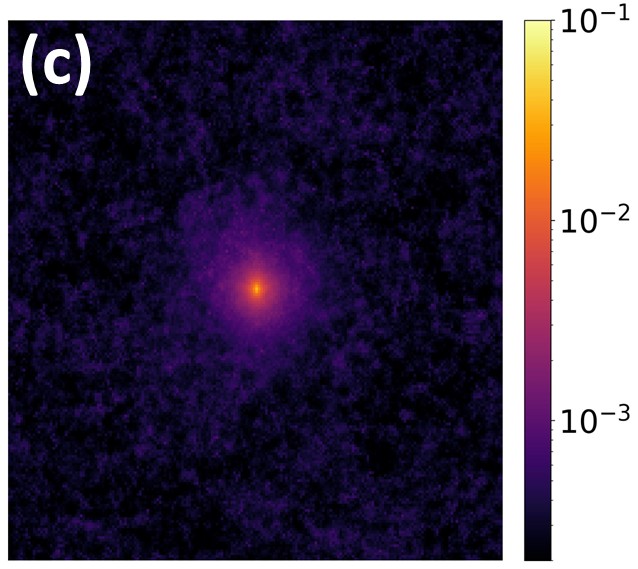}
\includegraphics[width=0.6\columnwidth]{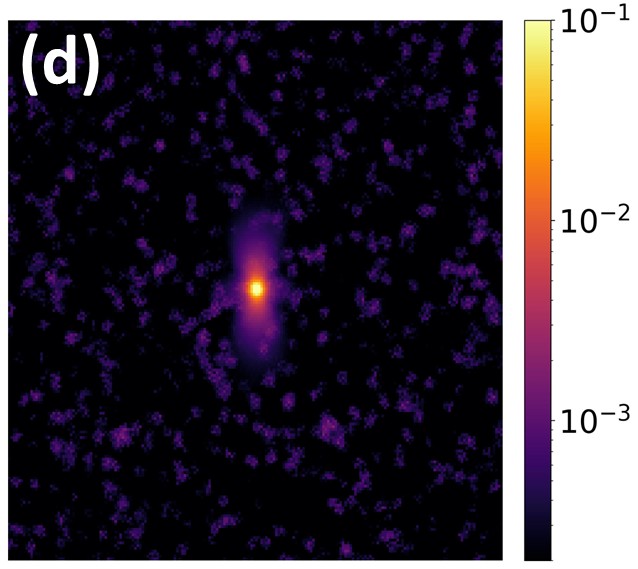}
\includegraphics[width=0.6\columnwidth]{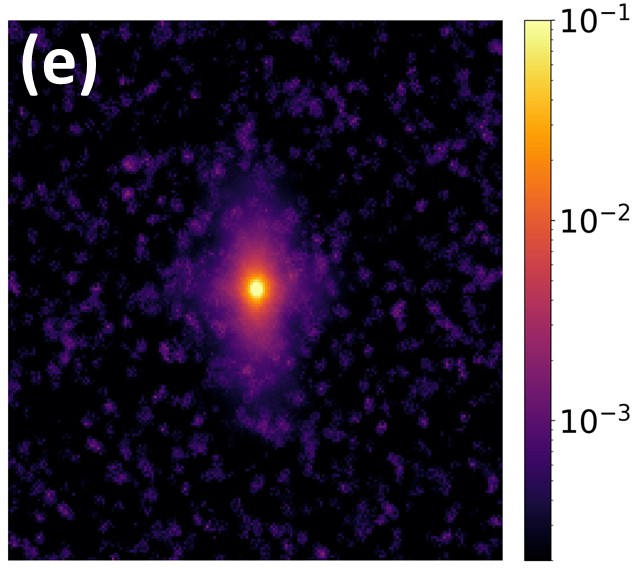}
\includegraphics[width=0.6\columnwidth]{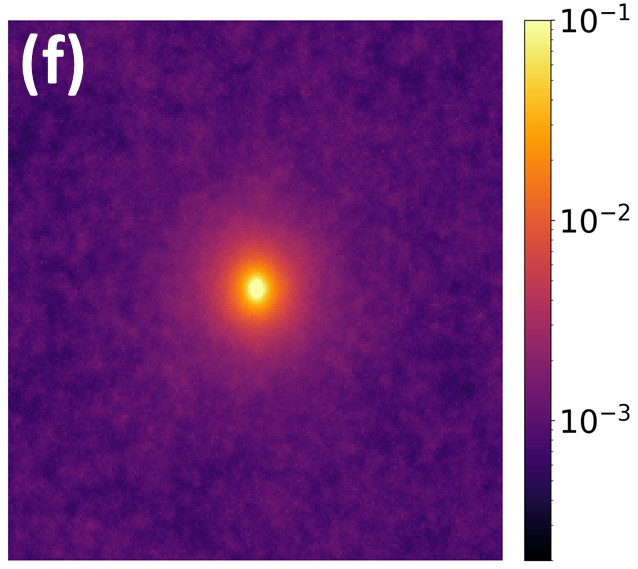}
\caption{(a--c) Plastic response function $R_2({\bf r},\Delta\ell)$ at
$\Delta\ell=0.5$ and (d--f) at $\Delta\ell=2.0$ for
(a,d) stable, (b,e) moderately annealed, and
(c,f) poorly annealed glasses, respectively.}
\label{fig:plastic_response_function}
\end{figure*}

To characterize this behavior statistically, we introduce the
plastic response function~\cite{puosi2016plastic},
\begin{align}
R_2({\bf r},\Delta\ell)
=
\left\langle
\overline{D^2_{\rm min}}({\bf r},\Delta\ell)
\right\rangle ,
\label{eq:R2}
\end{align}
where the brackets denote an average over independent pinching
realizations. Figure~\ref{fig:plastic_response_function} shows $R_2({\bf r},\Delta\ell)$ at $\Delta\ell=0.5$, corresponding to the elastic regime, and at $\Delta\ell=2.0$, corresponding to the plastic regime. The spatial structure of the plastic response depends strongly on glass stability. At $\Delta\ell=0.5$, the non-affine displacements in stable and moderately annealed glasses are confined to the central region around the pinched particles, as shown in Figs.~\ref{fig:plastic_response_function}(a) and (b), respectively. In poorly annealed glasses, by contrast, plastic activity already spreads outward in the elastic regime, forming a corona around the center (Fig.~\ref{fig:plastic_response_function}(c), consistent with the behavior discussed above.

In the plastic regime at $\Delta\ell=2.0$, the stable glasses exhibit strong plastic activity concentrated within the plastic core directly induced by pinching (Fig.~\ref{fig:plastic_response_function}(d)). Additional plastic events remain largely confined to the vicinity of the core and tend to propagate along the direction parallel to the pinching axis. The moderately annealed glasses exhibit a more delocalized plastic response, while retaining a preferential propagation of plastic activity parallel to the pinching axis, as shown in Fig.~\ref{fig:plastic_response_function}(e). In the poorly annealed glasses, by contrast, plastic events spread over a much broader region, and the response becomes increasingly delocalized and isotropic, with no clear preferred direction (Fig.~\ref{fig:plastic_response_function}(f)).

\section{Conclusion and Discussions}
\label{sec:conclusion}

We numerically studied the mechanical response of amorphous solids to pinching,
in which a force dipole is applied to a pair of neighboring particles.

Within linear elasticity, we established the connection between pinching
and a shear transformation. A shear transformation can be represented
as a pair of perpendicular force dipoles, and, in the incompressible
limit, the displacement field generated by a single pinching force
dipole is exactly one half of that generated by the corresponding
pinching--contraction event representing a shear transformation.

We performed molecular simulations of athermal quasistatic pinching,
in which the imposed pinching extension $\Delta\ell$ is increased
incrementally and the energy is minimized after each increment, in
analogy with athermal quasistatic shear simulations. We systematically
investigated the response as a function of glass stability, system
size, and the magnitude of the imposed pinching extension.

For sufficiently small $\Delta\ell$, the system responds predominantly
elastically. The resulting displacement fields are quantitatively
consistent with the predictions of linear elasticity, except for
poorly annealed glasses, where inelasticity and occasional plastic
rearrangements lead to deviations from a purely elastic response.
Remarkably, the continuum prediction of linear elasticity remains
accurate down to distances of only a few particle diameters, despite
the highly localized nature of the applied perturbation.

For sufficiently large $\Delta\ell$, pinching induces a plastic
rearrangement through cage breaking around the two pinched particles.
The nature of the resulting plastic response depends strongly on glass
stability. In stable glasses, plastic activity remains predominantly
localized around the pinching core, whereas in poorly annealed glasses
the primary plastic event can trigger additional rearrangements over
much larger distances, resulting in a more delocalized and
avalanche-like response. We quantified this effect using the local
non-affine displacement measure $D^2_{\rm min}$ and the corresponding plastic
response function, showing how plastic activity generated by the
primary rearrangement at the pinching core propagates throughout the
system as glass stability decreases.

Our study provides a detailed picture of the elementary mechanism of
pinching and complements previous works that have used pinching to probe
the local softness of glasses and the activation barriers of supercooled
liquids. Pinching therefore offers an alternative way to drive amorphous
solids, distinct from thermal activation, global shear deformation, or
self-propulsion in active matter.

An interesting direction for future work is to directly perturb the
four-particle key cores recently identified by Hu and Tanaka~\cite{hu2025unveiling} as the
elementary defects underlying quasi-localized modes. By imposing a
controlled local deformation on these four-particle key cores and
monitoring the resulting response, one could test whether they indeed
play a central role in plastic rearrangements and determine how
efficiently they generate long-range elastic fields and trigger
subsequent plastic activity.

Another interesting direction for future work is to study the response to
multiple pinching events and the resulting collective dynamics, in order
to understand how repeated local driving gives rise to relaxation
processes in amorphous solids. Such a protocol may also provide a simple
model for glasses under irradiation, where the accumulation of radiation-induced local
rearrangements or cage-breaking events act as localized sources of
mechanical perturbations~\cite{ruta2017hard,dallari2023stochastic,PhysRevX.13.041031}. In this perspective, the analogy drawn here between pinching events and the shear transformations induced by a mechanical loading would support the notion of "radiation induced yielding" introduced in Ref.~\cite{PhysRevX.13.041031}

\begin{acknowledgments}
We thank Anaël Lemaitre, Giulio Monaco,  Noël Jakse, and Beatrice Ruta for insightful discussions.
MO thanks the support by MIAI@Grenoble Alpes and the Agence Nationale de la Recherche under France 2030 with the reference ANR-23-IACL-0006. All the computations presented in this paper were performed using the GRICAD infrastructure (https://gricad.univ-grenoble-alpes.fr), which is supported by Grenoble research communities.
\end{acknowledgments}

\appendix

\section{Force-controlled protocol}
\label{sec:force_control}

In addition to the length-controlled protocol presented in the main text,
we also consider a force-controlled protocol. In this protocol, the
magnitude of the applied pinching force is directly controlled. The effective
energy is given by
\begin{equation}
E({\bf r}^N)
=
U({\bf r}^N)
-
F_{\rm p}\left|{\bf r}_i-{\bf r}_j\right|,
\label{eq:potential_force}
\end{equation}
where $F_{\rm p}$ denotes the externally imposed pinching force.
Starting from $F_{\rm p}=0$, we increase $F_{\rm p}$ in increments of
$dF_{\rm p}=0.1$. After each increment, we minimize the energy given by
Eq.~(\ref{eq:potential_force}) using the FIRE algorithm.

\begin{figure*}
\includegraphics[width=0.66\columnwidth]{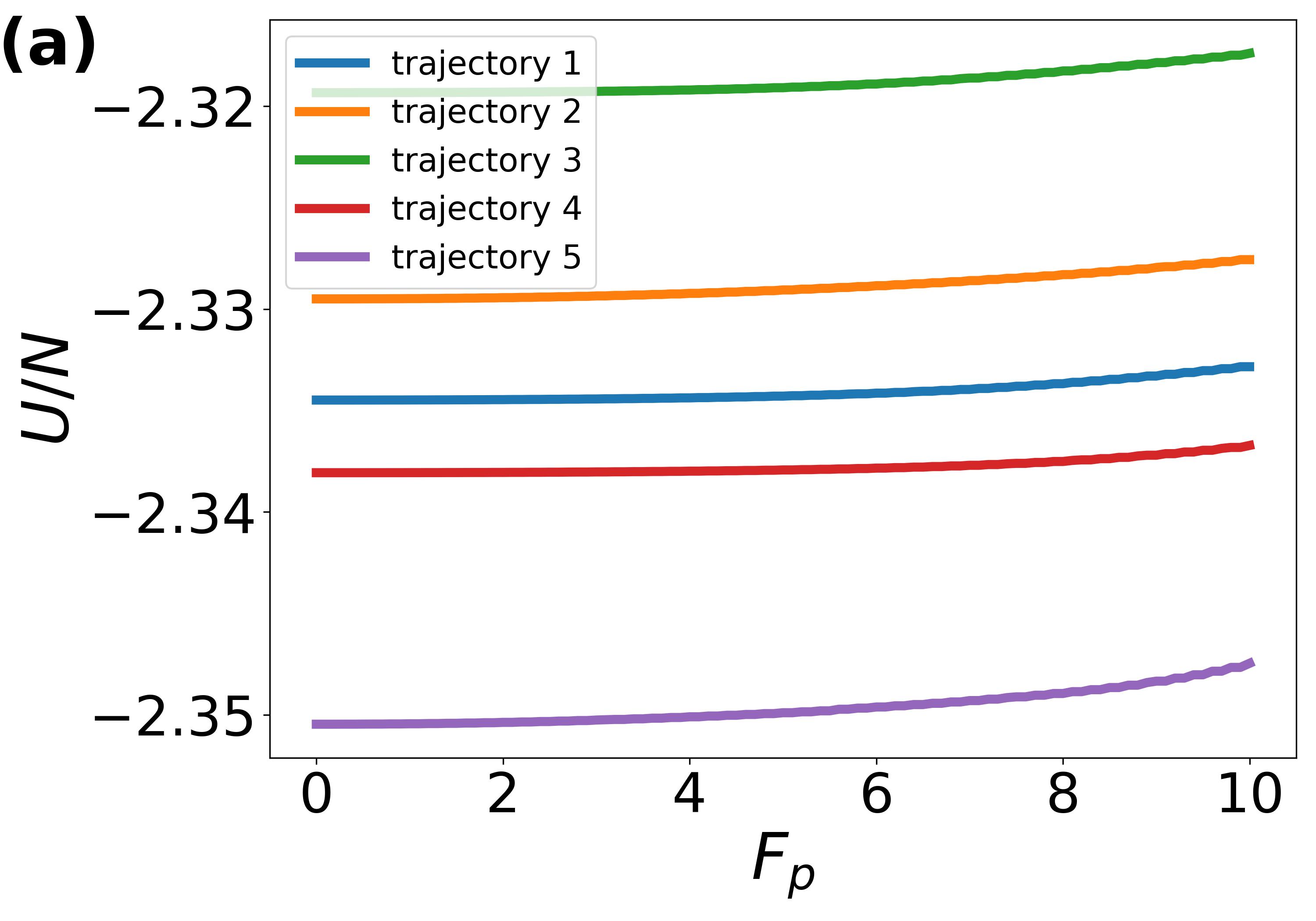}
\includegraphics[width=0.66\columnwidth]{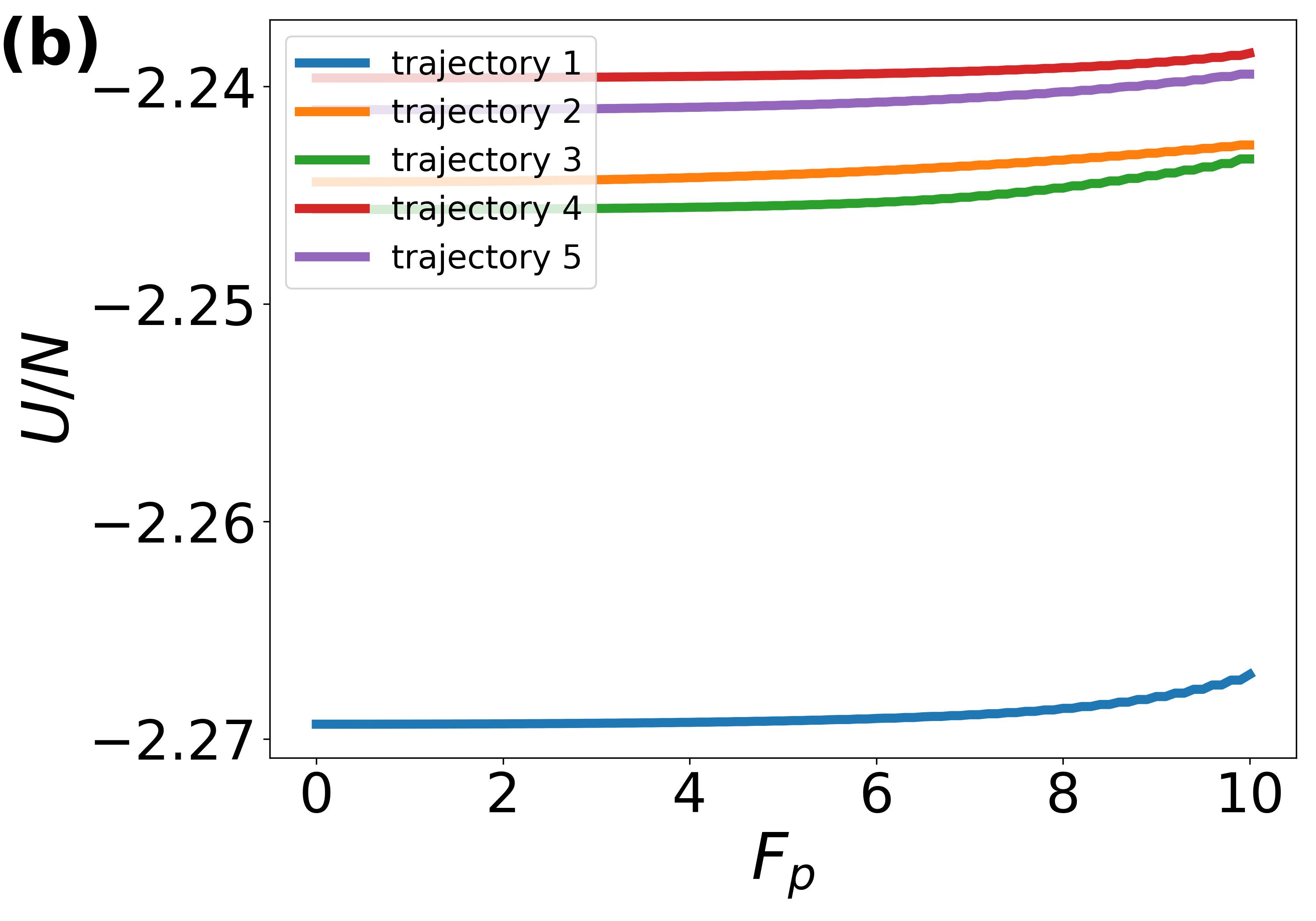}
\includegraphics[width=0.66\columnwidth]{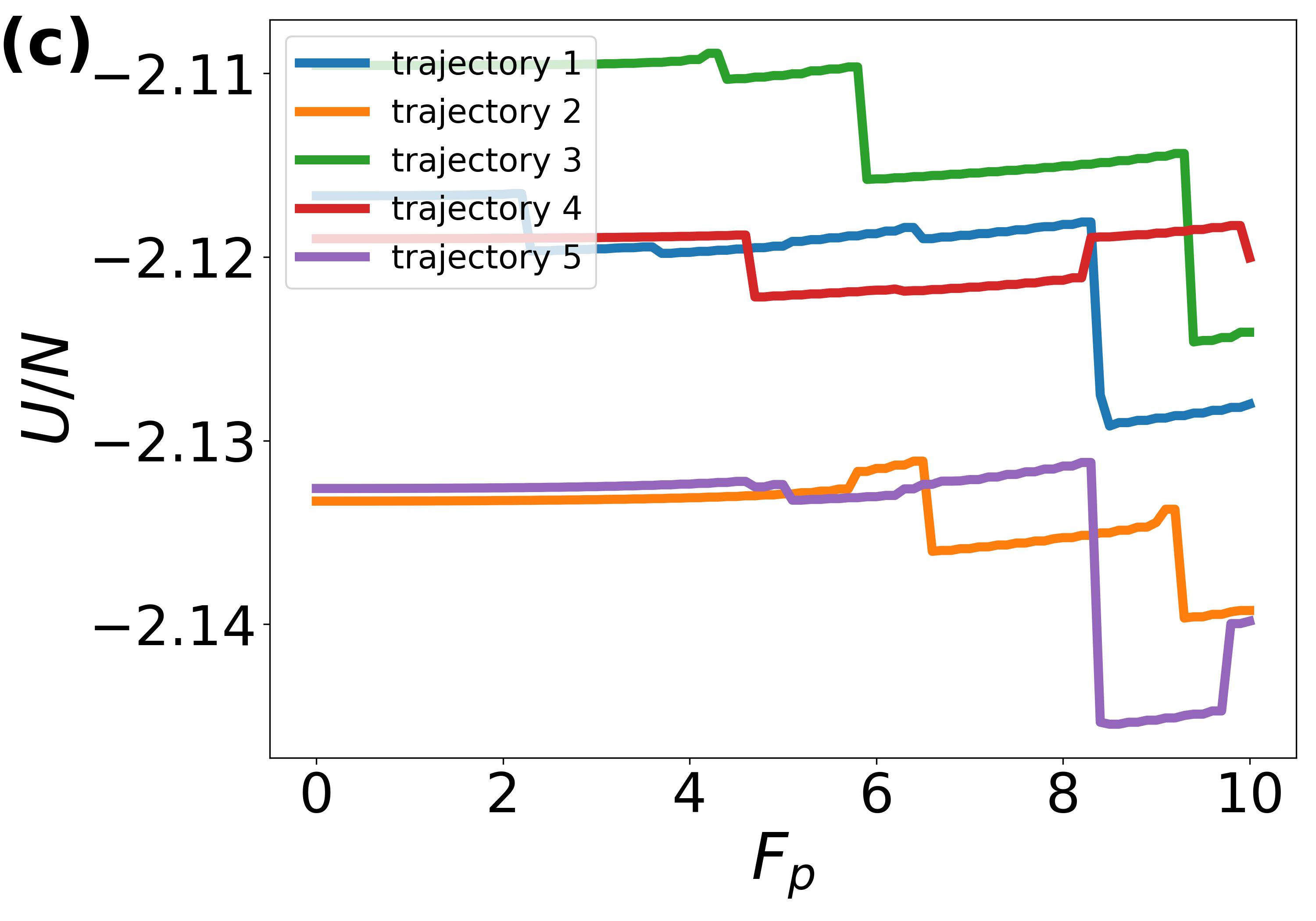}
\includegraphics[width=0.66\columnwidth]{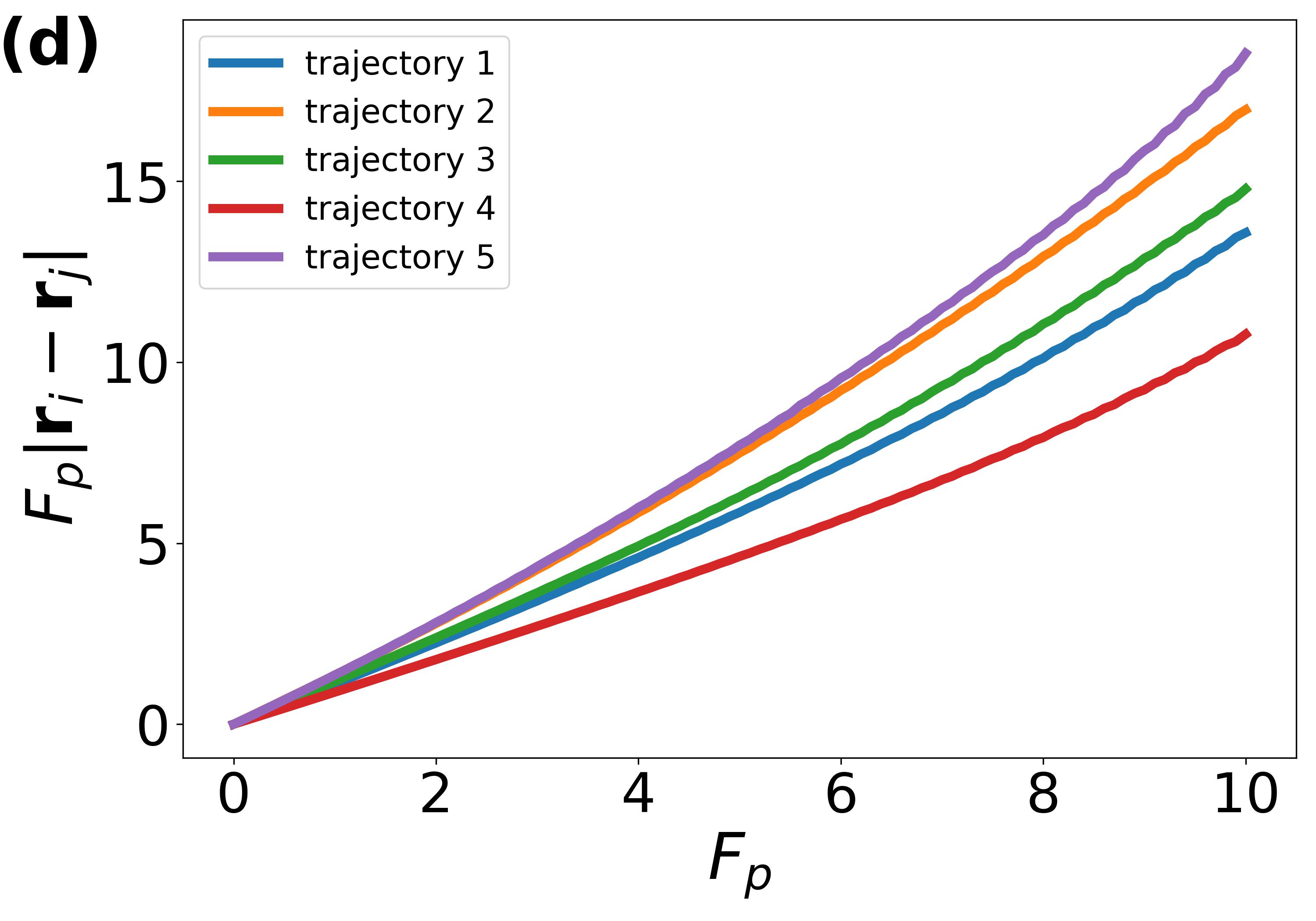}
\includegraphics[width=0.66\columnwidth]{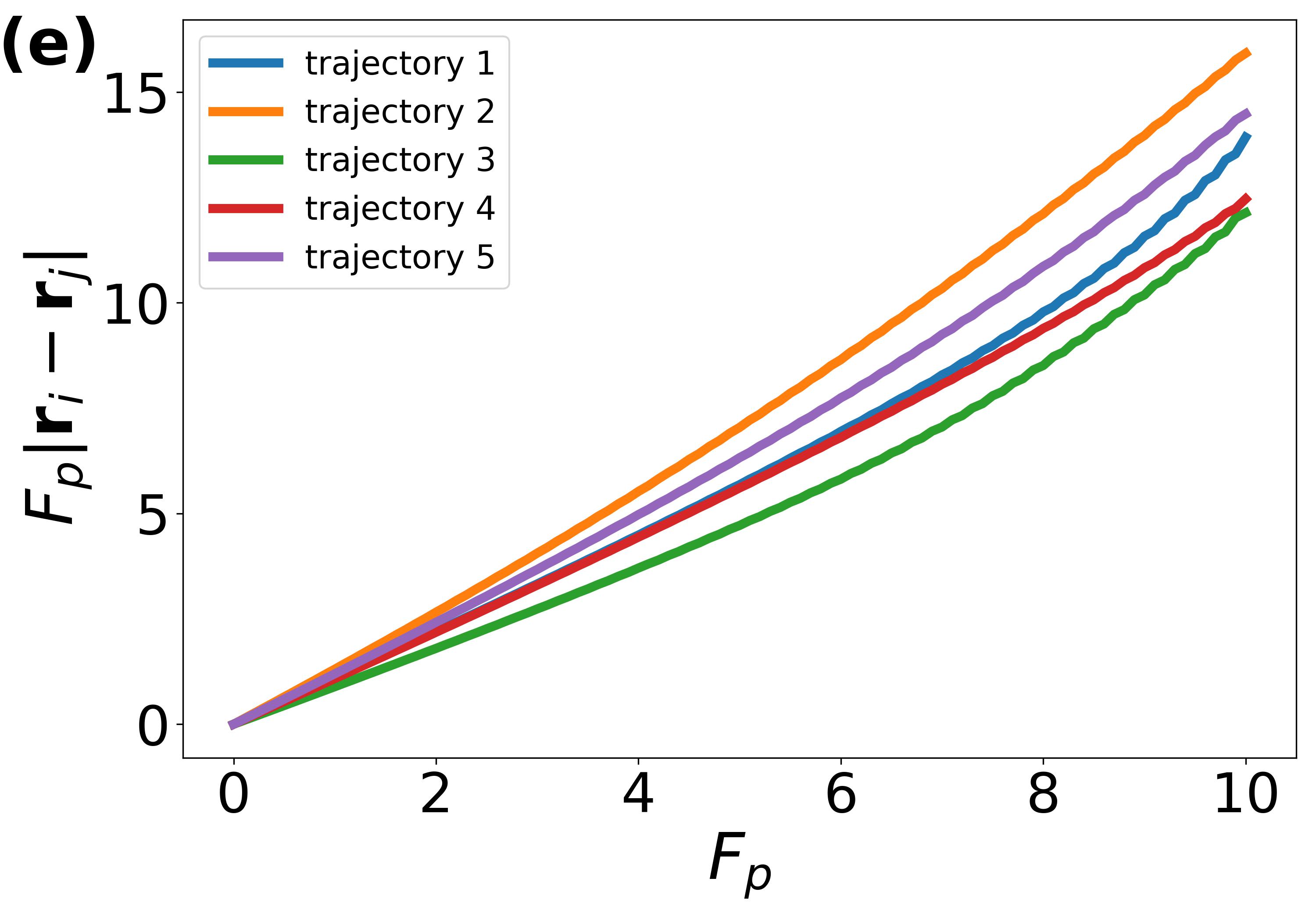}
\includegraphics[width=0.66\columnwidth]{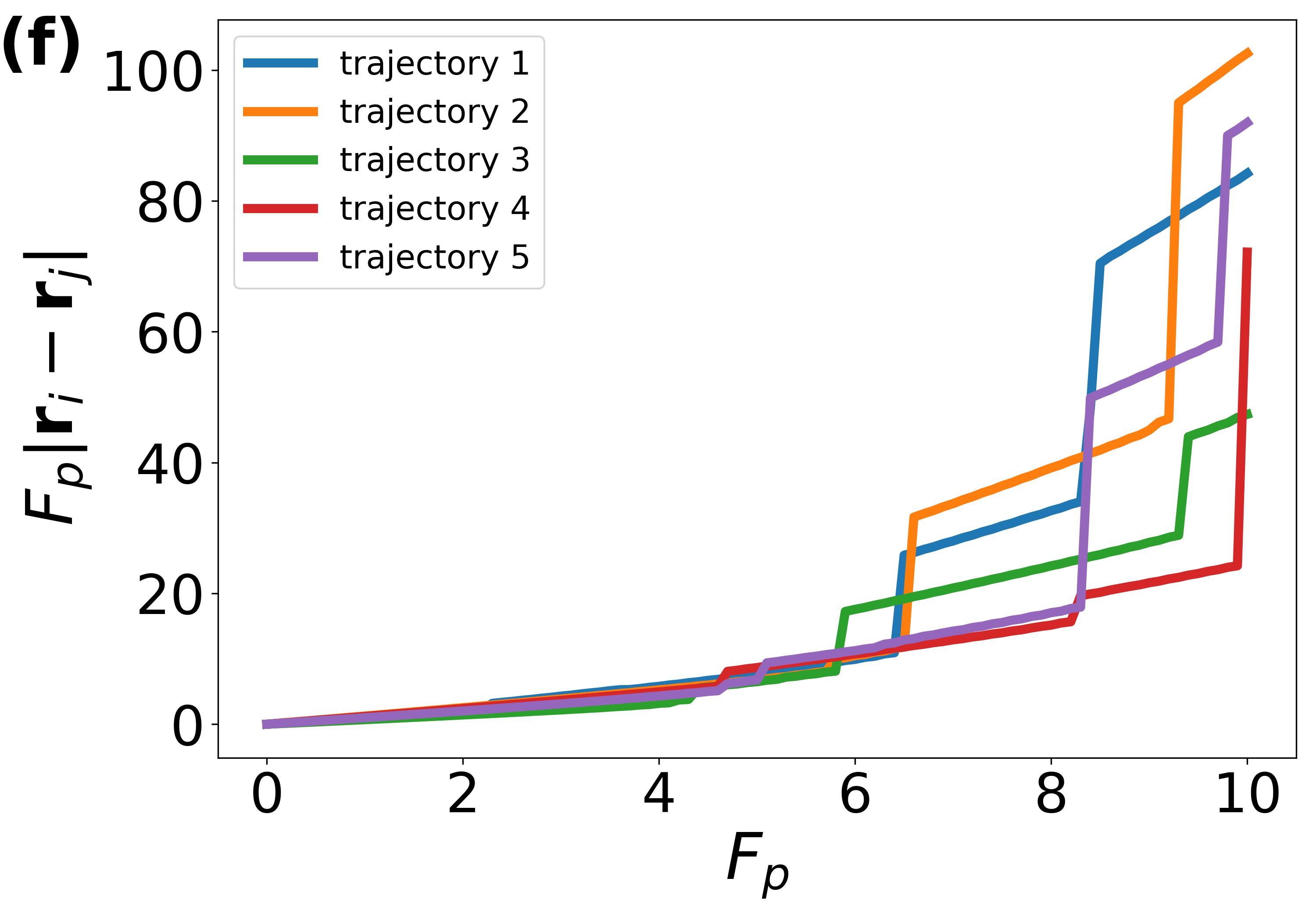}
\includegraphics[width=0.66\columnwidth]{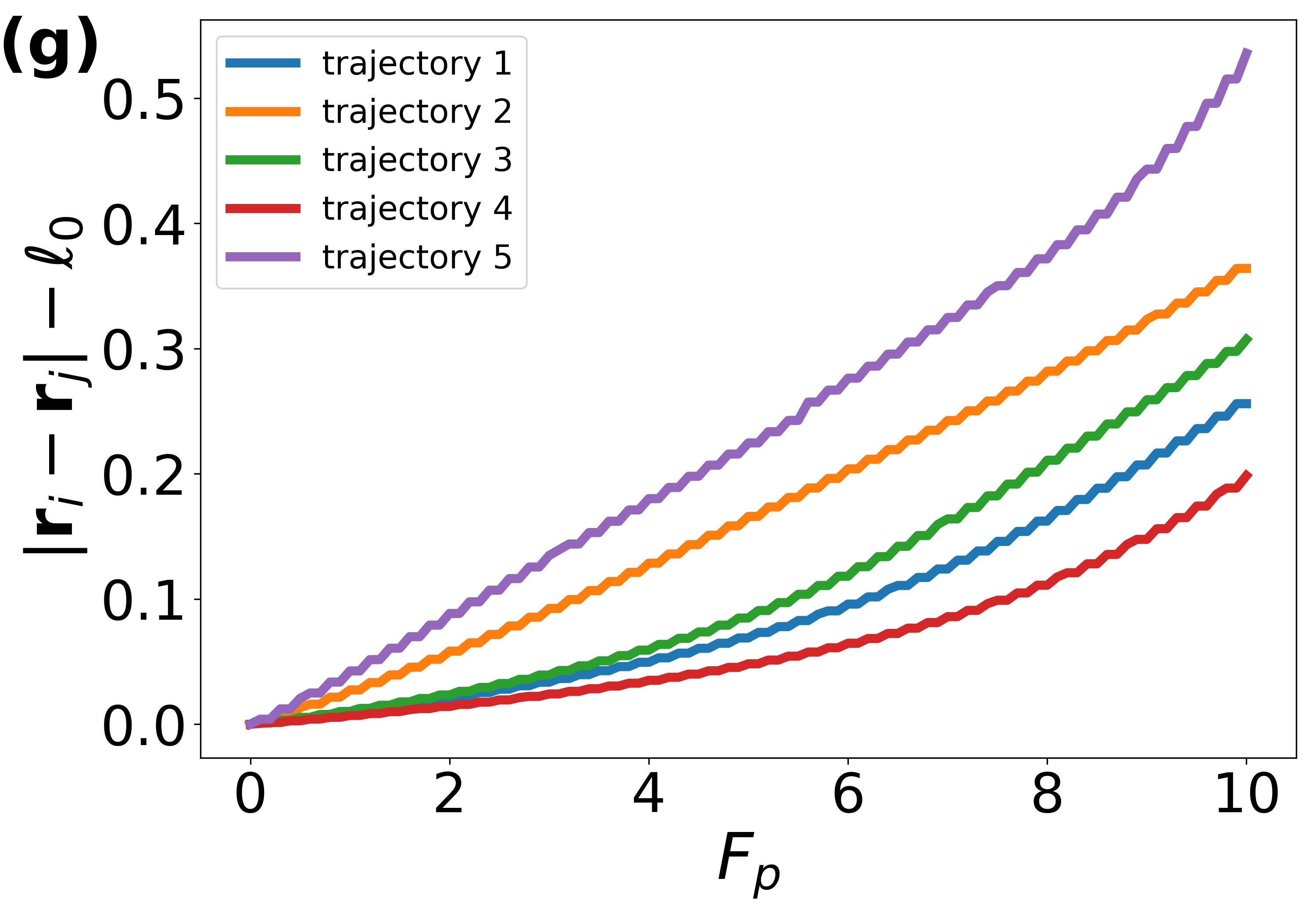}
\includegraphics[width=0.66\columnwidth]{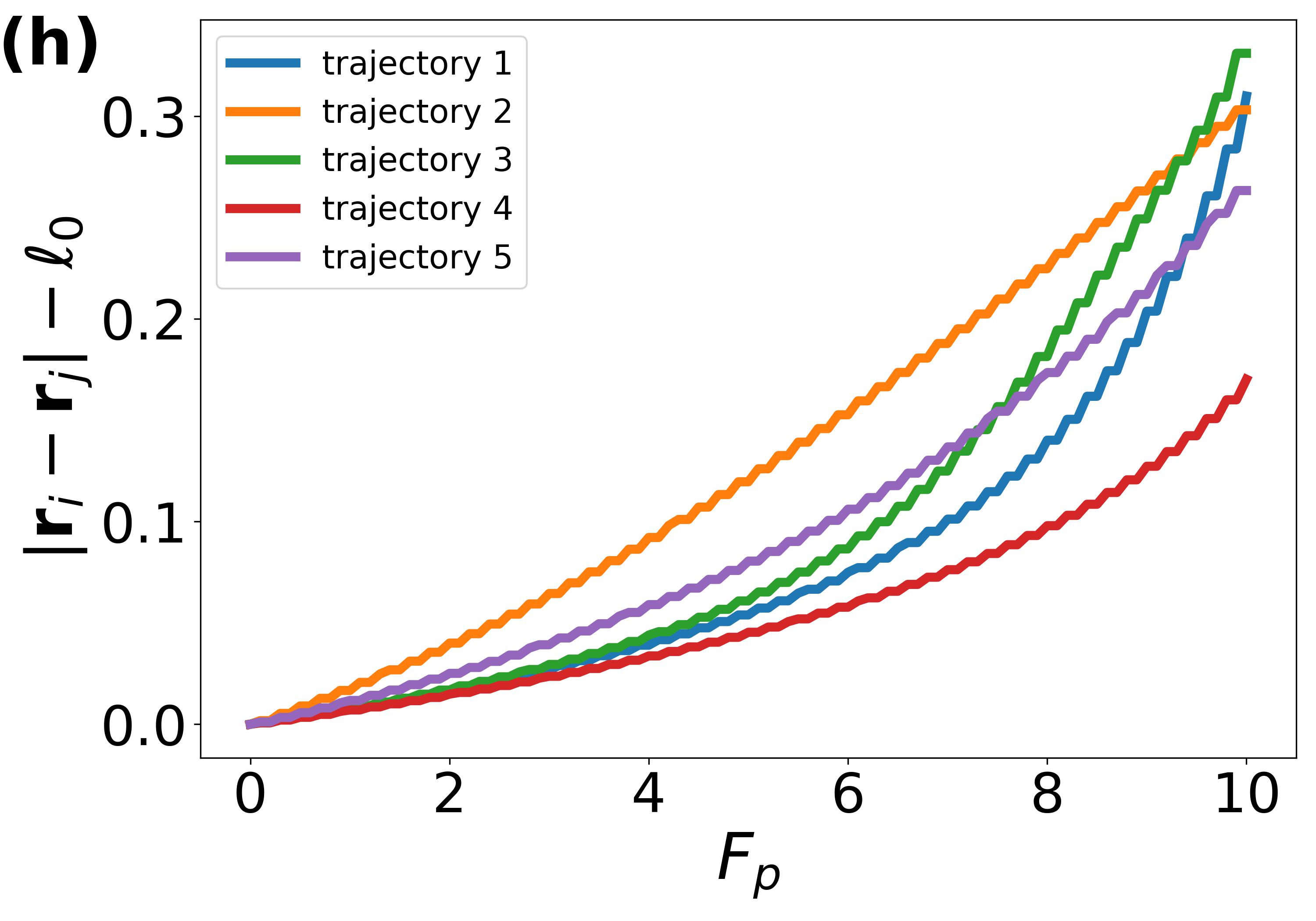}
\includegraphics[width=0.66\columnwidth]{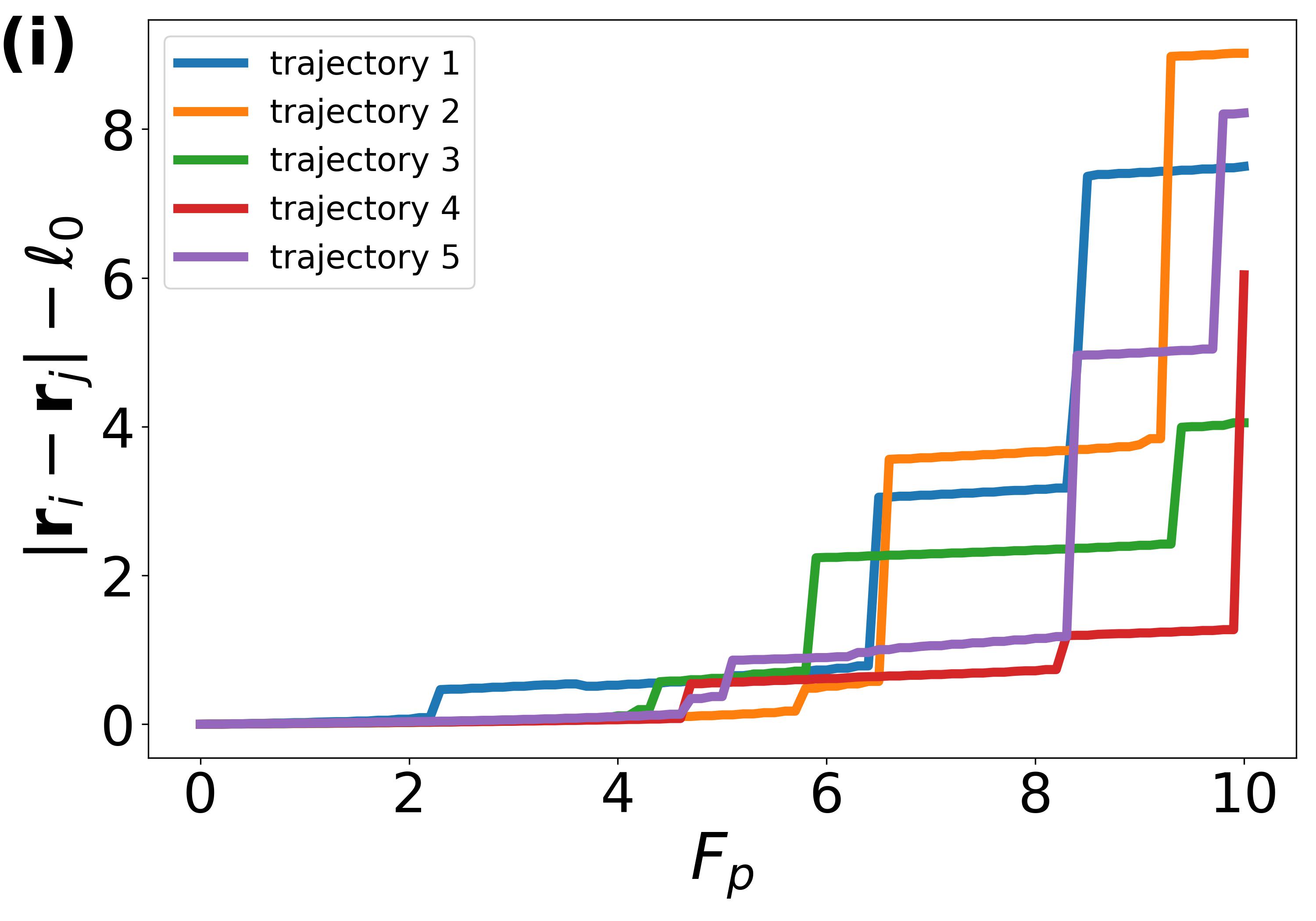}
\caption{(a--c) Potential energy $U/N$ as a function of the externally imposed
pinching force $F_{\rm p}$ for five pinching trajectories obtained with
the force-controlled protocol at $T_{\rm ini}=0.25$ (a), $0.35$ (b),
and $1.0$ (c).
(d--f) External pinching energy
$F_{\rm p}|{\bf r}_i-{\bf r}_j|$ as a function of $F_{\rm p}$ for
five pinching trajectories at $T_{\rm ini}=0.25$ (d), $0.35$ (e),
and $1.0$ (f).
(g--i) Corresponding actual separation increment
$|{\bf r}_i-{\bf r}_j|-\ell_0$ as a function of $F_{\rm p}$ for
$T_{\rm ini}=0.25$ (g), $0.35$ (h), and $1.0$ (i).
}
\label{fig:individual_force_control}
\end{figure*}

Figure~\ref{fig:individual_force_control} shows five individual
pinching trajectories obtained using the force-controlled protocol
for stable (a,d,g), moderately annealed (b,e,h), and poorly annealed (c,f,i) glasses, respectively.

For stable glasses, the potential energy $U/N$ (a), the
external-force contribution $F_{\rm p}|{\bf r}_i-{\bf r}_j|$
(d), and the actual extension
$|{\bf r}_i-{\bf r}_j|-\ell_0$ (g) all vary smoothly up to
$F_{\rm p}=10$. Over this force range, the actual extension remains
relatively modest and is comparable to that observed in the elastic
regime of the length-controlled protocol. If $F_{\rm p}$ is increased
further, however, the separation between the two pinched particles
undergoes a very large jump and can exceed the typical nearest-neighbor
distance by a substantial amount. We therefore restrict our analysis
to $F_{\rm p}\leq 10$ in the present study.
Moderately annealed glasses exhibit essentially the same behavior
within this range of $F_{\rm p}$. 

In contrast, poorly annealed glasses
show multiple discontinuities in the potential energy $U/N$ (c), indicating plastic rearrangements. Discontinuities are also visible in the external-force contribution
$F_{\rm p}|{\bf r}_i-{\bf r}_j|$ (f) and in the actual
extension $|{\bf r}_i-{\bf r}_j|-\ell_0$ (i), reflecting
plastic rearrangements in the vicinity of the pinched particles.
For relatively
small forces, approximately $F_{\rm p}\lesssim 5$, the extension
remains below unity, indicating that the two pinched particles remain
within a distance comparable to the local cage size. For
$F_{\rm p}\gtrsim 5$, however, their separation exhibits large,
discontinuous jumps and can extend beyond the nearest-neighbor scale.
This behavior suggests that the surrounding matrix of poorly annealed
glasses is mechanically softer and more susceptible to plastic
rearrangements than that of stable and moderately annealed glasses~\cite{richard2020predicting}.

\begin{figure*}
    \includegraphics[width=0.6\columnwidth]{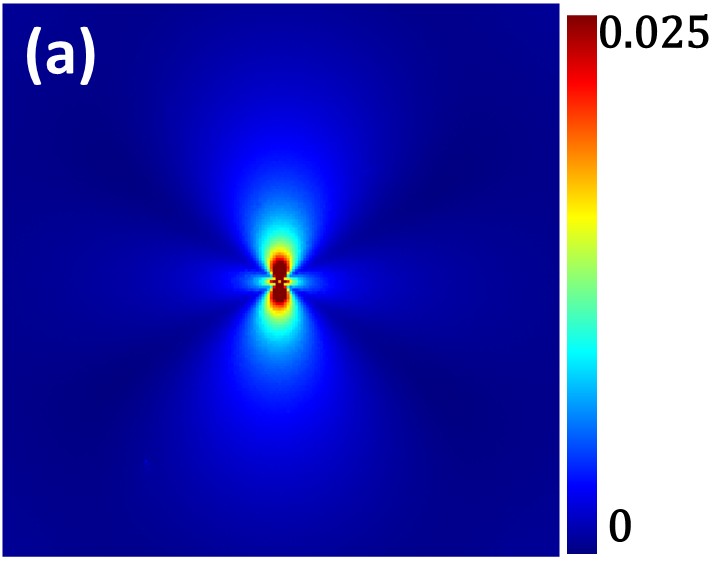}
    \includegraphics[width=0.6\columnwidth]{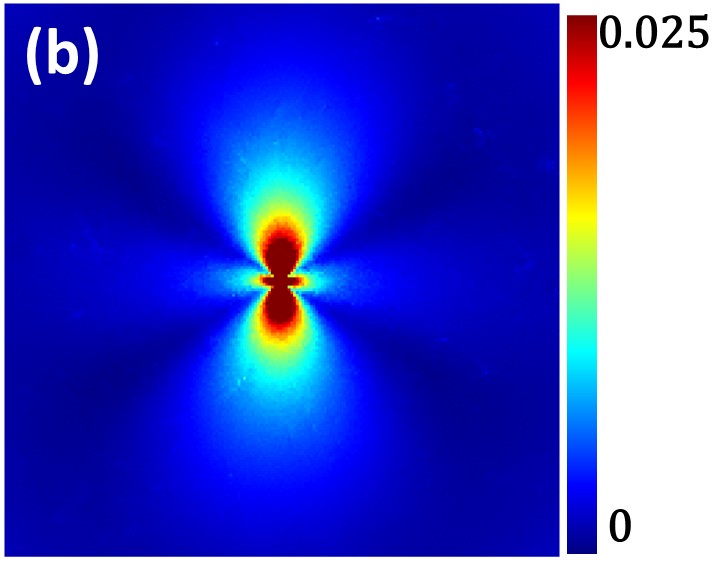}
    \includegraphics[width=0.6\columnwidth]{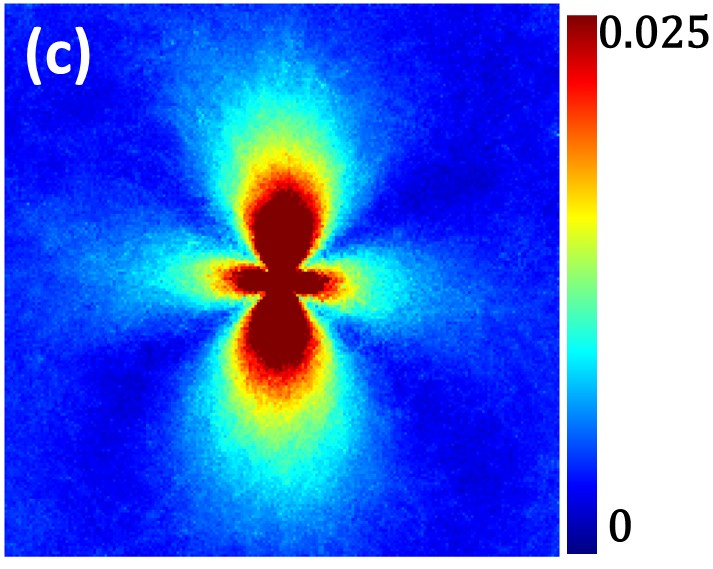}
    \includegraphics[width=0.6\columnwidth]{fig/TheoryEshelbyTp025.jpg}
    \includegraphics[width=0.6\columnwidth]{fig/TheoryEshelbyTp035.jpg}
    \includegraphics[width=0.6\columnwidth]{fig/TheoryEshelbyTp10.jpg}
\caption{(a--c) Magnitude of the displacement field induced by athermal quasistatic pinching using the force-controlled protocol at $F_{\rm p}=10$ for (a) stable glasses, $T_{\rm ini}=0.25$, (b) moderately annealed glasses, $T_{\rm ini}=0.35$, and (c) poorly annealed glasses, $T_{\rm ini}=1.0$. The results were obtained for $N=64000$ and averaged over $1000$ independent pinching realizations for (a), (b) and $5000$ for (c).
(d--f) Corresponding predictions from linear elasticity theory for (d) stable, (e) moderately annealed, and (f) poorly annealed glasses. The Poisson ratio used in each theoretical calculation was measured independently from athermal quasistatic compression and shear simulations, and its value is indicated in white.}
\label{fig:displacement_force_control}
\end{figure*}

We next examine the angular dependence of the averaged displacement
magnitude in Fig.~\ref{fig:displacement_force_control} and compare the
simulation results with the predictions of linear elasticity, using the
numerically measured Poisson ratio as an input parameter. We observe two
pairs of lobes: two larger lobes oriented parallel to the pinching
direction and two smaller lobes oriented perpendicular to it. This
anisotropic structure is consistent with that observed for the
length-controlled protocol in the elastic regime and with the prediction
of linear elasticity.

\begin{figure*}
\includegraphics[width=0.66\columnwidth]{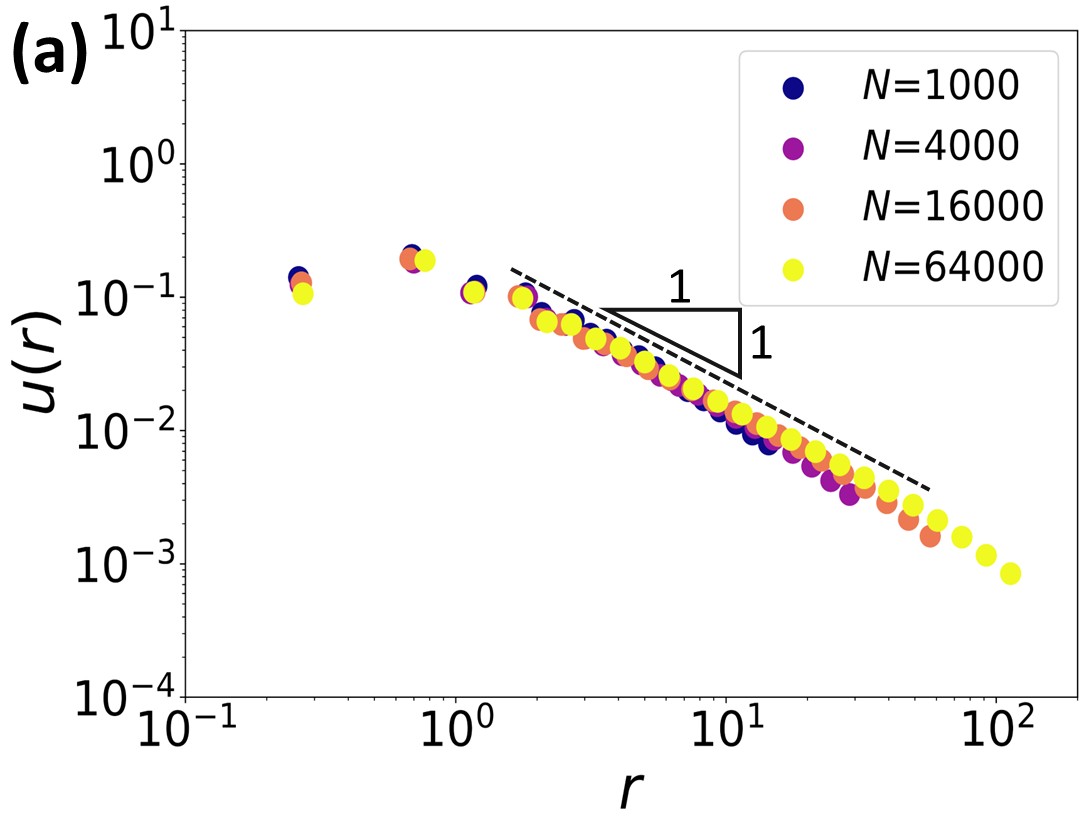}
\includegraphics[width=0.66\columnwidth]{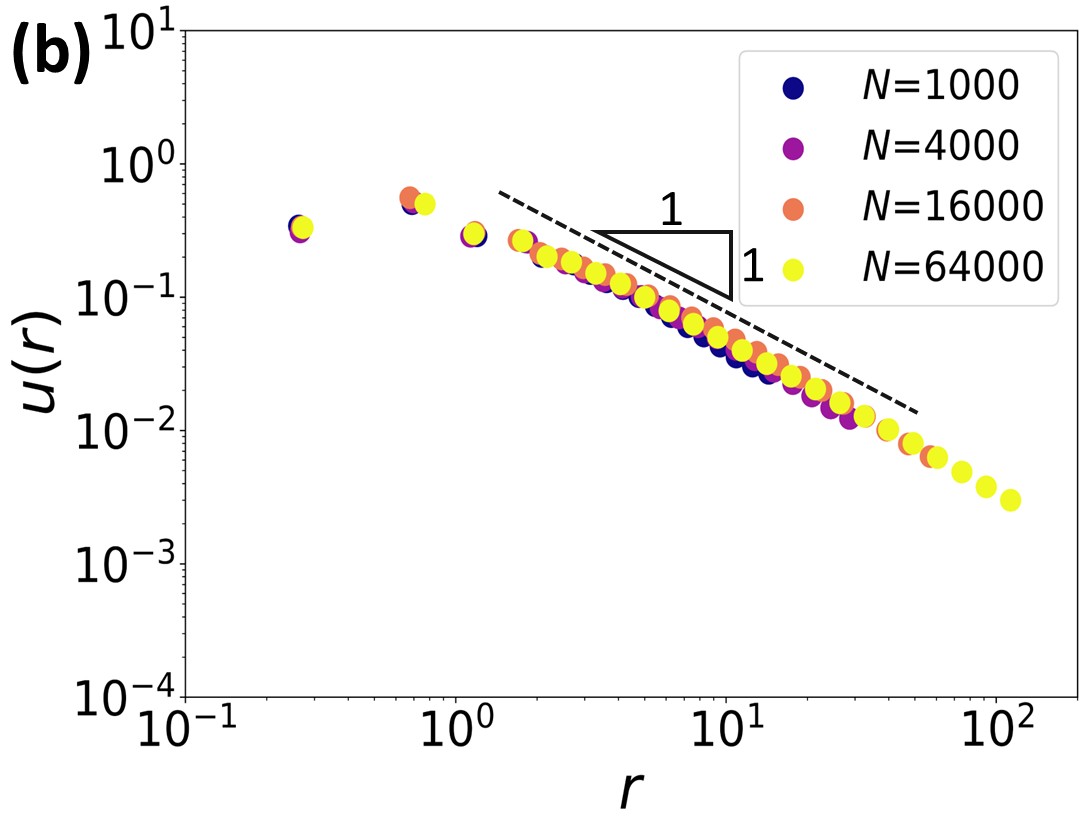}
\includegraphics[width=0.66\columnwidth]{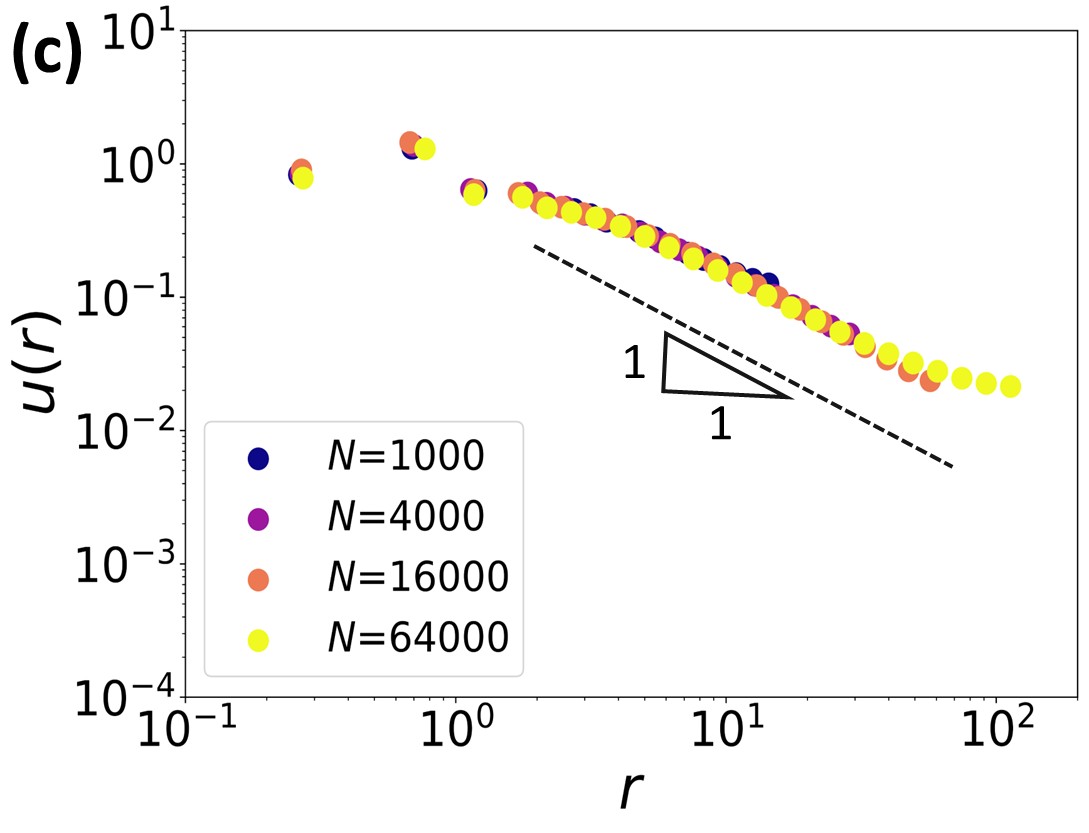}
\includegraphics[width=0.66\columnwidth]{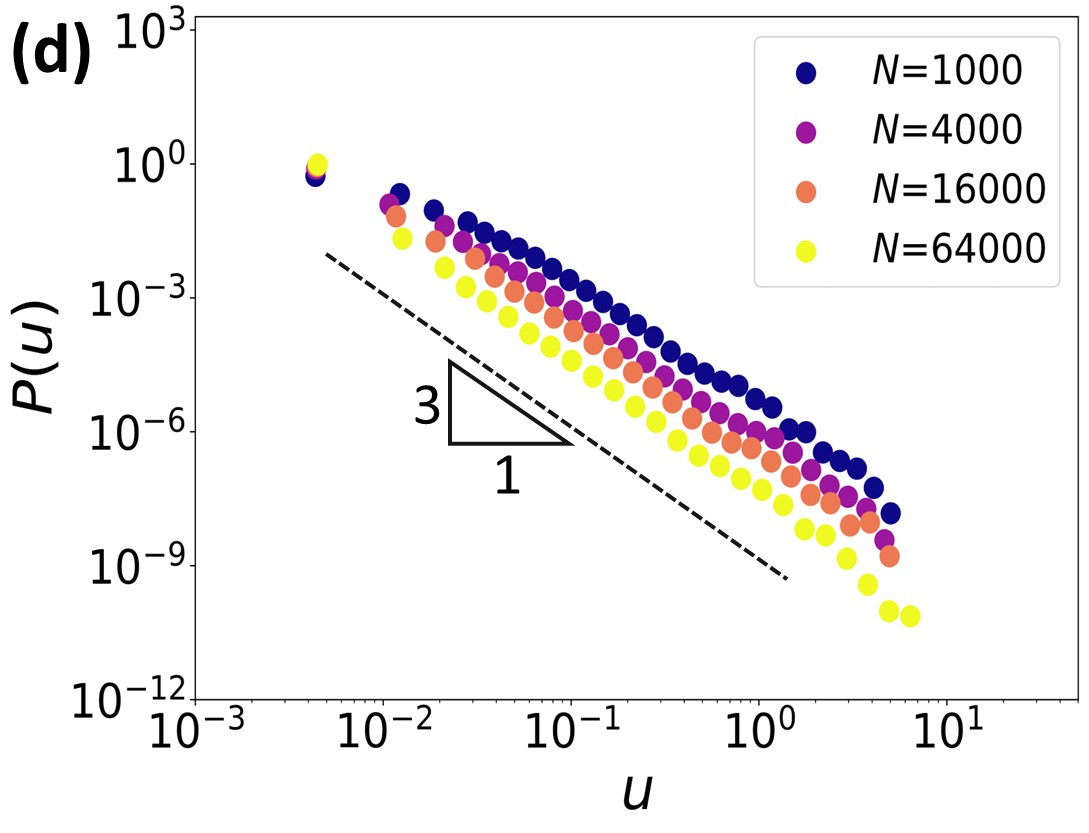}
\includegraphics[width=0.66\columnwidth]{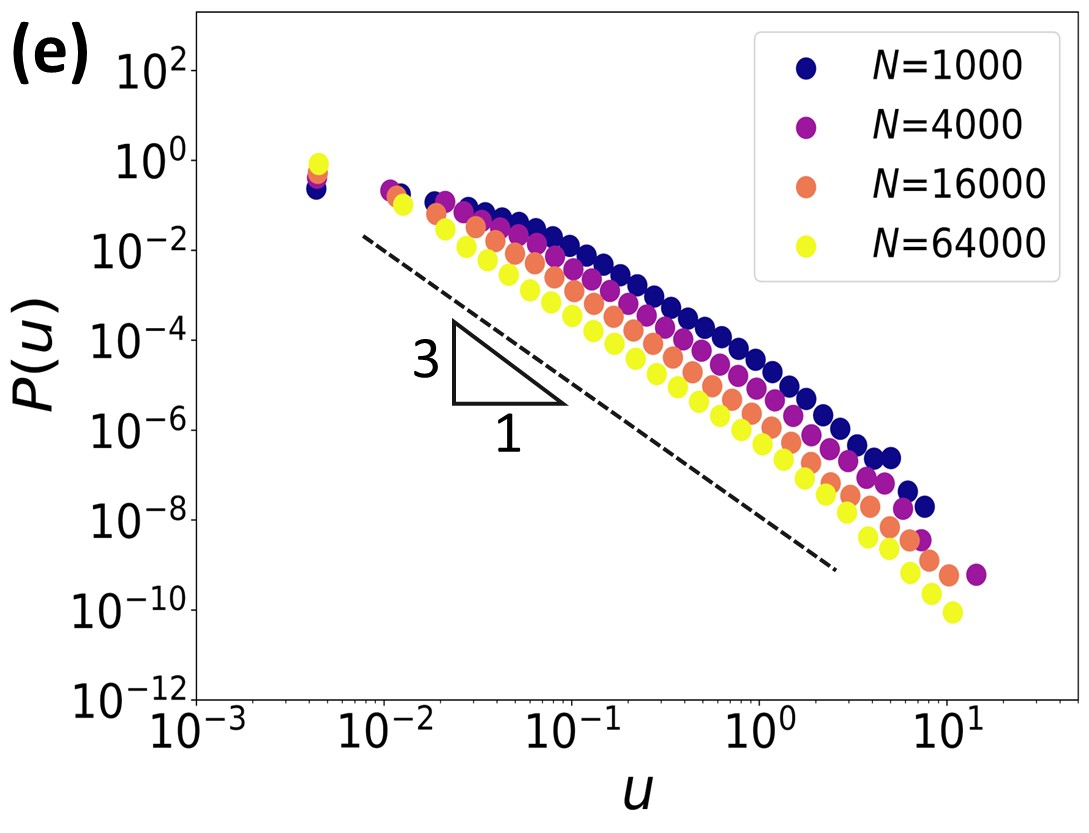}
\includegraphics[width=0.66\columnwidth]{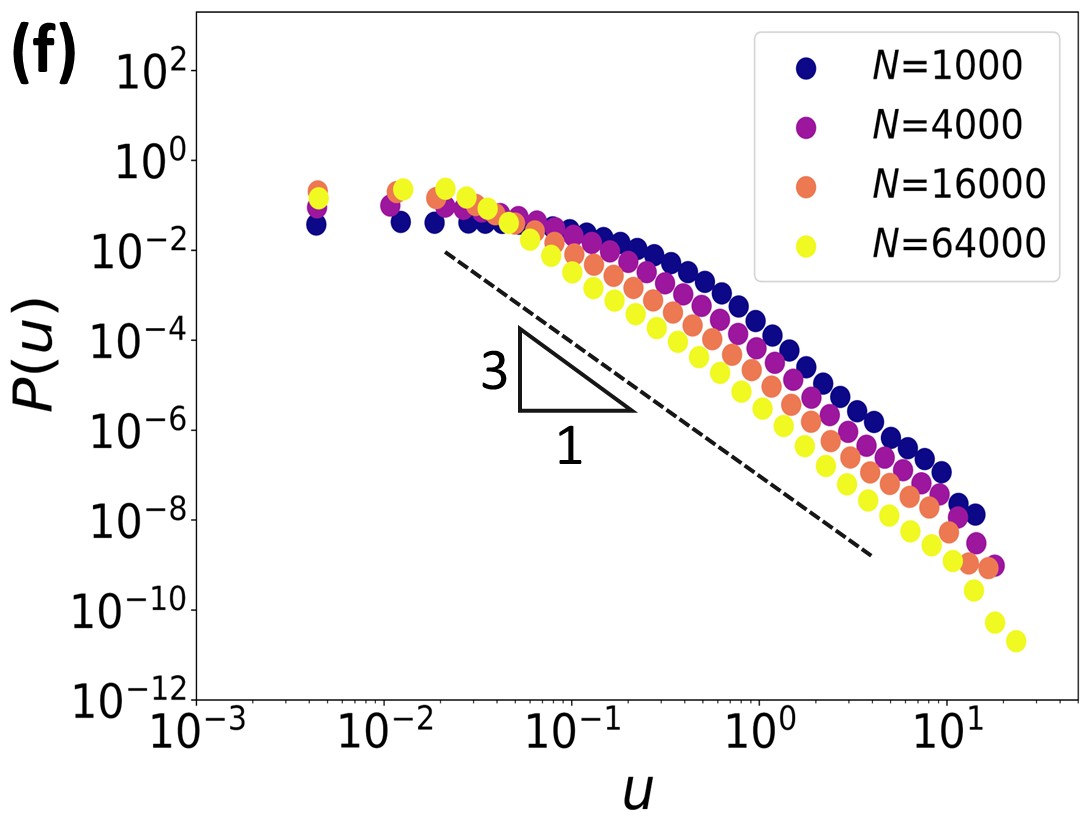}
\caption{(a--c) Mean radial decay of the displacement magnitude, $u(r)$, induced by pinching using the force-controlled protocol at
$F_{\rm p}=xx$ for (a) stable glasses, (b) moderately annealed
glasses, and (c) poorly annealed glasses, for system sizes
$N=1000$, $4000$, $16000$, and $64000$. The dashed line indicates
the linear-elasticity prediction, $u(r)\sim r^{-1}$.
(d--f) Corresponding probability distributions of the displacement
magnitude, $P(u)$, for (d) stable glasses, (e) moderately annealed
glasses, and (f) poorly annealed glasses, for the same system sizes.
The dashed line indicates the predicted scaling, $P(u)\sim u^{-3}$.}
\label{fig:decays_force_control}
\end{figure*}

\begin{figure*}[!t]
\includegraphics[width=0.6\columnwidth]{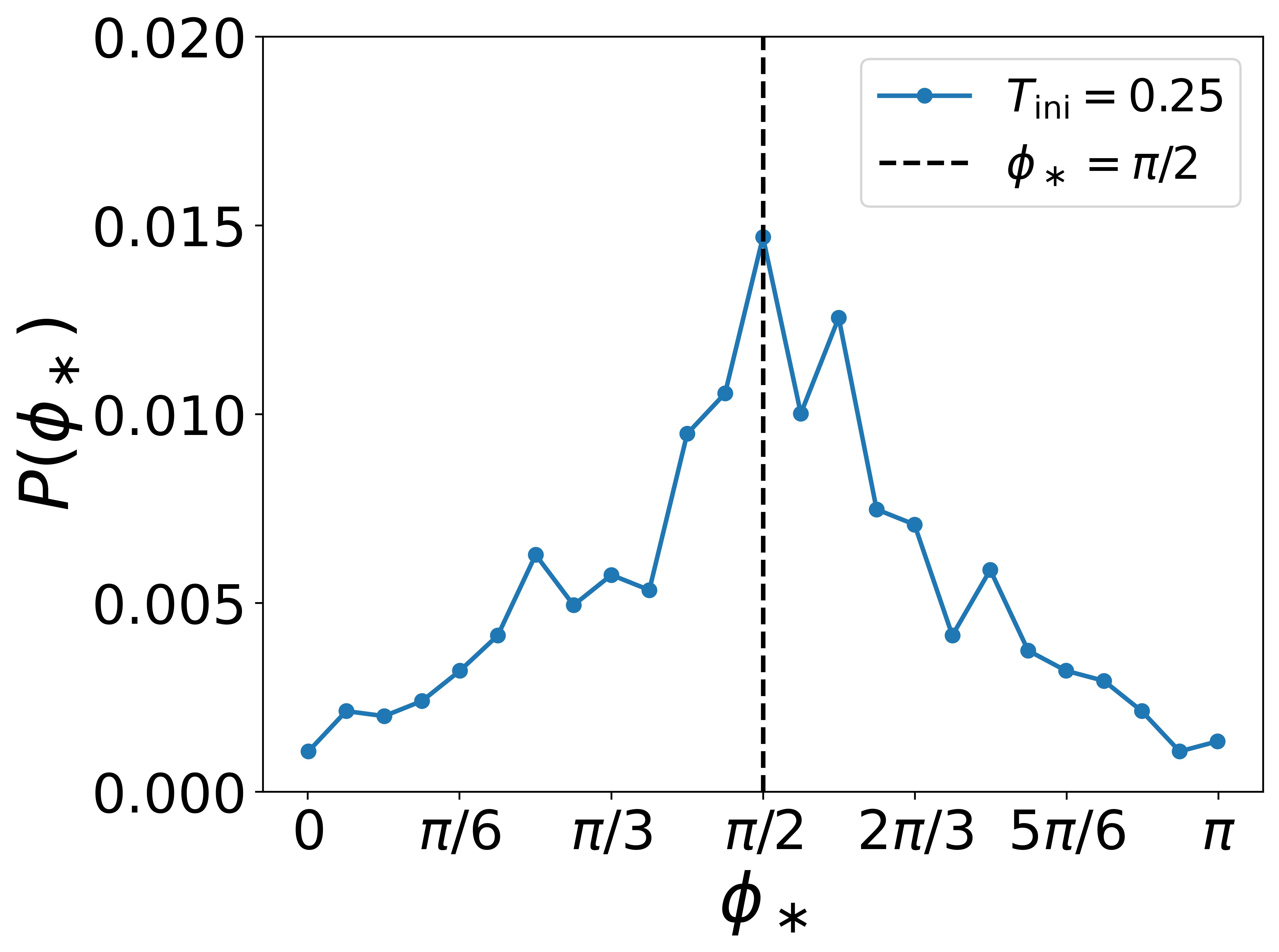}
\includegraphics[width=0.6\columnwidth]{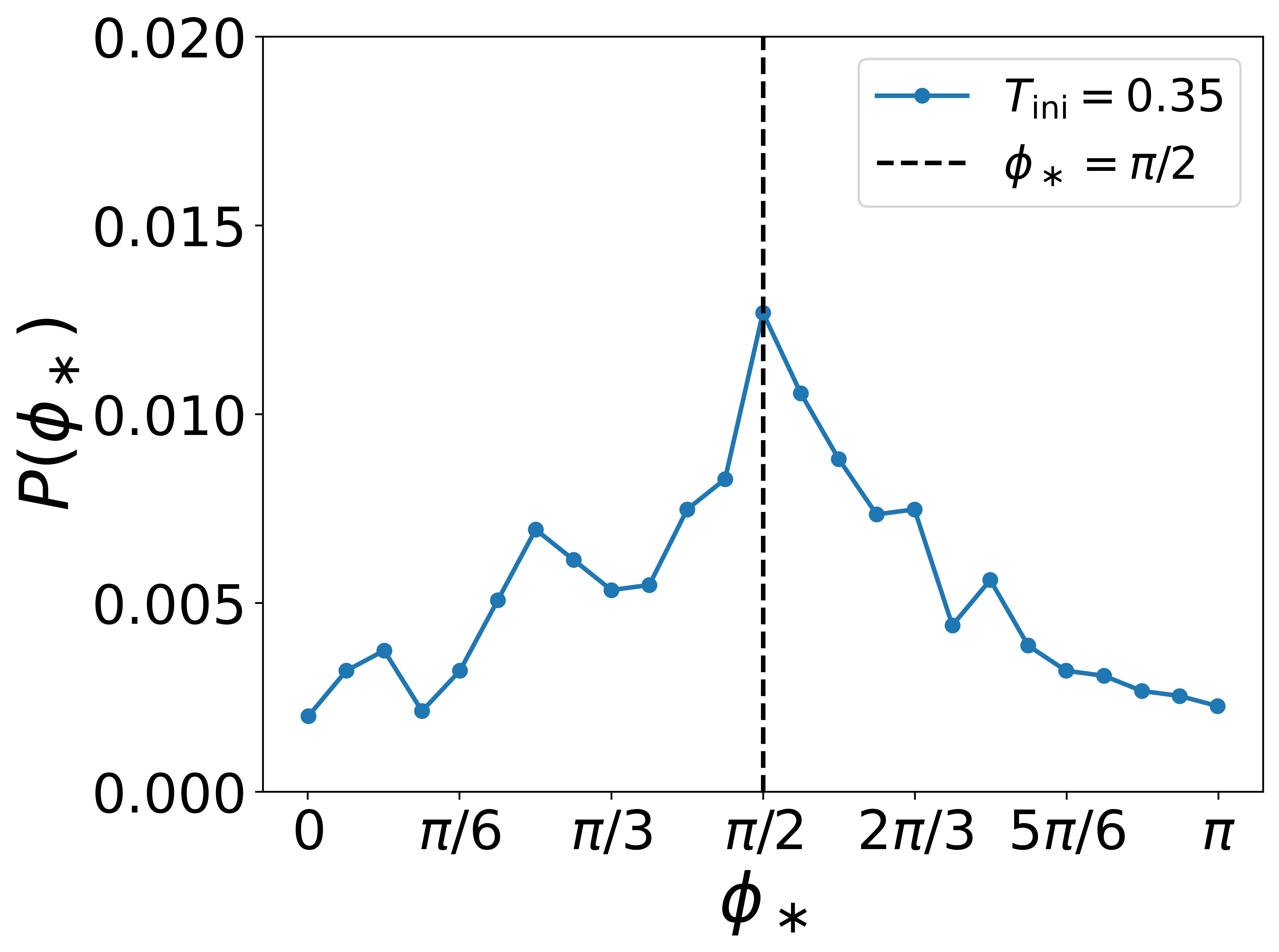}
\includegraphics[width=0.6\columnwidth]{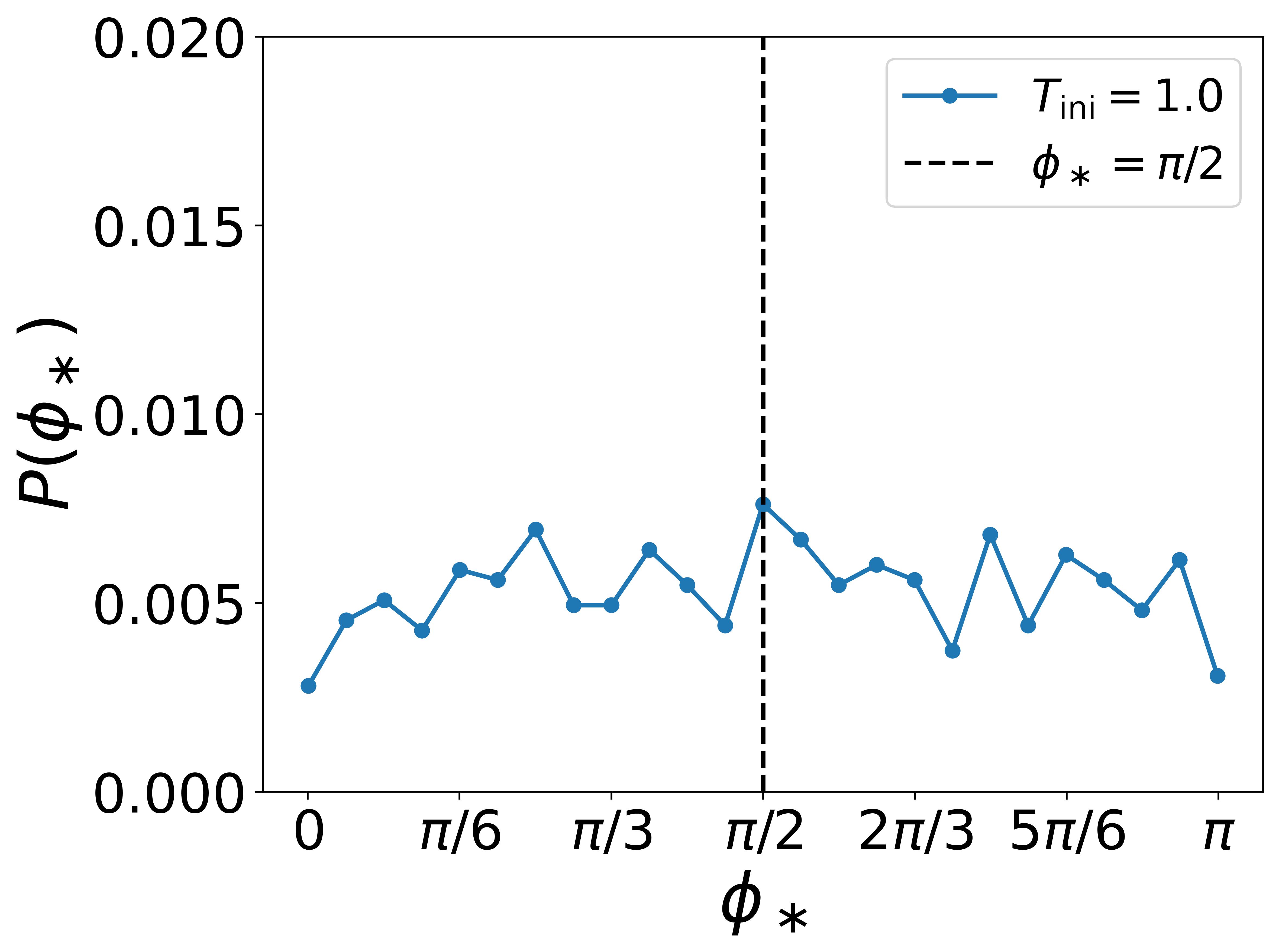}
\caption{(a--c) Probability distributions of the estimated principal-axis angle $\phi_*$ for (a) stable, (b) moderately annealed, and (c) poorly annealed glasses, respectively. The vertical dashed line indicates $\phi_*=\pi/2$.}
\label{fig:principal_direction}
\end{figure*}

Finally, in Fig.~\ref{fig:decays_force_control}, we examine the mean
radial decay of the displacement magnitude, $u(r)$, and the probability
distribution of the displacement magnitude, $P(u)$, at $F_{\rm p}=10$
for several system sizes in stable, moderately annealed, and poorly
annealed glasses. In all cases, we observe the expected scaling behaviors,
$u(r)\sim r^{-1}$ and $P(u)\sim u^{-3}$. These results are consistent
with those obtained with the length-controlled protocol in the elastic
regime and with the predictions of linear elasticity.

\section{Determination of the principal axis in the plastic regime}
\label{sec:principal_axis}

We determine the principal axis directly from the
displacement field measured at $\Delta \ell = 2.0$, rather than from
the initial pinching direction. The procedure is as follows.
We first divide the system into a circular core region containing the
pinched particles, with radius $R_{\rm cut}=12$, and the surrounding
matrix, denoted by $\mathcal{M}$. We confirm that the results don't depend significantly above $R_{\rm cut}=12$. To avoid the strongly nonlinear
displacements in the immediate vicinity of the pinched particles, only
particles in the surrounding matrix, $i\in\mathcal{M}$, are used to
determine the principal axis.

For each particle, we define the displacement vector $\Delta \mathbf{r}_i=
\mathbf{r}_i(\Delta\ell)-\mathbf{r}_i(0)$.
We then introduce the unit vector $\mathbf{e}_{\phi}=(\cos\phi,\sin\phi)$
and measure the total squared projection of the displacement field
along $\mathbf{e}_{\phi}$,
\begin{align}
    \chi(\phi)
    =
    \sum_{i\in\mathcal{M}}
    \left|
        \mathbf{e}_{\phi}^{T} \, \Delta \mathbf{r}_i
    \right|^{2} .
    \label{eq:chi_principal}
\end{align}
The principal-axis angle $\phi_{*}$ is then defined as
\begin{align}
    \phi_{*}
    =
    \underset{\phi}{\operatorname{arg\,max}}\,
    \chi(\phi).
    \label{eq:phi_principal}
\end{align}
Since $\chi(\phi+\pi)=\chi(\phi)$, it is sufficient to consider
$0\leq\phi<\pi$.

Figure~\ref{fig:principal_direction} shows the probability distribution of the estimated principal-axis angle $\phi_*$ for stable glasses (a), moderately annealed glasses (b), and poorly annealed glasses (c), respectively.
For stable glasses, the most probable value of $\phi_*$ is located around $\pi/2$, which corresponds to the pinching direction in the initial configuration, and $P(\phi_*)$ decays approximately symmetrically away from $\pi/2$. A qualitatively similar trend is observed for moderately annealed glasses, although the peak around $\pi/2$ is somewhat smaller. For poorly annealed glasses, by contrast, $P(\phi_*)$ is essentially flat within the resolution of our determination method. This behavior reflects the fact that, in poorly annealed glasses, inelastic responses and additional plastic events induced by pinching make the overall displacement field more complex, so that a single principal direction becomes less well defined.

\bibliography{references}

\end{document}